\documentclass[%
reprint,
superscriptaddress,
nofootinbib,
 amsmath,amssymb,
 aps,
]{revtex4-2}

\usepackage[english]{babel}
\usepackage[svgnames]{xcolor}
\usepackage[linktocpage, colorlinks, citecolor=blue, linkcolor=blue, urlcolor=blue]{hyperref} 
\usepackage{enumerate}  
\usepackage{float}
\usepackage{calc}
\usepackage{dutchcal}
\usepackage{revsymb}
\usepackage{comment}
\usepackage{upgreek}
\usepackage{IEEEtrantools}
\usepackage{MyCommands}
\usepackage[normalem]{ulem}
\usepackage{csquotes}

\newcommand{\orcid}[1]{\href{https://orcid.org/#1}{\includegraphics[width=8pt]{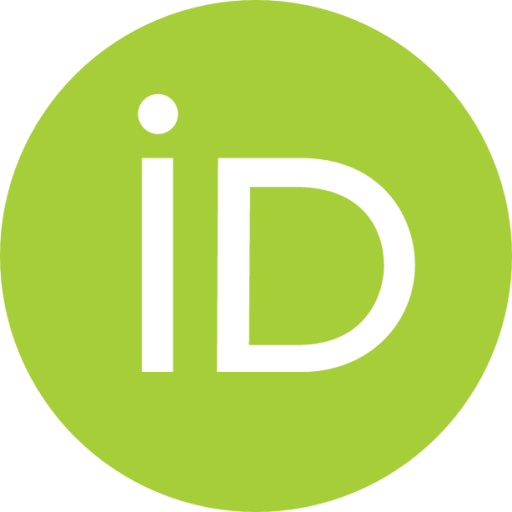}}}

\DeclareSymbolFont{cmbrightop}{OT1}{cmbr}{m}{n}
\DeclareMathSymbol{\sfPsi}{\mathalpha}{cmbrightop}{9}

\usepackage{amsfonts, amssymb, amsthm, mathtools, mathrsfs}
\usepackage{physics}

\usepackage{subcaption}
\usepackage{graphicx}
\usepackage{dcolumn}
\usepackage{bm}

\begin{document}

\title{Constraining superluminal Einstein-\AE{}ther gravity through gravitational memory}

\author{Lavinia Heisenberg}
\affiliation{Institut f\"{u}r Theoretische Physik, Philosophenweg 16, 69120 Heidelberg, Germany}
\author{Benedetta Rosatello\orcid{0009-0001-1227-8080}}
\affiliation{Dipartimento di Fisica e Astronomia “Galileo Galilei”, Università di Padova, Via Marzolo 8, I-35131 Padova, Italy}
\affiliation{Institut f\"{u}r Theoretische Physik, Philosophenweg 16, 69120 Heidelberg, Germany}
\author{Guangzi Xu\orcid{0009-0002-1276-9012}}
\affiliation{Institute for Theoretical Physics, ETH Z\"{u}rich, Wolfgang-Pauli-Strasse 27, 8093, Z\"{u}rich, Switzerland}
\author{Jann Zosso\orcid{0000-0002-2671-7531}}
\email{jann.zosso@nbi.ku.dk}
\affiliation{Center of Gravity, Niels Bohr Institute, Blegdamsvej 17, 2100 Copenhagen, Denmark}
\affiliation{Albert Einstein Center, Institute for Theoretical Physics, University of Bern, Sidlerstrasse 5, 3012, Bern, Switzerland}

\date{\today}

\begin{abstract}

Every emission of radiation in gravity also includes a nonwavelike component that leaves a permanent change in proper distances of the spacetime it travels through. This phenomenon is known as gravitational displacement memory. Building up on a recently developed computation framework that harnesses Isaacson’s insights on a fundamental definition of gravitational waves, we compute the leading displacement memory formula in Einstein-\AE{}ther gravity. Our analysis represents the first direct calculation of gravitational memory in a metric theory with nontrivial asymptotic vector field value. We find that an emission of scalar and vector \ae{}ther waves at a propagation speed greater than the speed of tensor radiation features unprotected causal directions with a priori unbound memory build-up. Based on the results and the existing constraint of luminally propagating tensor waves, we conjecture a stringent exclusion of the superluminal parameter space of Einstein-\AE{}ther gravity.

\end{abstract}

\maketitle


\section{\label{sec:Intro}Introduction}
Some of the biggest open questions in physics are found within the realm of gravity, especially in the branches of cosmology and its embedding into quantum gravity. A better understanding and questioning of the foundational principles of modern gravity theories, therefore, naturally suggests itself, together with an exploration of the theory space beyond general relativity (GR). This provides the backbone of a search for signatures past the current horizon of knowledge. 

A particularly well motivated guiding principle is provided by the notion of \emph{metric theories of gravity} \cite{Dicke:1964pna,poisson2014gravity,papantonopoulos2014modifications,Heisenberg:2018vsk,Will:2018bme,YunesColemanMiller:2021lky,Zosso:2024xgy} (see also Eq.~\eqref{eq:MetricTheoryAction} below), which describe the principle part of gravity theories on a manifold that comply with the Einstein equivalence principle, while allowing for additional dynamical gravitational degrees of freedom (dofs). In particular, these theories assume that any additional gravitational fields ought to act exclusively within the gravitational sector, while matter remains universally and minimally coupled to a physical metric. At the classical level, such a requirement is decisive for the existence of an observer-independent notion of dynamical spacetime, as well as the definition of energy-momentum conserving matter \cite{Will:2018bme,Zosso:2024xgy}. Furthermore, and crucially for the present work, this also implies that the geodesic deviation equation at the basis of the interpretation of current gravitational wave (GW) observations retains its universal validity.
The structure of metric theories of gravity beyond GR naturally proposes the possibility of breaking with principles of symmetries by restricting their breakdown to the empirically less precisely explored gravity sector, by coupling their breakdown to the additional gravitational fields. In particular, although numerous observations strongly back the Lorentz symmetry assumption among the standard model fields \cite{Will:2005va}, the option of pure gravitational local Lorentz breaking may be explored, as we shall do in this work.

The possibility of local Lorentz breaking through the existence of a preferred rest frame may in fact arise from the quantum gravity vacuum and would provide a natural short-distance cutoff \cite{Mattingly_2005,Jacobson:2007veq}. In the light of such a quantum gravity motivation one might however question the metric theory perspective advocated above, since a strict distinction between the gravity and matter sectors is expected to be challenged by general quantum corrections (see also Refs.~\cite{Padmanabhan:2004xk,papantonopoulos2014modifications}). While it is possible to conceive a general protection of the matter sector from such Lorentz-violating quantum corrections through an assumed hierarchy of scales \cite{Pospelov:2010mp}, in this work we will not address such questions and simply study the implication of gravitational local Lorentz breaking detached from any UV motivations. We will therefore be concerned with studying a well defined covariant classical gravity theory, in which a fixed preferred frame is imposed through local conditions. From this point of view, the motivation for a metric theory perspective is on the same footing as the assumption of validity of the Einstein equivalence principle in GR.

A simple and well-studied possibility of such a local Lorentz-violating classical gravity theory is to consider a dynamical gravitational vector field on the manifold that carries a nonlinear representation of the local Lorentz group, whose nontrivial value breaks the symmetry down to a rotational subgroup in all states \cite{Jacobson:2000xp}. Such a vector field effectively acts like a fluid in the gravitational sector that permeates everywhere, as it defines a congruence of timelike curves filling all of spacetime. The corresponding metric theory, which arises as the most general action of a spacetime metric and the vector field up to two powers of derivatives, is known as Einstein-\AE{}ther (E\AE{}) gravity \cite{Jacobson:2000xp,Jacobson:2004ts,Eling:2005zq,Foster:2006az,Jacobson:2007veq,Bonvin:2007ap,Yagi:2013ava,Will:2018bme}. Through the additional vector field on the manifold, E\AE{} theory describes 3 additional gravitational degrees of freedom distributed between the vector and scalar sector of the spatial rotational subgroup, which all may propagate at different but fixed nondispersive velocities. Due to local Lorentz violation, these speeds are not bound by the luminal propagation value of light and may, in principle, take any positive value. 

The emission of corresponding scalar and vector waves from astrophysical systems will inevitably source an additional component in the radiation. Indeed, a purely linear treatment of gravitational waves inherently overlooks an additional leading contribution that leaves a permanent mark in the form of geodesic displacement onto the asymptotic spacetime, which
corresponds to the phenomenon of \emph{gravitational displacement memory} \cite{Christodoulou:1991cr,Blanchet:1992br,FrauendienerJ,Ludvigsen:1989cr,Thorne:1992sdb,PhysRevD.44.R2945,Favata:2009ii,Favata:2010zu,Strominger:2014pwa,Bieri:2015yia,Garfinkle:2022dnm,Heisenberg:2023prj}. To be precise, gravitational memory is not a purely nonlinear effect, but is sourced by any unbound energy-momentum in an otherwise localized system. In fact, the effect was discovered within GR in a linearized context of unbound asymptotic matter \cite{Zeldovich:1974gvh,Turner:1977gvh,Braginsky:1985vlg, Braginsky:1987gvh}. However, the gravitational memory sourced by the energy-momentum carried by asymptotic gravitational waves themselves, generally dominates over its linear counterpart and represents an effect that is intrinsically present in all emissions of radiation.

While the effect is well studied within GR \cite{Christodoulou:1991cr,Ludvigsen:1989cr,Blanchet:1992br,FrauendienerJ,Thorne:1992sdb,PhysRevD.44.R2945,Favata:2008yd,Favata:2009ii,Barnich:2009se,Favata:2010zu,Bieri:2013ada,Ashtekar:2014zsa,Pasterski:2015tva,Nichols:2017rqr,Nichols:2018qac,Compere:2019gft,DAmbrosio:2022clk,Garfinkle:2022dnm}, and was also already described in various scalar-tensor metric theories \cite{Heisenberg:2023prj,Zosso:2024xgy,lang_compact_2014,lang_compact_2015,du_gravitational_2016,koyama_testing_2020,hou_gravitational_2021,tahura_brans-dicke_2021,tahura_gravitational-wave_2021,hou_conserved_2021,hou_gravitational_2021_2,hou_gravitational_2022,Hou:2021bxz,Heisenberg:2024cjk} the consequences of Lorentz-violating propagation speeds in the emitted radiation that sources gravitational memory were not yet explored. This work will fill this gap by computing for the first time the gravitational displacement memory within E\AE{} gravity. Based on these theoretical results, we conjecture severe constraints on any superluminal propagation of nondispersive energy-momentum-carrying waves.

Our methods build up on a new formalism for understanding and computing gravitational wave memory in general metric theories of gravity developed in Refs.~\cite{Zosso:2024xgy,Heisenberg:2023prj,Heisenberg:2024cjk} that is based on an \emph{Isaacson approach} of defining gravitational waves on arbitrary spacetime backgrounds \cite{Isaacson_PhysRev.166.1263,Isaacson_PhysRev.166.1272}. We generalize this framework for nontrivial background vector field values within the asymptotically flat assumption. Moreover, we explore the simplifying possibility of computing the memory equation through a reduced effective second-order action that solely depends on the relevant dynamical degrees of freedom of the theory. While the asymptotic structure of E\AE{} gravity with corresponding possibility of supporting gravitational memory was already explored in Refs.~\cite{Hou:2023pfz,Hou:2024exz}, the linearized treatment\footnote{Linearity was imposed to facilitate the choice of suitable boundary conditions for the dynamical fields that can be delicate as the existence of solutions crucially depends on them.} used in these works is not able to determine the displacement memory formula through its connection to the corresponding asymptotic supertranslations.  In contrast, the method employed in our work does not rely on the formulation of exact asymptotic symmetries and their induced transformation rules, which circumferences these technical difficulties.

Section~\ref{sec:BasicsEAtheory} sets the stage by solving the linearized Einstein-\AE{}ther theory in the asymptotically flat region. We do so in a manifestly gauge-invariant way by computing the relevant second-order action of the theory. This allows us to arrive at the fully gauge-invariant second-order action written exclusively in terms of dynamical degrees of freedom of the theory in Sec.~\ref{sSec:LinearizedAE} [Eq.~\eqref{gaugeS2}]. As far as we are aware, this purely dynamical second-order action of E\AE{} gravity has not been computed before. Section~\ref{sSec:GravPol} is then dedicated to a summarized analysis of the polarization content of the theory. The linearized regime is left behind in Sec.~\ref{Sec:MemoryGeneral} by introducing the Isaacson approach of formulating leading order equations of motion in Sec.~\ref{sSec:Isaacson}. Based on the purely dynamical second-order E\AE{} action, a second variation approach discussed in Sec.~\ref{sSec:AsymptoticEMGen} enables the derivation of the asymptotic Isaacson energy momentum tensor in terms of fully gauge invariant variables of the theory. This finally allows the formulation of the tensor memory equation of  E\AE{} gravity in Sec.~\ref{sSec:TensorMemoryGen}. Sec.~\ref{sec:MemoryinEA} contains the main results of this work by analyzing the solution of the E\AE{} tensor memory formula in the appropriate asymptotic limit. Through a simplifying thought experiment of computing the memory for individual \ae{}ther energy pulses in Sec.~\ref{sec:pulse}, we gain insight into the causal structure of the memory buildup, which, together with an additional constraint of luminally propagating tensor waves, finally culminates in a conjecture of strongly motivated exclusion of the superluminal parameter space of the theory in Sec.~\ref{ssSec:ConstrainigSuperluminalEA}. The complete formula for gravitational null displacement memory in Einstein-\AE{}ther gravity is written out in Sec.~\ref{sec:MemoryinLuminalEA}. The discussion is concluded in Sec.~\ref{sec:Discussion}.

We choose a mostly plus $(-,+,+,+)$ metric signature, define the symmetrization of indices as $T_{(\alpha\beta)}\equiv\frac{1}{2}\left(T_{\alpha\beta}+T_{\beta\alpha}\right)$ and denote spacetime indices from $0$ to $3$ by Greek letters, $\alpha,\,\beta,\,\mu,\,\nu\,,\,...$, while spatial indices from $1$ to $3$ are denoted by Latin letters, $a,\,b,\,i,\,j,\,...$. Moreover, we will choose natural units in which the speed of light $c=1$. Throughout the document, we will use the term ``luminal'' as denoting this vacuum speed of low energy electromagnetic radiation.


\section{The Basics of Einstein-\AE{}ther gravity }\label{sec:BasicsEAtheory}

\subsection{Action and equations of motion}
Einstein-\AE{}ther gravity \cite{Jacobson:2000xp,Jacobson:2004ts,Eling:2005zq,Foster:2006az,Jacobson:2007veq,Bonvin:2007ap,Yagi:2013ava,Will:2018bme} is defined as the most general metric theory up to two powers of derivative operators, that includes an additional vector field $A$ that explicitly breaks local Lorentz symmetry in the gravity sector through a constraint on its norm. In Appendix~\ref{Appendix:Action} we offer a derivation of this statement and comment on the status of E\AE{} gravity as a metric theory. In the mostly plus signature, the action can be written as
\begin{align}\label{eq:Action EA}
     S=\frac{1}{2\kappa_0}&\int \dd^4 x\sqrt{-g}\Big\{R-K\ud{\alpha\beta}{\mu\nu}\nabla_\alpha A^\mu\nabla_\beta A^\nu\nonumber\\
     &+\lambda (g_{\mu\nu}A^\mu A^\nu +\bar{A}^2)\Big\}+S_\text{m}[g,\sfPsi_\text{m}]\,,
\end{align}
where
\begin{equation}
    K\ud{\alpha\beta}{\mu\nu} = c_1\, g_{\alpha\beta}g_{\mu\nu}+c_2 \,\delta^\alpha_\mu\delta^\beta_\nu+c_3 \,\delta^\alpha_\nu\delta^\beta_\mu-c_4\,A^\alpha A^\beta g_{\mu\nu}\,,
\end{equation}
and $\kappa_0\equiv 8\pi G$ defines the bare Newton's constant.
Moreover, the $c_i$ are dimensionless coupling constants and $\lambda $ is a scalar field that serves as a Lagrange multiplier to impose the Lorentz-breaking holonomic constraint $g_{\mu\nu}A^\mu A^\nu =- \mathbb{A}^2$, where $\mathbb{A}$ is a constant that is often set to $1$. Moreover, as a metric theory, the gravitational \ae{}ther field does not directly couple to the matter fields $\sfPsi_m$, which are themselves only minimally coupled to the physical metric $g$.

The full metric equations of motion read (see also Ref.~\cite{Foster:2006az})
\begin{equation}
    G_{\mu\nu}-S_{\mu\nu}=\kappa_0\, T_{\mu\nu}^m
\end{equation}
where $T_{\mu\nu}^m$ is the matter energy-momentum tensor and
\begin{align}
    G_{\mu\nu}=&R_{\mu\nu}-\frac{1}{2}Rg_{\mu\nu}, \\
    S_{\mu\nu}=&\nabla_\sigma(K\ud{\sigma}{(\mu}A_{\nu)}+K_{(\mu\nu)}A^\sigma-K_{(\mu}{}^\sigma A_{\nu)}) \nonumber \\
    &+c_1(\nabla_\mu A_\sigma \nabla_\nu A^\sigma -\nabla_\sigma A_\mu \nabla^\sigma A_\nu) \nonumber \\
    &+c_4(A^\sigma\nabla_\sigma A_\mu)(A^\rho \nabla_\rho A_\nu) \nonumber\\
    &+\lambda A_\mu A_\nu -\frac{1}{2}g_{\mu\nu}(K\ud{\sigma}{\rho} \nabla_\sigma A^\rho)\,,\label{eq:EAenergymomentum}
\end{align}
with
\begin{equation}
    K\ud{\mu}{\nu}\equiv K\ud{\mu\alpha}{\nu\beta}\nabla_\alpha A^\beta.
\end{equation}
The \ae{}ther field satisfies the equation of motion
\begin{equation}
    \nabla_\mu K^\mu_\nu=-\lambda A_\nu -c_4(A^\sigma \nabla_\sigma A_\mu)\nabla_\nu A^\mu\,,
    \label{fieldeq}
\end{equation}
while the Lagrange multiplier imposes the constraint
\begin{equation}
    g_{\mu\nu}A^\mu A^\nu=-\mathbb{A}^2.
    \label{Aconstraint}
\end{equation}
Contracting Eq.~\eqref{fieldeq} with $A^\nu$, the Lagrange multiplier $\lambda$ can be eliminated through
\begin{equation}\label{eq:LagrangeMultip}
    \lambda=\frac{1}{\mathbb{A}^2}\left\{A^\nu \nabla_\mu K^\mu_\nu+c_4(A^\sigma \nabla_\sigma A^\mu)(A^\rho\nabla_\rho A_\mu)\right\}\,.
\end{equation}

\subsection{Linearized dynamics and second-order action}\label{sSec:LinearizedAE}

Next, we review the linearized treatment of the theory in asymptotically flat spacetime. In doing so, we provide for the first time a manifestly gauge-invariant second-order action of E\AE{} gravity that completely determines the linear dynamics of the theory. 

In this work, we will thus consider the $\mathcal{O}(1/r)$ perturbations $\{h_{\mu\nu},a^\mu\}$ of the metric and the vector field on top of a Minkowski-like background that preserves spatial rotational invariance. We assume the existence of a source-centered coordinate system $\{t,r,\theta,\phi\}$ in which the background metric reduces to the Minkowski metric $\eta_{\mu\nu}$ and the constant background vector $\nabla_\mu \bar{A}^\mu=0$ is purely temporal, as follows
\begin{equation}
    \bar A^\mu=(\bar{A},0,0,0)\,,
\end{equation}
where we choose $\bar{A}=\mathbb{A}$, such that the background already complies with the constraint in Eq.~\eqref{Aconstraint}. For simplicity, we therefore assume here that the source-centered coordinate frame coincides with the preferred frame set by the \ae{}ther field. In other words, we neglect any peculiar velocity effects. Hence, up to $\mathcal{O}(1/r)$, we decompose the physical metric and the aether field as
\begin{align}
   g_{\mu\nu} & = \eta_{\mu\nu}  + h_{\mu\nu}\,,\nonumber\\
    A^\mu & = \bar{A}^\mu + a^\mu\,.
\label{eqn:procadecomp}
\end{align}
Moreover, for simplicity, we will disregard any asymptotic matter contributions and therefore impose the vacuum condition $T^\text{m}_{\mu\nu}=0$.

\subsubsection{gauge-invariant SVT decompositions}\label{sec:SVTdec}

To decouple the linearized equations of motion in the presence of a nontrivial vector background that preserves spatial rotations, it is useful to resort to a so-called \emph{scalar-vector-tensor (SVT) decomposition} \cite{Lifshitz:1945du}. More precisely, due to Helmholtz's theorem, the irreducible parts of the perturbations under the $SO(3)$ rotational group can further be split into scalar, vector and tensor components under the $SO(2)$ subgroup of the rotations about a given direction. The resulting scalar, vector and tensor sectors then naturally decouple, which allows for a convenient description of the corresponding dynamical degrees of freedom at the leading-order. The nonlocality of such a decomposition \cite{carroll2019spacetime} poses no problems because the final observables, in particular the perturbed Riemann tensor dictating the geodesic deviation equation, are local objects \cite{Flanagan:2005yc,Zosso:2024xgy}

In practice, given the above setup, the components of the metric perturbation can be uniquely decomposed as \cite{Flanagan:2005yc}
\begin{subequations}
\begin{align}
        h_{00}=&2S,\\
        h_{0i}=&B_i^T+\partial_i B,\\
        h_{ij}=&h_{ij}^{TT}+\partial_{(i}E^T_{j)}+\left(\partial_i\partial_j-\frac{1}{3}\delta_{ij}\partial^2\right)E+\frac{1}{3}\delta_{ij}D,\label{eq:Decombmetric}
\end{align}
\end{subequations}
where $D=\delta^{ij}h_{ij}$ is the trace over the spatial components, $\partial^iB_i^T=\partial^i E_i^T=0$ satisfy the transversality condition and $\partial^ih_{ij}^{TT}=\delta^{ij}h^{TT}_{ij}=0$ is a transverse-traceless tensor. Analogously, the vector perturbation is decomposed as
\begin{equation}
    a^{\mu }=(a^{0 }, a^{T}_ i+\partial_i a),
\end{equation}
where $ \partial^ia^{T}_i=0$ is transverse.

Working in perturbation theory about a fixed background, the perturbed fields are subject to unphysical gauge redundancies parametrized by an infinitesimal vector field $\xi^\mu$ that need to be taken into account.\footnote{This gauge freedom of the second kind \cite{Nakamura:2020pre} arises due to the freedom in choosing the identification of spacetime points between the background manifold and the physical spacetime captured to linear order by a one-parameter family of diffeomorphisms generated by a small vector field $\xi^\mu$ \cite{carroll2019spacetime,Zosso:2024xgy}.} More precisely, physical observables ought to be invariant under the transformations
\begin{align}
    h_{\mu\nu} &\to h_{\mu\nu}+\mathcal{L}_{\xi}\eta_{\mu\nu}= h_{\mu\nu}+\partial_\mu\xi_\nu+\partial_\nu\xi_\mu\,,\label{eq:gaugetrM}\\
    a^\mu &\to a^\mu+\mathcal{L}_{\xi}\bar{A}^\mu=a^{\mu }-\bar{A}\dot\xi^\mu\,,
    \label{eq:gaugetrA}
\end{align}
where an overdot denotes a derivative with respect to the time coordinate $t$. Note that this transformation depends on our choice of defining the \ae{}ther as a vector field rather than a covector field, a choice that is to be consistently respected. 

However, instead of choosing a particular gauge, as it is customary in most of the literature on E\AE{} gravity, we will define a set of physical variables that are manifestly gauge-invariant, rendered possible by the 3+1 SVT decomposition introduced above (see also \cite{Gong:2018cgj}). This has the advantage that ambiguities in comparing different expressions in different gauges are avoided, and the final results capture all possible gauge choices at once. In particular, the determination of the gravitational polarizations of the theory in Sec.~\ref{sSec:GravPol} below will thus considerably gain in clarity. 

The gauge freedom generated by the vector $\xi^\mu$ indicates that only $10$ out of the initial $14$ perturbation field components contain the physical degrees of freedom of the theory. Identifying the corresponding gauge-invariant variables can be straightforwardly achieved by also decomposing $\xi^{\mu}$ its scalar and vector parts,
\begin{align}
\xi^{\mu}=(\xi^0, \xi^{T}_{i}+\partial_i\xi),
\end{align}
where $\partial^i\xi^{T}_{i}=0$, such that the gauge transformations in Eqs.~\eqref{eq:gaugetrM} and \eqref{eq:gaugetrA} become
\begin{subequations}
\begin{align}
     h_{ij}^{TT} &\to h_{ij}^{TT},\\
     E_i^T &\to E_i^T +2\xi_i^T,\\
    B_i^T&\to B_i^T+\dot{\xi_i},\\
    S&\to S-\dot{\xi^0},\\
    B&\to B-\xi^0+\dot{\xi},\\
    D&\to D+2\partial^2\xi,\\
    E&\to E+2\xi,\\
    a^{T}_i &\to a^{T}_i - \bar{A} \dot \xi^T_i\,,\\
    a^{0 } &\to a^{0 } - \bar{A} \dot\xi^0 \,, \\
    a &\to a -\bar{A} \dot \xi \,.
\end{align}
\end{subequations}
While the TT part of the metric $h_{ij}^{TT}$ is already gauge-invariant, the remaining $8$ gauge-invariant physical degrees of freedom can be chosen as 
\begin{subequations}\label{eq:ggaugepot}
\begin{align}
    \Phi&\equiv S -\dot{B}+\frac{1}{2}\ddot{E},\label{ggaugepot1}\\
    \Theta&\equiv \frac{1}{3}(D-\partial^2E),\\
    \Xi_i&\equiv B_i^T-\frac{1}{2}E_i^T,\\
     \Omega &\equiv  a^{0} - \bar{A}\left(\dot{B}-\frac{1}{2}\ddot{E}\right) \,, \\
   \Upsilon &\equiv  a +\frac{1}{2}\bar{A} \dot{E} \,, \\
     \Sigma_i& \equiv  a^T_i + \frac{1}{2}\bar{A}\dot{E^T_i}\,.
    \label{ggaugepot2}
\end{align}
\end{subequations}

\subsubsection{Linear equations of motion}\label{sSec:linearEOM}

In order to concisely study the linearized equations of motion of the dynamical degrees of freedom of the theory, we now want to compute and analyze the relevant second-order action ${}_{\myst{(2)}}S$.\footnote{As long as the background equations at zeroth order in perturbations of the theory are satisfied, it is sufficient to consider the second-order action to study the linearized dynamics.} Throughout this work, a subscript in parentheses preceding an operator ${}_{\myst{(i)}}O$ denotes the $i$th order in perturbation of that operator. What is more, using the gauge-invariant construction above, we can once and for all compute a manifestly gauge-invariant second-order action for Einstein-\AE{}ther gravity up to total derivatives. Indeed, using Eq.~\eqref{eq:LagrangeMultip} to eliminate the Lagrange multiplier, the E\AE{} action [Eq.~\eqref{eq:Action EA}] to second-order in perturbation fields can be written in terms of the gauge-invariant variables in Eq.~\eqref{eq:ggaugepot} as\footnote{We keep the explicit volume factor $\sqrt{-\eta}$ of the determinant of the metric, even though for Minkowski spacetime it of course reduces to unity, because the presence of this factor will become important in later sections.}
\begin{equation}\label{eq:SecondOrderAction}
    {}_{\myst{(2)}}S=\frac{1}{2\kappa_0}\int d^4x\sqrt{-\eta}\Big({}_{\myst{(2)}}\mathcal{L}_\text{T}+{}_{\myst{(2)}}\mathcal{L}_\text{V}+{}_{\myst{(2)}}\mathcal{L}_\text{S}\Big)\,,
\end{equation}
where
\begin{align}
{}_{\myst{(2)}}\mathcal{L}_\text{T}=&\,\frac{1}{4}\left(1-\bar{A}^2 c_{13} \right)\dot{ h}^{T T}_{a b}\dot{h}_{TT}^{ a b}-\frac{1}{4}\partial^c h_{TT}^{ ab }\partial_c h^{T T}_{ab } , \\
{}_{\myst{(2)}}\mathcal{L}_\text{V}=&\,c_{14}\left(\dot{\Sigma}_{a} \dot{\Sigma}^{a}+2 \bar{A}\dot{\Xi}^a\dot{\Sigma}_{a}+\bar{A}^2 \dot{\Xi}_a \dot{\Xi}^a \right) \nonumber \\
&-c_1\partial_b \Sigma_a\partial^b \Sigma^a-\bar{A}\left(c_1-c_3\right)\partial_b \Sigma_a\partial^b \Xi^a \nonumber  \\
&+\frac{1}{2}\left(1-\bar{A}^2\left(c_1-c_3\right)\right)\partial_b \Xi_a\partial^b \Xi^a \,,
\end{align}
\begin{align}
{}_{\myst{(2)}}\mathcal{L}_\text{S}=&\,-\frac{3}{2}\left(1+\frac{1}{2} \bar{A}^2\left(c_{123}+2 c_2\right)\right) \dot{\Theta}^2 +\frac{1}{2}\partial_a\Theta \partial^a \Theta \nonumber \\
&+c_{1234}\left( \dot{\Omega}^2+\bar{A}^2\dot{\Phi}^2-2 \bar{A} \dot{\Omega} \dot{\Phi}\right) -c_1 \partial_{a} \Omega \partial^a \Omega\nonumber\\
&+ \bar{A}^4c_4 \partial_a \Phi \partial^a \Phi+2 \bar{A}c_1\partial_{a} \Phi\partial^{a} \Omega -2\partial_a \Phi \partial^a \Theta \nonumber \\
&+c_{14}\partial_a\dot{\Upsilon}\partial^a\dot{\Upsilon}-\bar{A}(c_{123}+2c_2)\partial_a\dot{\Upsilon}\partial^a\Theta \nonumber \\
& -2\bar{A} c_{14}\partial_a\dot{\Upsilon}\partial^a\Phi -c_{123} \,\partial^2\Upsilon\partial^2 \Upsilon \,. 
\end{align}
Here, we have introduced the standard shortcut notation
\begin{subequations}
\begin{align}
    c_{13}\equiv&\, c_1+c_3\,,\\
    c_{123}\equiv&\, c_1+c_2+c_3\,,\\
    c_{14}\equiv&\, c_1+\bar{A}^2c_4\,,\\
     c_{1234}\equiv&\, c_1+c_2+c_3+\bar{A}^2c_4\,.
\end{align}
\end{subequations}
Thus, the second-order action can be written exclusively in terms of gauge-invariant variables, and the tensor, vector and scalar sectors decouple. This action can now be used to determine the corresponding linear equations of motion by varying with respect to the physical variables of the theory of each sector. The details of the equations can be found in Appendix~\ref{Appendix:lineareq}. Here we shall only present the final results.

\emph{Tensor modes.}
The tensor equation of motion reads
\begin{align}\label{eq:TTeom}
-\ddot h_{ij}^{TT} +V_\text{T}^2\,\partial^2h_{ij}^{TT}  = 0\,,
\end{align}
where we have assumed that $\bar{A}^2c_{13} \neq 1$ to define the constant Lorentz-breaking phase velocity of the tensor dofs
\begin{align}\label{eq:velT}
    V_\text{T}^2 \equiv \frac{1}{1-\bar{A}^2c_{13}} \,.
\end{align}
Hence, as expected, the tensor modes are dynamical and propagate at a given constant speed. Moreover, since it is nondispersive, this phase velocity also corresponds to the physical group velocity of the waves, which we denote as
\begin{equation}\label{eq:relPhaseGroup}
    \beta_\text{T}=V_\text{T}\,.
\end{equation}
In particular, the group velocity is the velocity of energy propagation of the corresponding wave packets. Such a distinction between phase and group velocity will become important when comparing the present results with previous work \cite{Heisenberg:2024cjk}. By requiring a real radial velocity, we will always assume that $V^2>0$ for the velocities in each sector and choose the positive solution to match the convention of positive radial velocities. Moreover, luminality of the tensor radiation is recovered when imposing
\begin{equation}\label{eq:LuminalityT}
    \text{Luminality Condition 1:} \quad c_{13}=0\,.
\end{equation} 
Equation~\eqref{eq:TTeom} further implies that the asymptotic solution of the tensor radiation satisfies a plane-wave form
\begin{align}\label{eqn:waveformT}
    h^{TT}_{ij} (x) &= \frac{1}{r} f^{TT}_{ij}(t-r/V_\text{T},\theta,\phi)  
\end{align}
for some transverse and traceless function $f^{TT}_{\text{T}ij}$ that depends only on time and radial coordinates through a specific combination $t-r/V_\text{T}$. Due to the direct relation between the phase velocity and the group velocity [Eq.~\eqref{eq:relPhaseGroup}] this combination will in this case also correspond to the physical asymptotic retarded time of the tensor modes. 
It follows that to leading-order in $r$ the solutions of the asymptotic tensor radiation satisfy
\begin{align}\label{eq:tensor_radiation}
    \partial_i h^{TT}_{kl} = -\frac{1}{V_\text{T}} n_i \dot h^{TT}_{kl}\,,
\end{align}
where
\begin{equation}\label{eq:n}
    n_i=(\sin\theta\cos\phi,\sin\theta\sin\phi,\cos\theta)\,,
\end{equation}
is the unit vector indicating the radial direction of propagation. 

\emph{Vector modes.}
In the vector sector, we obtain on the one hand the constraint
\begin{align}\label{eq:ConstraintXi}
    \Xi_i = -\bar{A}c_{13} \Sigma_i\,.
\end{align}
and, on the other hand, the dynamical equation
\begin{align}
    -\ddot \Sigma_i +V_\text{V}^2\,\partial^2\Sigma_i  = 0\,,
   \label{Vectoreq}
\end{align}
where
\begin{align}\label{eq:velV}
   V_\text{V}^2=\beta_\text{V}^2 \equiv \frac{ c_1 -\frac{1}{2}\bar{A}^2(c_1^2-c_3^2)}{(1-\bar{A}^2c_{13})c_{14}}\,.
\end{align}
We thus further assume that $c_{14}\neq0$. In this case, the equations of motion of E\AE{} gravity dictate that only 2 of the 4 physical vector degrees of freedom are dynamical. Analogously to the tensor case, this implies the asymptotic form
\begin{align}\label{eqn:waveformV}
   \Sigma_i (x) &= \frac{1}{r} f^{T}_{i}(t-r/\beta_\text{V},\theta,\phi)  \,.
\end{align}
for some transverse function $f^{T}_{i}$.
As a consequence, we also have the relation
\begin{align}\label{eq:vector_radiation}
        \partial_i \Sigma_j = -\frac{1}{V_\text{V}}n_i \dot \Sigma_j \,.
\end{align}
The vector radiation recovers the speed of light if, in addition to Eq.~\eqref{eq:LuminalityT}, we require
\begin{equation}\label{eq:LuminalityV}
   \text{Luminality Condition 2:}\quad c_4=0\,.
\end{equation}

\emph{Scalar modes.}
In the scalar sector on the other hand, assuming $c_{123}\neq 0$, we obtain three constraint equations of the form
\begin{subequations}\label{eq:ConstraintScalars}
    \begin{align}
        \Phi&=-\frac{1}{\bar{A}}\dot\Upsilon-\frac{1}{\bar{A}^2 c_{14}} \Theta\, ,\label{eq:ConstPhi}\\
        \Omega&=-\,\bar{A}\,\Phi\, ,\label{eq:ConstOmega}\\
        \partial^2\Upsilon&=-\frac{1+\frac{1}{2}\bar{A}^2(c_{123}+2c_2)}{\bar{A} c_{123}} \Theta\,,\label{eq:ConstPsi}
    \end{align}
\end{subequations}
together with the wave equation
\begin{align}
    -\ddot\Theta + V_\text{S}^2\partial^2\Theta =0\,,
    \label{theq}
\end{align}
where
\begin{align}\label{eq:velS}
     V_\text{S}^2=\beta_\text{S}^2 \equiv \frac{c_{123}(2-\bar{A}^2 c_{14})}{(1-\bar{A}^2 c_{13})(2+\bar{A}^2(c_{123}+2c_2))c_{14}}\,.
\end{align}
Thus, as expected, the scalar sector only supports 1 dynamical degree of freedom. Observe, however, that one could also have chosen $\Upsilon$, which is associated with the spatial scalar component of the \ae{}ther field, to play the role of the dynamical variable. The choice of $\Theta$ or $\Upsilon$ as the propagating dof is, in this sense, arbitrary.

In the scalar sector, we can play the same game as above and define the asymptotic behavior of the dynamical scalar mode to be
\begin{align}\label{eqn:waveformS}
   \Theta (x) &= \frac{1}{r} f(t-r/V_\text{S},\theta,\phi)  \,,
\end{align}
for some scalar function $f$,
together with the relation
\begin{align}\label{eq:scalar_radiation}
        \partial_i \Theta = -\frac{1}{V_\text{S}}n_i \dot \Theta \,.
\end{align}
Moreover, luminality now requires on top of Eqs.~\eqref{eq:LuminalityT} and \eqref{eq:LuminalityV} the additional constraint
\begin{equation}\label{eq:LuminalityS}
    \text{Luminality Condition 3:}\quad c_2=\frac{-c_3}{1+2\bar{A}^2 c_3}\,.
\end{equation}
Finally, at the level of the solutions, we can use Eq.~\eqref{theq} in the constraint Eq.~\eqref{eq:ConstPsi} to arrive at the following direct relations between the physical degrees of freedom
\begin{subequations}
    \begin{align}
        \dot{\Upsilon}&=-\frac{2-\bar{A}^2c_{14}}{2\bar{A}(1-\bar{A}^2c_{13})c_{14}} \Theta\,, \label{eq:psith}\\
        \Phi&=-\frac{c_1+2c_3-\bar{A}^2c_4}{2(1-\bar{A}^2c_{13})c_{14}} \Theta\, ,\label{eq:phith}\\ 
        \Omega&=-\frac{\bar{A}(c_1+2c_3-\bar{A}^2c_4)}{2(1-\bar{A}^2c_{13})c_{14}} \Theta\,. \label{eq:omth}
    \end{align}
\end{subequations}

\subsubsection{Gauge-invariant second-order action of dynamical degrees of freedom}
Ultimately, the constraint equations in the vector and scalar sectors can be used to get rid of 5 out of the 10 physical degrees of freedom in the leading-order description of Einstein-\AE{}ther gravity. More precisely, and unlike in a cosmological setting, the physics of the asymptotically flat vacuum depends only on the 5 dynamical degrees of freedom of the theory. Concretely, Eqs.~\eqref{eq:ConstraintXi} and \eqref{eq:ConstraintScalars} allow the cancellation of 5 of the 10 variables within the second-order action [Eq.~\eqref{eq:SecondOrderAction}]. Together with the identifications of the following theory-dependent constants,
\begin{align}
\bar C_\text{V}&=2(1-\bar{A}^2c_{13})(2c_1-\bar{A}^2 c_{13}(c_1-c_3))\,,\label{eq:CV}\\
\bar C_\text{S}&=\frac{4-2\bar{A}^2c_{14}}{\bar{A}^2c_{14}}\,,\label{eq:CS} 
\end{align}
the second-order action [Eq.~\eqref{eq:SecondOrderAction}] of E\AE{} gravity reduces to the particularly simple form
\begin{align}\label{gaugeS2}
{}_{\myst{(2)}} S=\,\frac{1}{8\kappa_0}\int d^4x&\sqrt{-\eta}\Bigg\{\frac{1}{V_\text{T}^2}\dot{h}^{TT}_{ab}\dot{h}_{TT}^{ab}-\partial_c h^{TT}_{ab}\partial^c h_{TT}^{ab} \nonumber\\
&\quad+\bar C_\text{V}\left(\frac{1}{V_\text{V}^2}\dot{\Sigma}_a\dot{\Sigma}^a-\partial_b\Sigma_a\partial^b\Sigma^a\right) \nonumber\\
&\quad+\bar C_\text{S}\left(\frac{1}{V_\text{S}^2}\dot{\Theta}^2-\partial_a\Theta\partial^a\Theta\right)\Bigg\}\,.
\end{align}
This form of the second-order action will especially be useful in our endeavor of computing gravitational wave memory sourced by the asymptotically propagating waves in E\AE{} theory below.
\subsection{Gravitational polarizations}\label{sSec:GravPol}
To conclude this section on the review of Einstein-\AE{}ther gravity, we want to discuss the \emph{gravitational polarizations} of the theory. The introduction of this concept will form the basis for the discussion of memory in the next sections. In what follows, we will first define the gravitational polarizations in generic metric theories of gravity. Then, we will apply these results to the Einstein-\AE{}ther theory and derive its polarization content. In contrast to a large part of the existing literature (see e.g. Refs.~\cite{Jacobson:2004ts,Schumacher:2023jxq}), we will do so in a manifestly gauge-invariant manner, with the advantage of using a universal language that does not depend on convenient gauge choices of specific metric theories. We show the correspondence between our gauge-invariant and the gauge-fixed results for the polarizations in Appendix~\ref{Appendix:Literature}.

\subsubsection{Gravitational polarizations in metric theories}\label{sSec:polmetricth}

It is important to stress that the widely used notion of gravitational polarizations is tied to the consideration of the so-called \emph{metric theories of gravity}. As already mentioned, based on the Einstein equivalence principle, these theories are defined through the assumption that matter is universally and minimally coupled to a physical metric, which in particular implies that the energy-momentum tensor of matter is locally covariantly conserved and the physical metric can be viewed as describing an observer-independent spacetime \cite{Will:2018bme,Zosso:2024xgy}. The key point here is that due to the universal and minimal-coupling assumption, the direct interaction of any gravitational radiation with matter, hence also any detector, is entirely determined by the six independent physical components of the spacetime metric. This is the crucial insight behind the statement that a generic metric theory of gravity only admits up to six gravitational polarizations \cite{Eardley_PhysRevLett.30.884,Eardley_PhysRevD.8.3308}. 

Note that these gravitational polarizations are therefore conceptually different from the propagating or \emph{dynamical degrees of freedom} of a theory. On one hand, as discussed above, only the six independent modes associated with the physical metric can directly interact with matter fields and contribute to a detector response in terms of gravitational polarizations; on the other hand, there is a priori no upper limit on the number of possible physical modes satisfying wave-type equations of motion of a linearized metric theory. The link between the two is that the presence of dynamical degrees of freedom is needed to excite a particular set of gravitational polarizations. Which of the six polarizations is excited depends on the couplings between the physical metric and the additional fields in the gravitational sector. We will precisely answer this question in the case of E\AE{} gravity below.

But first, let us define in full generality the six independent gravitational polarizations in a generic metric theory. One way of doing so is to consider the geodesic deviation equation, the basic principle for current GW observations, to leading-order in the asymptotically Minkowski background \cite{maggiore2008gravitational,Zosso:2024xgy}
\begin{equation}\label{eq:GeodesicDeviation}
    \ddot{s}^i=-\phantom{}_{\myst{(1)}}R\ud{i}{ 0 j 0} \,s^j +\mathcal{O}\left( \frac{1}{r^2}\right),
\end{equation}
where $s^i$ denotes the proper spatial distance between two freely falling test masses. As already mentioned, subscript in parenthesis $\phantom{}_{\myst{(1)}}$ preceding the Riemann tensor denotes its first order in perturbations given by an expansion in $1/r$. The modification in the separation of two timelike geodesics is therefore entirely governed by the six independent components of the electric part of the linearized Riemann tensor. These can be associated to the manifestly gauge-invariant metric components defined in Sec.~\ref{sec:SVTdec} through \cite{Flanagan:2005yc,Heisenberg:2024cjk,Zosso:2024xgy}
\begin{align}\label{eq:gauge-invariant Riemann first}
    \phantom{}_{\myst{(1)}}R_{0i0j} =- \frac{1}{2}\left( \ddot h_{ij}^{TT} -2  \partial_{(i} \dot\Xi_{j)}+ \delta_{ij} \ddot \Theta +2\partial_i \partial_j \Phi\right)\,.
\end{align}

In order to identify the individual gravitational polarizations within this expression, we need to further use the knowledge that only dynamical degrees of freedom that are part of the asymptotic radiation will contribute to the asymptotic Riemann tensor. Hence, we assume that each gauge-invariant mode, which we collectively denote as $\Psi$, propagates at a constant asymptotic phase velocity $V_\psi$. To remain general and to avoid confusion with previous works, we will write all general results in terms of a generic phase velocity. Only for the E\AE{} specific results, we will use the nondispersive assumption and set the phase velocity equal to the group velocity $V_\psi=\beta_\psi$. In particular, as already stated in the case of E\AE{} gravity above [see Eqs.~\eqref{eq:tensor_radiation}, \eqref{eq:vector_radiation} and \eqref{eq:scalar_radiation}], each dof in the asymptotic radiation satisfies up to $\mathcal{O}(1/r)$ the relation
\begin{equation}\label{eq:WavecondGen}
     \partial_{i} \Psi = -\, \frac{n_i}{V_\psi}\, \dot \Psi\, .
\end{equation}
\newline

Returning to Eq.~\eqref{eq:GeodesicDeviation}, the electric part of the Riemann tensor can then be rewritten as
\begin{align}
 \phantom{}_{\myst{(1)}}R_{0i0j} \equiv- \frac{1}{2}\ddot P_{ij}\,,
\end{align}
where
\begin{equation}\label{eqn:Apolarization}
    P_{ij}=h_{ij}^{TT} + \frac{2}{V^2_\Xi} n_{(i} \Xi_{j)} + \delta_{ij} \Theta+\frac{2}{V_\Phi^2} n_i n_j \Phi\,.
\end{equation}
This is the response matrix $P_{ij}$, which characterizes the six independent gravitational polarizations governing the physical stretch in proper distances between two test masses. Indeed, integrating Eq.~\eqref{eq:GeodesicDeviation} now yields
\begin{align}\label{eq:integratedGeodesicDeviation}
    \Delta s_i  = \frac{1}{2} P_{ij}\, s_0^j  \,,
\end{align}
where
\begin{equation}
    \Delta s^i \equiv s^i -s^i_0\,,
\end{equation}
where $s^i_0$ denotes the reference proper distance separation before the presence of any gravitational radiation.

Given a longitudinal direction vector $n_i$ [Eq.~\eqref{eq:n}] for the asymptotic radial radiation, it is customary to construct the basis of the transverse space \cite{poisson2014gravity}
\begin{subequations}\label{eq:Def Transverse Vectors u v}
\begin{align}
u_i&=(\cos\theta \cos\phi,\,\cos\theta \sin\phi,\,-\sin\theta )\,,\\
v_i&=(-\sin\phi,\,\cos\phi,\,0)\,,
\end{align}
\end{subequations}
which completes the orthonormal spatial basis with the relation
\begin{equation}\label{eq:polcompleteness}
\delta_{ij}=n_in_j+u_iu_j+v_iv_j\,.
\end{equation} 
In terms of this basis, one defines a spatial polarization basis
\begin{subequations}\label{eq:PolTensors}
\begin{align}
e^+_{ij}&\equiv u_iu_j-v_iv_j\,,\quad e^\times_{ij}\equiv u_iv_j+v_iu_j\,, \label{eqn:crossandplusdef}\\
e^u_{ij}&\equiv n_iu_j+u_in_j\,,\quad e^v_{ij}\equiv n_iv_j+v_in_j\,,\\
\quad e^b_{ij}&\equiv u_iu_j+v_iv_j\,,\quad e^l_{ij} \equiv n_in_j\,,
\end{align}
\end{subequations}
in terms of which the polarization matrix can be expanded as
\begin{align}
    P_{ij} \equiv\, & P_+ \,e^+_{ij} + P_\times\, e^\times_{ij} + P_u\, e^u_{ij} + P_v \,e^v_{ij}\nonumber \\
    &+ P_b\, e^b_{ij} + P_l\, e^l_{ij}  \,.\label{eq:PolM 2}
\end{align}
By direct comparison with Eq.~\eqref{eqn:Apolarization}, we finally arrive at the most general expression for the gravitational polarization in terms of the gauge-invariant metric components of a generic metric theory of gravity
\begin{align}\label{eq:Polarization Modes}
P_+  &=  \frac{e^{ij}_+}{2}h^{TT}_{ij}\,, & P_u  &= \frac{u^i}{V_\Xi} \Xi_i\,, &P_l  &= \Theta + \frac{2}{V_\Phi^2} \Phi \,,\nonumber \\
P_\times  &= \frac{e^{ij}_\times}{2}h^{TT}_{ij}\,, & P_v  &= \frac{v^i}{V_\Xi} \Xi_i\,, &P_b &= \Theta \,.
\end{align}
A generic metric theory of gravity therefore admits up to six polarizations, two tensor polarizations $(+,\times)$, two vector polarizations $(u,v)$ and two scalar polarizations $(l,b)$, depending on whether the corresponding metric variables are dynamical or not. We will now examine which of the polarizations are excited by the propagating degrees of freedom in the E\AE{} theory of gravity.

\subsubsection{Gravitational polarizations in Einstein-\AE{}ther gravity}\label{ssSec:GravPolAE}

Given the analysis of the linearized dynamics of Einstein-\AE{}ther gravity in Sec.~\ref{sSec:LinearizedAE}, it is now straightforward to determine its gravitational polarizations. First of all, recall that E\AE{} has a total of five propagating degrees of freedom, two within the tensor field, two within the vector field and one scalar field. Indeed, due to gauge invariance, only 10 out of the 14 independent components in the metric $g$ and the vector field $A$ are physical. Among these physical degrees of freedom, only 5 are dynamical, as dictated by the equations of motion through the 5 constraints Eqs.~\eqref{eq:ConstraintXi}, \eqref{eq:psith}, \eqref{eq:omth}, \eqref{eq:phith}. 

To determine the gravitational polarization content, all that is left to do is to apply the equations of motion, in particular the constraints, to the general formulas in Eqs.~\eqref{eq:Polarization Modes} and express everything in terms of the most natural dynamical variables of the theory, which in the case of Einstein-\AE{}ther gravity are $h^{TT}_{ij}$, $\Sigma_i$ and $\Theta$, as follows
\begin{subequations}\label{eq:polarizations}
\begin{align}
P_+  &=  \frac{e^{ij}_+}{2}h^{TT}_{ij}\,,\\
P_\times  &= \frac{e^{ij}_\times}{2}h^{TT}_{ij}\,,\\
 P_u  &=-   \frac{\bar{A}c_{13}}{\beta_V} \Sigma_iu^i=\frac{1/\beta_T^2-1}{\beta_V} \Sigma_iu^i\,, \\
 P_v  &= -\frac{\bar{A}c_{13}}{\beta_V} \Sigma_iv^i=\frac{1/\beta_T^2-1}{\beta_V} \Sigma_iv^i\,, \\
 \quad P_l  &= \left(1- \frac{c_1+2c_3-\bar{A}^2c_4}{\beta_S^2(1-\bar{A}^2c_{13})c_{14}}\right) \Theta\,,\\
 P_b& = \Theta\,.
\end{align}
\end{subequations}
Note that in addition, we have used the fact that in E\AE{} theory the velocities are nondispersive and therefore the phase velocities correspond to the group velocities $V_\Xi=\beta_V$ and $V_\Phi=\beta_S$. Thus, in the most general case, the five dynamical dofs of E\AE{} theory excite all six gravitational polarizations. In particular, so long as the scalar dof remains dynamical, the breathing mode is always excited, while the longitudinal scalar polarization is present whenever
\begin{equation}
    \left(1- \frac{c_1+2c_3-\bar{A}^2c_4}{\beta_S^2(1-\bar{A}^2c_{13})c_{14}}\right)\neq0\,.
\end{equation}

Finally, it is interesting to examine what happens to the gravitational polarizations as the luminality conditions in Eqs.~\eqref{eq:LuminalityS}, \eqref{eq:LuminalityV}, and \eqref{eq:LuminalityT} are imposed. In this case, we have
\begin{subequations}
\begin{align}
 P_u  &=0\,, \\
 P_v  &= 0\,, \\
 \quad P_l  &= 2P_b=2\Theta\,.
\end{align}
\end{subequations}
In words, the presence of the vector polarizations requires a departure from the speed of light of the tensor dofs, and their magnitude scales with the degree of nonluminality. The same is, however, not true for the scalar polarizations. Even when a luminal propagation of all dofs is imposed, both scalar polarizations $P_b$ and $P_l$ prevail. In particular, it is notable that the longitudinal scalar mode does not vanish as soon as the modes propagate at the speed of light, since this is a behavior observed for instance in Horndeski gravity \cite{Hou:2017bqj,Heisenberg:2024cjk}. The Lorentz violation of the theory is therefore such that the longitudinal mode can survive even as luminality is imposed.

\section{An Efficient Method for Computing Gravitational Memory}\label{Sec:MemoryGeneral}

In this section, we present our method for computing the memory equation in Einstein-\AE{}ther gravity. The memory equation is the equation in asymptotically flat spacetime whose solution yields the memory component sourced by the asymptotic radiation. This is an optimization of the framework developed in Refs.~\cite{Heisenberg:2023prj,Heisenberg:2024cjk,Zosso:2024xgy} based on Isaacson's formulation of the leading-order equations of motion of gravitational waves. In this work, we additionally exploit the simplifying power of purely dynamic second-order action. The considerations in this section are valid for any metric theory whose asymptotic radiation involves linearly decoupled dynamical degrees of freedom with a nondispersive, constant asymptotic velocity. Readers who are only interested in the final result of gravitational memory in E\AE{} gravity and the conjectured constraints on its parameter space can directly skip to Sec.~\ref{sec:MemoryinEA}.

\subsection{Gravitational memory in metric theories}\label{sSec:GWMemory}

We begin with a proper definition of displacement memory in metric theories of gravity. To simplify the notation, we will present most formulas in this section by collectively denoting all additional gravitational fields with a common symbol $\sfPsi$. Consider, therefore, the explicit definition of a generic covariant metric theory action (see e.g. Ref.~\cite{Zosso:2024xgy}),
\begin{equation}\label{eq:MetricTheoryAction}
        S=\frac{1}{2\kappa_0}\int d^4x\,\sqrt{-g} \,\mathcal{L}_\text{G}[g,\sfPsi] +S_\text{m}[g,\sfPsi_\text{m}]\,,
    \end{equation}
with gravitational Lagrangian $\mathcal{L}_\text{G}$ covariantly depending on the physical metric $g$, as well as possibly on a set of additional nonminimally coupled gravitational fields $\sfPsi$. As already stated, a metric theory of gravity is then defined through the assumption of the validity of the Einstein equivalence principle by imposing that the matter fields $\sfPsi_\text{m}$ only couple minimally to the spacetime metric $g$ within a separate matter action $S_\text{m}[g,\sfPsi_\text{m}]$ \cite{papantonopoulos2014modifications,Will:2018bme,Zosso:2024xgy}. 

In the framework of metric theories, the gravitational displacement memory effect is well defined as arising from components in asymptotic radiation that are not wavelike, but induce a permanent displacement between test masses in an idealized GW detector. More precisely, gravitational memory is characterized by a nonzero leading-order spatial proper distance displacement between two timelike geodesics within the solution to the geodesic deviation equation in Eq.~\eqref{eq:integratedGeodesicDeviation}, that persists after the passage of a burst of gravitational waves in terms of asymptotic retarded time $u$, as follows
\begin{equation}\label{eq:memorydef}
    \lim_{u\rightarrow\infty}\Delta s\neq 0\,.
\end{equation}
Thus, memory manifests itself as a nonzero difference in one of the components of the polarization matrix before and after the gravitational wave has passed,
\begin{equation}
\lim_{u\rightarrow\infty}P_{ij}\neq 0\,.
\end{equation}
Consequently, through Eq.~\eqref{eq:PolM 2}, the displacement memory effect can naturally be associated with a specific gravitational polarization type. Accordingly, a generic metric theory of gravity can admit \emph{tensor}, \emph{vector}, and \emph{scalar memory}. We want to stress that this is a very general result, since a metric theory is defined through its property of universal and minimal coupling to matter. Hence, the geodesic deviation equation, which is independent of the dynamics of individual theories, universally characterizes the fundamental principle of detecting asymptotic gravitational radiation.

Yet, while the existence of such a nonzero permanent offset within the radiation is guaranteed through the nonlinear nature of gravity, one requires additional input in order to have access to the signal. In other words, Eq.~\eqref{eq:memorydef} is just the definition of a phenomenon, whose magnitude and characteristics in specific gravitational wave events still need to be computed. While there exist various approaches for doing so \cite{Christodoulou:1991cr,JFrauendiener_1992,Blanchet:1992br,PhysRevD.44.R2945}, we will present here an efficient method that is general enough to apply to a very broad set of viable metric theories of gravity. As mentioned, the basis of the approach that we want to advertise here was pioneered in Ref.~\cite{Heisenberg:2023prj} and further refined in Ref.~\cite{Heisenberg:2024cjk} and rests upon the insight that displacement memory is sourced by the presence of emitted unbound energy-momentum from an otherwise localized system and represents a low-frequency component in the radiation that is in principle indistinguishable from what one calls the background spacetime. Therefore, the effort of Isaacson \cite{Isaacson_PhysRev.166.1263,Isaacson_PhysRev.166.1272} in defining the backreaction of GW energy-momentum onto a background spacetime appears tailor-made for the derivation of displacement memory components. An additional advantage of this method is that it naturally provides a notion of clear distinction between the memory signal and its oscillatory counterpart through an averaging procedure that separates distinct frequency scales. In this way, the method not only provides the final zero-frequency offset of the displacement memory but actually computes its low-frequency time variation that determines the potential signal to be observed in current and future GW detectors. Before discussing the details of the method to compute memory that is novel to this work, we will, for convenience, first review the Isaacson approach to gravitational waves.

\subsection{The Isaacson approach to gravitational waves}\label{sSec:Isaacson}

For simplicity, and in order to focus on the aspects most relevant for beyond GR theories, we will, as already stated, omit any contributions from ordinary asymptotically unbound ordinary matter components and focus on memory that is sourced by the emission of gravitational waves. We will therefore review here the most important aspects of the Isaacson viewpoint on gravitational waves in vacuum, which is, for instance, also treated in Refs.~\cite{misner_gravitation_1973,Flanagan:2005yc,maggiore2008gravitational}. Yet, although Isaacson's initial work was carried out before the first description of the phenomenon of gravitational memory \cite{Zeldovich:1974gvh,Christodoulou:1991cr}, the Isaacson framework was not associated to the phenomenon of memory until the work in Ref.~\cite{Heisenberg:2023prj,Zosso:2024xgy}, which introduced the appropriate refinements to carve out this connection.

The Isaacson approach starts with the identification of the two key assumptions that are needed to define the concept of waves, and allow for a viable description of their energy content along with their leading-order backreaction onto the geometry of spacetime. First, a definition of waves requires the existence of an exact but arbitrary solution $\bar{g}_{\mu\nu}$ to the equations of motion, on top of which perturbations $h_{\mu\nu}$ can be defined.\footnote{For a more careful definition of the notion of perturbation theory on spacetime manifolds, see for instance \cite{carroll2019spacetime,Zosso:2024xgy}.} Second, a clear separation of scales between the wave perturbations and a background spacetime is required.

This second assumption is crucial for the concept of gravitational waves to make sense within a general dynamical spacetime. It entails that at any spacetime point, field variables can uniquely be split into a contribution to a slowly varying \emph{background} and a contribution to the \emph{waves} characterized by a faster (shorter) scale of variation. Such a separation is in practice defined via a spacetime average $\langle...\rangle$ that needs to satisfy the following criteria 
\cite{Isaacson_PhysRev.166.1272,misner_gravitation_1973,Flanagan:2005yc,maggiore2008gravitational,Zalaletdinov:2004wd,Stein:2010pn,Heisenberg:2023prj}: (i) the average of an odd number of wave operators vanishes; (ii) total derivatives of tensors average out to zero; (iii) integration by parts of covariant derivatives are allowed inside the average.\newline
The background contribution of any operator $O$ is then given by $\langle O\rangle$, while the wave part is included in $O-\langle O\rangle $.

For a given foliation of spacetime, these scales can be chosen to be described in terms of typical frequencies of variation, which we will denote as the low $L$ and high $H$ frequency scales. For instance, the assumed split in metric perturbations $h_{\mu\nu}$ is written as $h_{\mu\nu}=h_{\mu\nu}^L+h_{\mu\nu}^H$ with $h_{\mu\nu}^L=\langle h_{\mu\nu}\rangle$. Moreover, because we will specialize in this work to the concrete case of an exact static Minkowski solution $\bar{g}_{\mu\nu}=\eta_{\mu\nu}$, this part naturally only contributes to the slowly varying background. In summary, we end up with the following leading-order decomposition of the metric,
\begin{equation}\label{eq:PertAssumption}
g_{\mu\nu}=\eta_{\mu\nu}+h_{\mu\nu}^L+h_{\mu\nu}^H\,,
\end{equation}
where the full background spacetime is given by
\begin{equation}
\langle g_{\mu\nu}\rangle=\eta_{\mu\nu}+h_{\mu\nu}^L\,.
\end{equation}
In a generic metric theory of gravity [Eq.~\eqref{eq:MetricTheoryAction}], the same decomposition is assumed for all gravitational fields,
\begin{equation}\label{eq:IsaacsonSplit}
\sfPsi=\bar \sfPsi+\psi^L+\psi^H\,,
\end{equation}
where $\bar \sfPsi$ denotes the collection of fields that make up the static background solution.

The most notable consequences of the Isaacson assumptions for the definition of waves appear in the description of the leading-order equations of motion. Writing the generic vacuum equations for the physical metric as $\mathcal{G}_{\mu\nu}[g,\sfPsi]=0$, the parametric separation in scales of variation imply that the leading-order is described by two sets of equations in a bivariate expansion that govern dynamics at two distinct scales
\cite{Heisenberg:2023prj,Zosso:2024xgy}, as follows
\begin{align}
    \phantom{}_{\mys{(1)}}G_{\mu\nu}[h^H,\psi^H]&=0\,,\label{eq:EOMIISGR2}\\
    \phantom{}_{\mys{(1)}}G_{\mu\nu}[h^L,\psi^L]&=-\frac{1}{2}\,\big\langle\phantom{}_{\mys{(2)}}G_{\mu\nu}[h^H,\psi^H]\big\rangle\,.\label{eq:EOMISGR2}
\end{align}
As before, $\phantom{}_{\mys{(i)}}O$ denotes the $i$th order in the perturbative expansion of the operator.\footnote{Here, we follow \cite{Zosso:2024xgy} and define the second perturbation of the operator $O={}_{\mys{(0)}}O+{}_{\mys{(1)}}O+\frac{1}{2}{}_{\mys{(2)}}O+...$ with an explicit factor of $1/2$, whereas in \cite{Heisenberg:2023prj}, this factor is implicit in the definition of ${}_{\mys{(2)}}O$.} While the first equation naturally describes the propagation of high-frequency gravitational waves, the second equation is the crucial equation of this work that describes the backreaction of a coarse-grained high-frequency operator onto the background of the spacetime. As such, it can be viewed as defining the energy-momentum carried by wave perturbations of the theory
\begin{equation}\label{eq:DefPseudoEMTensor}
    {}_{\mys{(2)}}t_{\mu\nu}\equiv -\frac{1}{2\kappa_0} \big\langle\phantom{}_{\mys{(2)}}G_{\mu\nu}[h^H,\psi^H]\big\rangle\,.
\end{equation}
We want to stress here that the averaging that naturally appears in the Isaacson approach is a fundamental property of defining the local energy-momentum content of wavelike phenomena, both in the classical and the quantum picture (see also Ref.~\cite{maggiore2008gravitational}).

Now, in the asymptotically flat region of a spacetime with a localized source of gravitational waves, Eq.~\eqref{eq:EOMISGR2} precisely corresponds to the displacement memory equation, as shown in Ref.~\cite{Heisenberg:2023prj}. In other words, the backreaction of asymptotic waves onto the asymptotic background spacetime that naturally appears in the Isaacson approach signals the existence of displacement memory sources within the radiation.

\subsection{Asymptotic energy-momentum}\label{sSec:AsymptoticEMGen}

As a necessary requirement for the existence of a well defined solution to the memory equation in Eq.~\eqref{eq:EOMISGR2}, we must first of all show that the asymptotic energy-momentum tensor in Eq.~\eqref{eq:DefPseudoEMTensor} satisfies a set of crucial properties. The first two such basic properties are the covariant conservation $\nabla^\mu{}_{\mys{(2)}}t_{\mu\nu}=0$ and gauge invariance of the expression. While the former is automatically satisfied thanks to the property (ii) of the average introduced above and the fact that covariant derivatives and the average commute \cite{Heisenberg:2023prj}, the latter is an indispensable,\footnote{Gauge invariance of locally defined energy-momentum is a requirement to define it as a physical observable.} yet a rather nontrivial statement. This is because it is inherently an object that enters at second-order in the fields of a perturbative expansion of the equations of motion, where, in general, the nice properties of the linear equations are lost. Indeed, at second-order in the equations of motion, the notion of gauge-invariant perturbation theory is still an open research question \cite{Nakamura:2020pre}. 
\newline
However, within the Isaacson approach, the energy-momentum tensor of gravitational waves appears in the \emph{leading-order} equation, Eq.~\eqref{eq:EOMISGR2}. As we will now show, this implies that this particular operator at second-order in the fields actually keeps all the nice properties that are present at the linear level. In other words, assuming that the linear equations of motion for the propagating degrees of freedom of a metric theory are well defined, also its energy-momentum tensor and consequently also the memory equation Eq.~\eqref{eq:EOMISGR2} are sound. In GR, this statement can be shown to be true on an arbitrary background spacetime, by noting that all second-order fields vanish due to property (i) of the average, while also the leading-order gauge-dependent contribution of the first-order fields vanishes upon averaging over the high-frequency scales \cite{Isaacson_PhysRev.166.1272}. Since this last argument is based on the specific form of the energy-momentum tensor of gravitational waves in GR, we must resort to a different technique to make a statement for generic metric theories of gravity.
\newline
In the specific case of Einstein-\AE{}ther gravity, the asymptotic energy-momentum tensor has already been computed in Refs.~\cite{Foster:2006az,Eling:2005zq,Saffer:2017ywl}. However, these works do not employ a manifestly gauge-invariant approach, and, in particular, the computations of the explicit components of the energy-momentum tensor in Ref.~\cite{Saffer:2017ywl} are not gauge-invariant and different methods lead to different results. Based on the manifestly gauge-invariant and purely dynamical second-order action of E\AE{} gravity in Eq.~\eqref{gaugeS2}, we present here a straightforward method to compute a fully gauge-invariant energy-momentum tensor. 

\subsubsection{Second-variation method}\label{ssSec:SecondVariation}

Following the discussion above, we introduce here the so-called \emph{second-variation approach} \cite{Maccallum:1973gf}, which can be used to show that the Isaacson energy-momentum tensor [Eq.~\eqref{eq:DefPseudoEMTensor}] only depends on the second-order action of a theory that governs the linear equations of motion \cite{Heisenberg:2023prj}. More precisely, starting from the second-order action ${}_{\mys{(2)}}S[h^H_{\mu\nu},\psi^H]$ that depends only on the high frequency perturbation fields defined in the context of the Isaacson approach [Sec.~\ref{sSec:Isaacson}] as well on top of an exact background solution $\{\bar g_{\mu\nu},\bar \sfPsi\}$, one defines an effective second-order action ${}_{\mys{(2)}}S\rightarrow {}_{\mys{(2)}}S_\text{eff}$ by temporally treating the background solution $\{\bar g_{\mu\nu},\bar \sfPsi\}$ as independent field variables. The on shell energy-momentum tensor defined in Eq.~\eqref{eq:DefPseudoEMTensor} is then given by\footnote{Contrary to what is claimed in Ref.~\cite{Dong:2024zal}, this is the correct relation for an arbitrary background spacetimes $\bar g_{\mu\nu}$, since all additional terms associated to a variation of the nontrivial volume factor vanish upon imposing the background and linear equations of motion.}
\begin{equation}\label{SV}
    {}_{\mys{(2)}}t_{\mu\nu}[
\{\bar g_{\mu\nu},\bar \sfPsi\},\{h^H_{\mu\nu},\psi^H\}]=\frac{-2}{\sqrt{-\bar{g}}} \Biggl\langle \frac{\delta_{\mys{(2)}}S_\text{eff}}{\delta\bar{g}^{\mu\nu}}\Biggr\rangle\,.
\end{equation}
This statement is the key to understanding how the properties of the linearized theory automatically carry over to the a priori second-order operator defined as the energy-momentum tensor. Furthermore, the second-order action naturally provides the linear equations of motion by varying with respect to the corresponding perturbation fields \cite{Maccallum:1973gf,Heisenberg:2023prj}.

Although Eq.~\eqref{SV} holds for arbitrary exact solutions of the perturbative expansion $\bar g_{\mu\nu}$, in this work we want to give up this generality in favor of simplicity and only consider the case of $\bar g_{\mu\nu}=\eta_{\mu\nu}$ that is relevant for the computation of memory in asymptotically flat spacetimes. In this context, the second variation approach was already used to prove a theorem on the general form of the memory equation in the case of a local Lorentz-preserving background solution \cite{Heisenberg:2023prj}. With the use of the $3+1$ SVT decomposition presented in Sec.~\ref{sSec:LinearizedAE} we want to extend this work by loosening the assumption on local Lorentz invariance within the gravitational sector. Yet, for simplicity, will still assume that the background solution $\{\eta_{\mu\nu},\bar\sfPsi\}$ preserves invariance under spatial rotations, such that the decoupling of the tensor, vector and scalar sectors of a $3+1$ SVT decomposition presented in Sec.~\ref{sSec:LinearizedAE} is ensured. These are precisely the assumptions needed to discuss the case of Einstein-\AE{}ther gravity. Again, although the property of a clear separation of SVT spin sectors would be lost at the level of the second-order equations of motion, the Isaacson energy-momentum tensor will still exhibit this feature.

While the variation in Eq.~\eqref{SV} in principle assumes an arbitrary metric and the precise form of the background metric is only plugged after the variation has been carried out, it was explicitly proven in Lemma 1 of Ref.~\cite{Heisenberg:2023prj} that in the case of an asymptotically flat background solution $\{\eta_{\mu\nu},\bar\sfPsi\}$, the properties of the metric $\eta_{\mu\nu}$ in the Minkowski coordinates can explicitly be used \emph{before} performing the variation. This means that in computing the asymptotic energy-momentum tensor from Eq.~\eqref{SV}, one can in fact employ the flat version of the effective action
\begin{equation}\label{SV2}
    {}_{\mys{(2)}}t_{\mu\nu}[\{\eta,\bar\sfPsi\},\{h^H,\psi^H\}]=\frac{-2}{\sqrt{-\eta}} \Biggl\langle \frac{\delta_{\mys{(2)}}S_\text{eff}^{\mys{flat}}}{\delta \eta^{\mu\nu}}\Biggr\rangle\,,
\end{equation}
which is defined by inserting the properties $\partial_\alpha\eta_{\mu\nu}=\partial_\alpha\bar\sfPsi=0$ of the assumed background solution into the specific form of effective action
\begin{equation}\label{eq:SecondOrderActionFlat}
    _{\mys{(2)}}S_\text{eff}^{\mys{flat}}\equiv \phantom{}_{\mys{(2)}}S_\text{eff}[\{\eta,\bar\sfPsi\},\{h^H,\psi^H\}]\Big|_{\partial_\alpha\bar\sfPsi=0}\,.
\end{equation}
It is important to stress that the statements above crucially depend on the spacetime averaging over the high-frequency modes.

\subsubsection{gauge-invariant energy-momentum in E\AE{} Gravity}\label{ssSec:GaugeInvarianAsymptoticEMGen}

Given the gauge-invariant, flat and exclusively dynamical second-order action in Eq.~\eqref{gaugeS2} it is now straightforward to apply the $3+1$ decomposition to Eq.~\eqref{SV2} and compute the spatial components of the on shell asymptotic energy-momentum tensor of E\AE{} theory in each of the sectors as
\begin{subequations}\label{eq:self_stress_energy}
\begin{align}
    {}_{\mys{(2)}}t^\text{T}_{ij}&=\frac{1}{4\kappa_0} \,\Big\langle\partial_i h^{TT}_{Hab}\partial_jh_{TT}^{Hab}\Big\rangle\,,\label{eq:EmEAT}\\
    {}_{\mys{(2)}}t^\text{V}_{ij}&=\frac{\bar C_\text{V}}{4\kappa_0}  \Big\langle\partial_i \Sigma^H_{a}\partial_j\Sigma_H^{a}\Big\rangle\,,\\
     {}_{\mys{(2)}}t^\text{S}_{ij}&=\frac{\bar C_\text{S}}{4\kappa_0}  \Big\langle\partial_i \Theta^H\partial_j\Theta^H\Big\rangle\,,
\end{align}
\end{subequations}
where the coefficients $\bar C_\text{V}$ and $\bar C_\text{S}$ are defined in Eqs.~\eqref{eq:CV} and \eqref{eq:CS}, as follows
\begin{subequations}
\begin{align}
\bar C_\text{V}&=2(1-\bar{A}^2c_{13})(2c_1-\bar{A}^2 c_{13}(c_1-c_3))\,,\label{eq:CVs}\\
\bar C_\text{S}&=\frac{4-2\bar{A}^2c_{14}}{\bar{A}^2c_{14}}\,,\label{eq:CSs} 
\end{align}
\end{subequations}
Note that as a result of the Isaacson split in Eq.~\eqref{eq:IsaacsonSplit} the flat second-order action that enters the second variation approach [Eq.~\eqref{gaugeS2}] is defined in terms of high-frequency fields.

The plane-wave conditions [Eqs.~\eqref{eq:tensor_radiation}, \eqref{eq:vector_radiation} and \eqref{eq:scalar_radiation}] then imply that this result can be rewritten as
\begin{equation}\label{eq:EmGen2}
    {}_{\mys{(2)}}t_{ij}^\psi=\frac{1}{4\kappa_0}\sum_\psi \frac{\bar C_\psi}{V_\psi^2}\,\Big\langle\dot\psi_H^2\Big\rangle\,n_in_j\,,
\end{equation}
for $\psi=$ T, V, S. These represent the momentum flux of momentum $i$ in the direction $j$ (see also Appendix~\ref{App:EMTensorGen}). Due to the symmetry, these momentum fluxes can be characterized by a single momentum flux scalar $F_\psi$ per unit solid angle $d\Omega$ through
\begin{equation}\label{eq:EmGenFinal}
     {}_{\mys{(2)}}t^\psi_{ij}(t,r,\Omega)\equiv\frac{1}{r^2}F_\psi(t-r/\beta_\psi,\Omega)n_i(\Omega)n_j(\Omega)\,,
\end{equation}
with
\begin{subequations}\label{eq:self_stress_energyFlux}
    \begin{align}
    F_\text{T}(u_\text{T},\Omega)&=\frac{1}{4\kappa_0}\frac{r^2}{\beta_\text{T}^2}\langle \dot h^{TT}_{Hab}\dot h_{TT}^{Hab}\rangle\,,\\
     F_\text{V}(u_\text{V},\Omega)&=\frac{\bar C_\text{V}}{4\kappa_0}\frac{r^2}{\beta_\text{V}^2}\langle \dot \Sigma^{H}_a\dot \Sigma_{H}^{a}\rangle
     \,,\\
     F_\text{S}(u_\text{S},\Omega)&=\frac{\bar C_\text{S}}{4\kappa_0}\frac{r^2}{\beta_\text{S}^2}\langle \dot \Theta^{H}\dot \Theta^{H}\rangle\,,\label{eq:ScalarFlux}
\end{align}
\end{subequations}
where the velocities are defined in Eqs.~\eqref{eq:velT}, \eqref{eq:velV} and \eqref{eq:velS}, as follows
\begin{subequations}
\begin{align}
    \beta_\text{T}^2& = \frac{1}{1-\bar{A}^2c_{13}} \,,\\
    \beta_\text{V}^2& = \frac{ c_1 -\frac{1}{2}\bar{A}^2(c_1^2-c_3^2)}{(1-\bar{A}^2c_{13})c_{14}}\,,\\
   \beta_\text{S}^2& = \frac{c_{123}(2-\bar{A}^2 c_{14})}{(1-\bar{A}^2 c_{13})(2+\bar{A}^2(c_{123}+2c_2))c_{14}}\,.
\end{align}
\end{subequations}

Observe that we have here switched from the phase velocities $V_\psi$ to the group velocities $\beta_\psi$ and defined the asymptotic retarded time with respect to it, as follows
\begin{equation}
    u_\psi\equiv t-\frac{r}{\beta_\psi}\,.
\end{equation}
Although in the case of Einstein-\AE{}ther gravity, in which the phase and group velocities are equivalent, the fact that the asymptotic momentum and energy fluxes depend on the retardation set by the group velocity is an important general result \cite{PhysRev.105.1129}.\footnote{This statement can also be understood as directly following from asymptotic energy-momentum conservation and the fact that the momentum density and flux are precisely related by a factor of the group velocity.} It represents the key reason for the group velocity to enter the discussion and is important to keep in mind when comparing the results to earlier work \cite{Heisenberg:2024cjk}. We elaborate on this point and the comparison to nontrivial dispersion relations in Appendix~\ref{App:EMTensorGen}. Moreover, from the considerations in this appendix, it can be understood that through the fundamental relation between energy and momentum of a propagating energy of a wave packet, the scalar momentum fluxes in Eq.~\eqref{eq:self_stress_energyFlux} can readily be associated to corresponding energy fluxes. From now on, we will thus denote all velocities in terms of physical group velocities $\beta_\psi$.

Through the second variation approach, we therefore relatively straightforwardly obtained a manifestly gauge-invariant expression for the spatial components of the asymptotic energy-momentum tensor in Einstein-\AE{}ther gravity. Moreover, as we show in Appendix \ref{Appendix:lineareq}, reducing these expressions to the corresponding specific gauges, they match earlier computations in the literature. The spatial components of the energy-momentum tensor in the special form in Eq.~\eqref{eq:EmGen2} are in fact all we need for the computation of the memory, which is why we will leave a technically more complex computation of the full expressions for future work. However, in principle, applying the second variation approach to the manifestly gauge-invariant second-order action is expected to yield a well defined energy-momentum in all components. Indeed, in general, if the linear equations of motion can be written in a manifestly gauge-invariant way, the second variation approach indicates that this property automatically carries over to the asymptotic Isaacson energy-momentum tensor as well.

\subsection{Tensor memory equation}\label{sSec:TensorMemoryGen}

With a well defined general form of an asymptotic energy-momentum tensor of high-frequency gravitational waves, we are ready to come back to the memory equation identified in Sec.~\ref{sSec:Isaacson} and discuss its general solution in the context of locally Lorentz-breaking theories of gravity. As mentioned, we will however focus on the so-called tensor memory and leave a discussion of potential contributions to other types of memory for future work.


In terms of the Isaacson formalism discussed in Sec.~\ref{sSec:Isaacson}, the high-frequency propagation of the tensor degrees of freedom describing gravitational waves [Eqs.~\eqref{eq:EOMIISGR2}] of E\AE{} gravity is governed by Eq.~\eqref{eq:TTeom}, as follows
\begin{align}\label{eq:HTEOM}
     \left[ \phantom{}_{\mys{(1)}}\mathcal{G}_{ij}[h^H,\psi^H] \right]^{TT}= \frac{1}{2} \left(\frac{1}{\beta_\text{T}^2}\ddot h^{TT}_{Hij} -\partial^2 h^{TT}_{Hij}\right)=0\,.
\end{align} 
Note that the $TT$ on the left-hand side defines here a projection onto the $TT$ part of the equations of motion that can be achieved through the application of the projection operator
\begin{align}\label{eq:Projector}
    \perp_{i j, k l} \,\equiv\, \perp_{i k} \perp_{j l}-\frac{1}{2} \perp_{i j} \perp_{k l}\,,
\end{align}
with
\begin{equation}\label{eq:ProjectionT}
    \perp_{ij}\equiv \delta_{i j}-n_i n_j=u_i u_j+v_i v_j\,.
\end{equation}
We have defined the spatial basis of asymptotic radiation $\{n,u,v\}$ back in Eqs.~\eqref{eq:n} and \eqref{eq:Def Transverse Vectors u v}. Crucially, for plane waves in the asymptotically flat limit, such a $TT$ projection is equivalent to the definition of the $TT$ part of a $3+1$ decomposition of a general tensor that enters in Eq.~\eqref{eq:TTeom}. Since such an equivalence is sometimes disputed in the literature \cite{Racz:2009nq,Frenkel:2014cra,Ashtekar:2017wgq}, we explicitly offer a derivation in Appendix~\ref{App:TTProjection} that this feature is not bound to GR, but is true for any asymptotic plane-wave radiation, regardless of the precise form of the propagation equation.  

Now, since the linearized operator $ \phantom{}_{\mys{(1)}}\mathcal{G}_{ij}$ always takes the same form for both the low- and high-frequency asymptotic perturbations we can combine Eqs.~\eqref{eq:DefPseudoEMTensor} and \eqref{eq:HTEOM} to write the tensor memory equation [Eq.~\eqref{eq:EOMISGR2}] as
\begin{align}\label{eq:MemEqGenFinal}
   -\frac{1}{\beta_\text{T}^2}\ddot h_{Lij}^{TT}+\partial^2 h_{Lij}^{TT} = -2 \kappa_{0} \perp_{ij,kl} {}_{\mys{(2)}}t_{kl}\left[h^H,\psi^H\right] \,,
\end{align}
where the asymptotic energy-momentum tensor of the high-frequency waves ${}_{\mys{(2)}}t_{ij}\left[\psi^H\right] $ is given by Eqs.~\eqref{eq:EmGenFinal} and \eqref{eq:self_stress_energyFlux}.

\section{\label{sec:MemoryinEA}Displacement Memory in Einstein-\AE{}ther Gravity}

We are finally able to compute the gravitational tensor memory in Einstein-\AE{}ther gravity governed by the memory equation in Eq.~\eqref{eq:MemEqGenFinal}.
Back in Sec.~\ref{sec:BasicsEAtheory} we identified the dynamical degrees of freedom of E\AE{} gravity in terms of manifestly gauge-invariant variables within the three distinct tensor, vector and scalar sectors. Consequently, the resulting tensor memory will receive a distinct contribution from the emitted energy-momentum fluxes carried by dynamical variables in each of these decoupled sectors [Eq.~\eqref{eq:EmGenFinal}].
The only remaining task is now to determine whether Eq.~\eqref{eq:MemEqGenFinal} admits a well defined solution, and solve it in the appropriate asymptotic limit in the specific theory of Einstein-\AE{}ther gravity.

\subsection{Solve the memory equation}\label{sSec:SolveMemoryEqAE}

It is straightforward to write down the general solution to the memory equation in Eq.~\eqref{eq:MemEqGenFinal},
\begin{align}
   h_{Lij}^{TT}(x) = -2\kappa_{0} \,\perp_{ij,kl}  \int d^{4}x' \; G(x-x') \,{}_{\mys{(2)}}t_{kl} (x')   \label{eqn:SolForm} \,,
\end{align}
with the aid of the general Green's function of the wave equation \cite{Jackson:1998nia},
\begin{align}
    G(x-x') =-\frac{\delta(t_{ret}-t')}{4\pi  \left | \vec{x}-\vec{x}' \right | } H(t-t')\, , \label{eq:Green's function}
\end{align}
where $H$ is the Heaviside step function imposing causality and $t_{ret}$ is the retarded time adapted to the asymptotic speed of the tensor perturbations,
\begin{equation}
    t_{ret}\equiv t-\frac{|\bar{x}-\bar{x}'|}{\beta_\text{T}}\,.
\end{equation}
As it is the case for any gravitational radiation, to extract the corresponding component in the asymptotic radiation as the leading-order signal that arrives at a GW detector, we need to take the limit of this expression to future null infinity. As we will see, the radiative component within Eq.~\eqref{eqn:SolForm} will precisely represent a memory contribution to the asymptotic radiation.
However, due to the possibility of a nontrivial velocity of tensor degrees of freedom $\beta_\text{T}$, such a limit to null infinity is slightly altered, in the sense that the asymptotic retarded time kept fixed during this process is now defined as 
\begin{equation}\label{eq:AsymptoticRetardedTime}
    u\equiv t-\frac{r}{\beta_\text{T}}\,.
\end{equation}
Hence, the limit to null infinity is defined as the large distance limit $r\rightarrow\infty$ at fixed retarded time $u$. Indeed, the nontrivial part of the computation will precisely reside within the definitions of the suitable limits and the implications of the presence of different propagation speeds.

Let us perform a change of coordinates to
\begin{align}
    (t,x,y,z) \rightarrow (u,r,\theta,\phi)\, ,
\end{align}
with $u$ defined in Eq.~\eqref{eq:AsymptoticRetardedTime}.
Similarly, for the tensor, vector and scalar components in the asymptotic energy-momentum tensor, which represent the source in Eq.~\eqref{eqn:SolForm}, it is useful to switch to the relevant asymptotic coordinates
\begin{align}
    (t',x',y',z') \rightarrow (u'_\psi,r',\theta',\phi') \,,
\end{align}
with asymptotic retarded time
\begin{align}
    u'_\psi=t'-\frac{1}{\beta_\psi} \,r' \, ,
\end{align}
where $\psi=\text{S, V, T}$ again stands for the scalar, vector and tensor sectors respectively.\newline
Under this transformation, the volume form in the integration becomes
\begin{align}
    d^{4}x' \rightarrow r'^{2}du'_\psi dr'd\Omega' \, .
\end{align}

At this point, it is standard to assume that at every spacetime point where the source in Eq.~\eqref{eq:MemEqGenFinal} is non-zero, the condition
\begin{equation}\label{eq:assumptionSource}
    r'\ll r\,,
\end{equation}
is satisfied \cite{Garfinkle:2022dnm,Heisenberg:2023prj,Heisenberg:2024cjk,Zosso:2024xgy}. In other words, the value of the memory component evaluated at a given spacetime point in the asymptotic limit only depends on the value of the source in a confined radius of spacetime, and can thus be evaluated ``outside of its own source".
Assuming this, allows an expansion of the source-to-observer distance
\begin{align}
    \left | \vec{x}-\vec{x}' \right | \simeq r-r'\vec{n}' \cdot \vec{n} \, ,
\end{align}
where we have written the vectors $\vec{x}$ and $\vec{x}'$ as $r\vec{n}$ and $r'\vec{n}'$, where $\vec{n}=\vec{n}(\Omega)$ [Eq.~\eqref{eq:n}] and $\vec{n}'=\vec{n}(\Omega')$ are the direction of the detector and the source with respect to the coordinate origin. With the assumption in Eq.~\eqref{eq:assumptionSource}, the Green's function then takes the form\footnote{Here we use $\delta(g(x))=  \sum_{i} \delta(x-x_{i})/ \left | g'(x_i) \right |$, 
where $x_i$'s are roots of $g(x)$. Moreover, the Heaviside step function in the definition of the Green's function in Eq.~\eqref{eq:Green's function} is dropped, because in the presence of an integration over the entire spacetime, the retardation condition $\delta(t_{ret}-t')$ implies $H(t-t_{ret})=H(|\vec x -\vec x'|/V_h)=1$.}
\begin{align}
    G(x-x') &=  -\frac{ \delta \left(u-u'-r'\left\{\frac{1}{\beta_\psi}-\frac{\vec{n}' \cdot \vec{n}}{\beta_\text{T}}\right\}\right) }{4\pi r}
    \nonumber \\
    &=  -\frac{\beta_\psi \,|\mathcal{V}_\psi|\,\delta (r'-(u-u')\beta_\psi \mathcal{V}_\psi)  }{4\pi r}
    \, ,\label{eq:GreensFunctionFinal}
\end{align}
where we have defined the dimensionless ratio
\begin{equation}\label{eq:DefRatio}
    \mathcal{V}_\psi\equiv 
    \frac{1}{1-\frac{\beta_\psi}{\beta_\text{T}} \vec{n}' \cdot \vec{n}}\,.
\end{equation}
Plugging this expression into the general solution in Eq.~\eqref{eqn:SolForm} and solving the $r'$ integration results in a formal solution of the memory equation in the limit to asymptotic infinity. The delta function in Eq.~\eqref{eq:GreensFunctionFinal} characterizing the retardation of the signal hereby mainly also gives a condition on the integration bounds over $u'$ through the requirement that $r'>0$. As we will discuss below, this bound crucially depends on the sign of the coefficient $ \mathcal{V}_\psi$ in Eq.~\eqref{eq:DefRatio}. We will however refrain at this point to perform this general computation of the E\AE{} memory. The reason is that the final result is not entirely well defined as the source term includes parameters in which our assumptions of the computation break down.

Indeed, whenever $\beta_\text{T}\leq\beta_\psi$, there exists a value of $\vec{n}' \cdot \vec{n}=\beta_\text{T}/\beta_\psi$, for which the factor $ \mathcal{V}_\psi$ diverges. In the limiting case $\beta_\text{T}=\beta_\psi$, the infinity is present when $\vec{n}' \cdot \vec{n}=1$; hence, the direction vector of the source points in the same direction as one evaluates the memory. Such an apparent divergence is also present in the non-linear memory for GR that is sourced by the luminal tensor waves. However, if $\vec{n}' = \vec{n}$, the $TT$ projection in Eq.~\eqref{eqn:SolForm} together with the general form of the asymptotic energy-momentum tensor in Eq.~\eqref{eq:EmGen2} ensures that the memory source vanishes in these regions. In the end, the zero from the $TT$ projection together with the divergence in Eq.~\eqref{eq:DefRatio} results in a finite value of the memory. Yet, there is no similar cancellation of infinities in the case $\beta_\text{T}<\beta_\psi$. 
At first sight, in these divergent directions the confined source assumption $r'\ll r$ breaks down \cite{Garfinkle:2022dnm}. Hence, as a preliminary conclusion, in retrospect our assumption in Eq.~\eqref{eq:assumptionSource} is not valid everywhere and the computation requires a generalization. 

In the following Sec.~\ref{sSec:Constraining}, we will however show that the infinite limit in Eq.~\eqref{eq:DefRatio} as well as the breakdown of the assumption Eq.~\eqref{eq:assumptionSource} have a common physical origin, which ultimately will allow us to argue in favor of an exclusion of the parameter space of E\AE{} gravity that satisfies $\beta_\text{T}<\beta_\psi$. To understand this statement, we will undertake a detour in Sec.~\ref{sec:pulse} and consider a simplifying thought experiment of emitting single pulses of localized energy-momentum. To unravel the essence of the argument even more cleanly, we will first review some of the most stringent existing constraints on E\AE{} theory in Sec.~\ref{ssSec:Existing Constraints} that will allow us to reduce the discussion to an already confined theory space.

\subsection{Constraining superluminal Einstein-\AE{}ther gravity}\label{sSec:Constraining}

As mentioned, we wish to start this subsection by reviewing the two most important existing constraints on the propagation speed of E\AE{} waves. For a more complete list of existing constraints on the parameter space, we refer to the small summary in Appendix~\ref{App:Existing Constraints}.

\subsubsection{Existing Constraints on Propagation Speeds}\label{ssSec:Existing Constraints}

 As is well known, the multimessenger observation of GWs from a neutron star binary merger and the simultaneous optical signal from a short gamma-ray burst placed stringent bounds on the propagation speed of tensor gravitational waves as compared to the propagation of light \cite{LIGOScientific:2017zic}. More precisely, the electromagnetic signal was observed around $1.74$ s after the merger, such that together with the assumption that the gamma-ray burst did occur between the merger and $10$ s after the merger\footnote{Most models expect a delay below $4$ s.} one obtains a constraint on the propagation speed of gravitational tensor modes of \cite{LIGOScientific:2017zic},
\begin{equation}
    -3\times 10^{-15}<\beta_\text{T} -1 < 7\times 10^{-16}\,.
\end{equation}
Through Eq.~\eqref{eq:velT}, this directly translates to the constraint of the E\AE{} parameter \cite{Schumacher:2023cxh}
\begin{equation}
    c_{13}\approx \mathcal{O}(10^{-15})\,.
\end{equation}
which effectively sets $c_{13}=0$, equivalent to imposing a luminal propagation of the tensor modes. Therefore, for simplicity, we will assume that $\beta_\text{T}=1$  in the remainder of this work. From now on, any statement of sub- and superluminality will therefore refer to a speed of propagation below, respectively above the velocity of gravitational tensor waves.

On the other hand, one can obtain another type of stringent constraint on the propagation speeds of E\AE{} modes through the phenomenon of the so-called \emph{Cherenkov radiation} \cite{Cherenkov:1934ilx,V_Jelley_1955,Ratcliff:2021ofp}. In short, Cherenkov radiation arises as soon as charged relativistic particles propagate at a speed greater than the phase velocity of the corresponding force-carrier field. It occurs in electromagnetism within dielectric media that lower the effective speed of light, and the amount of emitted radiation can be very large, as it is described as an apparent divergence caused by an unbounded energy content of the solution that is only tamed by finite size effects. 

In the context of E\AE{} gravity, this means that for subluminal propagation speeds of the dynamical gravitational modes, high-energy cosmic rays that travel at a higher velocity will spontaneously emit Cherenkov excitations of the corresponding gravitational scalar, vector or tensor modes.\footnote{The process is only kinematically allowed for modes with subluminal propagation in the preferred frame set by the nonvanishing vector background.} Through the absence of such Cherenkov radiation from cosmic rays, one can derive the following constraints on the propagation speeds of E\AE{} degrees of freedom \cite{Elliott_2005,Gupta:2021vdj,Sarbach:2019yso},
    \begin{align}
    \beta_\psi^2 \gtrsim 1-\mathcal{O}(10^{-15})\,,
    \end{align}
for $\psi=\text{T, V, S}$.
Again, for simplicity, we will take this stringent constraint as enough reason to rule out any subluminal propagation of gravitational E\AE{} modes and impose $\beta_\psi\geq 1$ in order to discuss the gravitational memory in full E\AE{} gravity in Sec.~\ref{sec:MemoryinLuminalEA} below.

\subsubsection{Displacement memory from energy pulse emission}\label{sec:pulse}

In order to describe the argument for a severe constraint on the superluminal parameter space of Einstein-\AE{}ther gravity, it is worth breaking down the complexity for the sake of clarity and considering the simplest possible scenario as a thought experiment. In this subsection, we will therefore consider the tensor memory created by the emission of a single energy pulse, which will subsequently allow us to analyze the interesting effects of local Lorentz violation. 

Let us thus assume the emission of a pulse of energy $E_\text{p}$ from a localized source. For concreteness, we will assume this pulse to be made out of scalar radiation. The pulse ought to be extended enough such that it is possible to assign a definite energy-momentum to the associated wave packet by averaging over the high-frequency scales. Yet, it is still localized in space and its motion can be described via its ``center of energy'' that follows a radial outward trajectory,
\begin{equation}\label{eq:outwardTrajectory}
    x^\text{p}_i(t')=\beta^\text{S}_i\,(t'-t_0)\,,
\end{equation}
where we have assumed that the pulse was emitted at $t_0$ at the coordinate origin.
Using $x^\text{p}_i(t')=r_\text{p}(t')n^\text{p}_i$ and $\beta^\text{S}_i=\beta_\text{S}\,n^\text{p}_i$, for a given direction $\Omega_p=\{\theta_p,\phi_p\}$, we have
\begin{equation}\label{eq:sourceRadius1}
    r_\text{p}(t')=\beta_\text{S}\,(t'-t_0)\,.
\end{equation}
Note that these trajectories correspond to lines of constant retarded times of the source,
\begin{equation}
    u'_\text{S}=t'-\frac{r'}{\beta_\text{S}}\,,
\end{equation}
and we define
\begin{equation}\label{eq:defup}
    u'_\text{p}\equiv t'-\frac{r_\text{p}}{\beta_\text{S}}=t_0\,.
\end{equation}

In the context of such a localized pulse, it is possible to assign a corresponding energy-momentum tensor as follows (see Appendix~\ref{App:EMTensorGen} for a derivation)
\begin{align}
    t_{\mu\nu}^\text{p}(x') = p^\text{p}_\mu \beta^\text{S}_\nu \,\delta^{(3)}(\vec{x}'-\vec{x}_\text{p}(t))  \, , \label{eq:mattersourcesimp}
\end{align}
with the group four-velocity vector
\begin{equation}
    \beta^\text{S}_\mu = (-1, \beta_\text{S} n_i)\,,
\end{equation}
and four-momentum
\begin{equation}
    p_\mu=(-E_\text{p},\frac{E_\text{p}}{\beta_\text{S}}n_i)\,.
\end{equation}
Moreover, the momentum current is naturally related to the energy flux per solid angle, whose ``time evolution'' is in a first approximation dictated by the value of the asymptotic retarded time with respect to the group velocity of the scalar waves, as follows 
\begin{align}
    \frac{dE^\text{p}}{du_\text{S}'d\Omega'}&=r'^2\beta_\text{S} \,t^\text{p}_{00}=r'^2\beta_\text{S} \,E_\text{p} \,\delta^{(3)}(\vec{x}'-\vec{x}_\text{p}(t))\,,\nonumber\\
    &=\beta_\text{S} \,E_\text{p} \,\delta(r'-r_\text{p}(t'))\delta^2(\Omega'-\Omega_p)\,,\nonumber\\
    &= E_\text{p}\,\delta(u'_\text{p}-u'_\text{S})\delta^2(\Omega'-\Omega_p)\,.
\end{align}
The momentum current therefore reads
\begin{align}
   t_{ij}^\text{p}(u'_\text{S},r',\Omega')& = t_{00}^\text{p}\,n^p_i n^p_j = \frac{1}{r'^2\beta_\text{S}}\frac{dE^\text{p}}{du_\text{S}'d\Omega'}\,n^p_i n^p_j \,,\nonumber\\
   &=\frac{1}{r'^2\beta_\text{S}}E_p\,\delta(u'_\text{p}-u'_\text{S})\delta^2(\Omega'-\Omega_p)\,n^p_i n^p_j\,.\label{eq:spatialEMTpf}
\end{align}

This energy-momentum tensor neglects the fact that the wave packet carrying the energy pulse extends over the high-frequency/small scales defined by the frequencies of the emitted waves. To account for this, we want to smooth out the delta function in the retarded time by replacing $\delta\rightarrow D$, where
\begin{equation}
    D(u'_\text{S}-u'_\text{p})\equiv \frac{1}{\Delta u'_\text{S}\sqrt{2\pi}}e^{-\frac{(u'_\text{S}-u'_\text{p})^2}{2\Delta u'^2_\text{S}}}\,
\end{equation}
is a Gaussian wave packet of width $\Delta u'_\text{S}\sim 1/f_L$ for concreteness, so that the momentum current reads
\begin{align}
   t_{ij}^\text{p}(u'_\text{S},r',\Omega')=\frac{1}{r'^2\beta_\text{S}}E_\text{p}\,D(u'_\text{S}-u'_\text{p})\delta^2(\Omega'-\Omega_\text{p})\,n^\text{p}_i n^\text{p}_j\,.\label{eq:spatialEMTp}
\end{align}
Observe that, through the definition of the momentum flux scalars in Eqs.~\eqref{eq:EmGenFinal} and \eqref{eq:ScalarFlux}, this expression is related to the scalar waves characterized by the dof $\Theta$ through
\begin{equation}
    F_\text{S}(u_\text{S}',\Omega')=\frac{E_\text{p}}{\beta_\text{S}}D(u_\text{S}'-u_\text{p})\delta^2(\Omega'-\Omega_\text{p})\,.
\end{equation}

We obtain the corresponding displacement tensor memory by plugging Eq.~\eqref{eq:spatialEMTp} into Eq.~\eqref{eqn:SolForm} and following the appropriate steps outlined above to perform the suitable limit to null infinity. Concretely, assuming $r'\ll r$ and in terms of the asymptotic retarded time\footnote{Recall that we set the speed of tensor modes to unity due to the existing constraints discussed in Sec.~\ref{ssSec:Existing Constraints} above.} $u=t-r$ and $u'_\text{S}=t'-r'/\beta_\text{S}$ of the scalar wave source, we have that [Eq.~\eqref{eq:GreensFunctionFinal}]
\begin{align}
    G(x-x') =&- \frac{\beta_\text{S} \,|\mathcal{V}_\text{S}|\,\delta (r'-(u-u'_\text{S})\beta_\text{S} \mathcal{V}_\text{S})  }{4\pi r} 
    \, ,\label{eq:GreensFunctionFinalParticle}
\end{align}
where we recall that
\begin{equation}\label{eq:Vfactor}
    \mathcal{V}_\text{S}=
    \frac{1}{1-\beta_\text{S}\, \vec{n}' \cdot \vec{n}}\,.
\end{equation}

In the context of our thought experiment of an energy pulse with definite direction $\Omega_\text{p}$ selected by the delta function within Eq.~\eqref{eq:spatialEMTp}, such that $n'_i=n^\text{p}_i$, then Eq. \eqref{eq:Vfactor} becomes
\begin{equation}\label{eq:VspecialP}
    \mathcal{V}^\text{p}_\text{S}=
    \frac{1}{1-\beta_\text{S}\, \vec{n}^\text{p} \cdot \vec{n}}\,.
\end{equation}

We can now separately consider two cases in which $\mathcal{V}^\text{p}_\text{S}$ carries a different sign depending on whether $\beta_\text{S}<1/\vec{n}' \cdot \vec{n}$ or $\beta_\text{S}>1/\vec{n}' \cdot \vec{n}$,
\begin{align}
    \text{Case I:}&\quad  \mathcal{V}^\text{p}_\text{S}>0\,,\\
    \text{Case II:}&\quad \mathcal{V}^\text{p}_\text{S}<0\,.
\end{align}
Recall that the latter case is excluded if the propagation speed of scalar waves is subluminal $\beta_\text{S}<1$ and luminal $\beta_\text{S}=1$, whereas a superluminal $\beta_\text{S}>1$ propagation speed inevitably entails the existence of directions for which $\mathcal{V}_\text{S}<0$. For convenience, we will in the following label the three different possibilities for the magnitude of the propagation velocity $\beta_\text{S}$ compared to the luminal speed of the tensor waves $\beta_\text{T}=1$ as three different ``theory types'' (i-iii) and assign to each a given color, as summarized in the table below. 
\begin{table}[H]
\centering
\begin{tabular}{ |c||c|c|c|  }
\hline
    Theory type & \textcolor{OliveDrab}{(i)} &  \textcolor{Orange}{(ii)} &   \textcolor{RoyalBlue}{(iii)} \\
    \hline
    $\beta_\text{S}$ & $\beta_\text{S}<1$ & $\beta_\text{S}=1$ & $\beta_\text{S}>1$ \\
    \hline
    Case & (I) & (I) & (I) \& (II)\\
    \hline
  \end{tabular}
\end{table}

If $\mathcal{V}^\text{p}_\text{S}>0$ (Case I), the computation parallels with the familiar derivation of tensor memory in Lorentz-preserving theories. The retardation dictated by the delta function in Eq.~\eqref{eq:GreensFunctionFinalParticle} implies
\begin{equation}\label{eq:rpC}
   0\leq r'= (u-u'_\text{S})\beta_\text{S}\mathcal{V}_\text{S}\,,
\end{equation}
which in turn sets an upper bound on the integration over the source retarded time $u'_\text{S}\leq u$. 
Hence, in the limit to null infinity, the resulting tensor memory becomes
\begin{align}\label{eq:PulseMem1}
     _{(\text{I})}h_{Lij}^{TT}(u,r,\Omega,\Omega_\text{p})&=\frac{\kappa_{0}E_\text{p}}{2\pi r}H_\text{G}(u,u'_\text{p})\left[\mathcal{V}^\text{p}_\text{S}\,n^p_in^p_j\right]^{TT}\,,
\end{align}
where 
\begin{equation}
    H_\text{G}(u,u'_\text{p})\equiv\int_{-\infty}^u D(u'_\text{S}-u'_\text{p})du'_\text{S}\,,
\end{equation}
is a smoothed out  ``Gaussian'' step function of asymptotic value $1$ with an effective high-frequency cutoff at $f_L$. Moreover, recall that $\Omega=\{\theta,\phi\}$ parametrizes the direction of the evaluation of the memory, while $\Omega_\text{p}=\{\theta_\text{p},\phi_\text{p}\}$ indicates the direction of emission of the energy pulse in the source-centered reference frame.

It is interesting to evaluate this expression for specific values. For instance, for an emission of the pulse in the $z$ direction of the source-centered coordinate system, hence $\theta_\text{p}=0$, evaluating the TT projection via Eq.~\eqref{eq:Projector} and extracting the $+$ and $\times$ polarizations through Eq.~\eqref{eq:Polarization Modes} such that $_{(\text{I})}h_{L}^{+}\equiv\frac{1}{2}e^{ij}_+\,{}_{(\text{I})}h_{Lij}^{TT}$ and $_{(\text{I})}h_{L}^{\times}\equiv\frac{1}{2}e^{ij}_\times 
    \,{}_{(\text{I})}h_{Lij}^{TT}$, the result reduces to the simple expression
\begin{subequations}\label{eq:MemPulseSpecialI}
\begin{align}
    _{(\text{I})}h_{L}^{+}(u,r,\Omega,\{0,\phi_\text{p}\})&=\frac{\kappa_{0}E_\text{p}}{4\pi r}\frac{H_\text{G}(u,u'_\text{p})\sin^2\theta}{1-\beta_\text{S} \cos\theta}\,,\\
    _{(\text{I})}h_{L}^{\times}(u,r,\Omega,\{0,\phi_\text{p}\})&=0\,.
\end{align}
\end{subequations}
Let us notice here that the characteristics of a linear polarization and the dependence on the angles of the above signal are typical for displacement memory. 

\begin{figure}[H]
  \centering
    \includegraphics[scale=0.5]{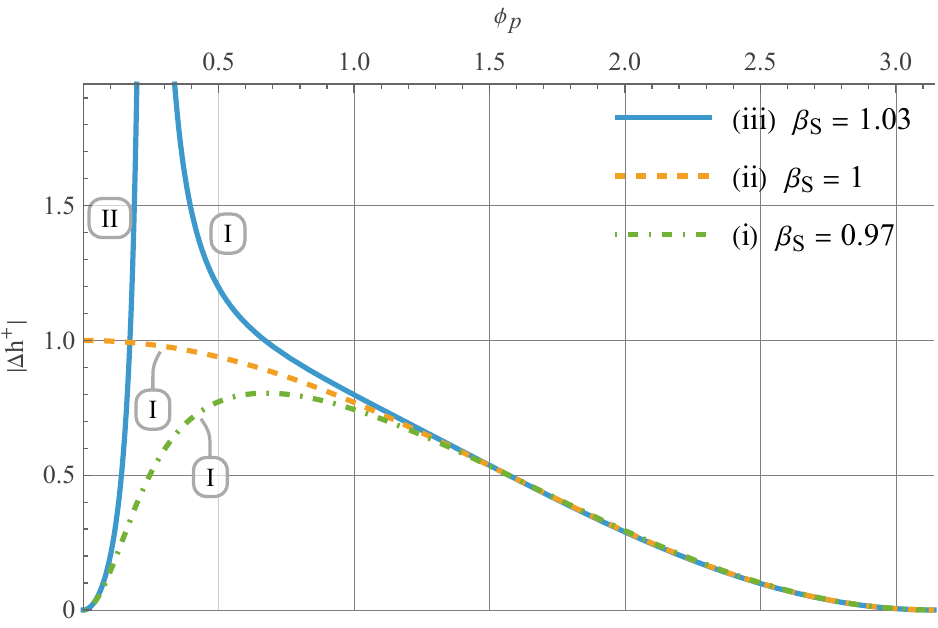}
  \caption{\small Total displacement memory offset in the $+$ polarization $\Delta h_{L}^{+}$ [Eq.~\eqref{eq:TotalMemOffset}], evaluated at $\Omega=\{\theta,\phi\}=\{\pi/2,0\}$ within the $x-y$ plane ($\theta_\text{p}=\pi/2$), as a function of the emission of a pulse of scalar waves in the direction $\phi_\text{p}\in\{0,\pi\}$ (symmetric results for $\{\pi,2\pi\}$). Depending on the sign of $\mathcal{V}^\text{p}_\text{S}$ [Eq.~\eqref{eq:VspecialP}] (Case I or II indicated in the plot), the memory is computed either through Eqs.~\eqref{eq:PulseMem1} or \eqref{eq:PulseMem2}. For illustrative purposes, we choose the values of $\kappa_0E_\text{p}/4\pi r=1$. Shown are representative curves of the three different theory types: superluminal scalar propagation speed $\beta_\text{S}>1$ (iii) (solid blue), luminal speed $\beta_\text{S}=1$ (ii) (dashed orange) and subluminal speed $\beta_\text{S}<1$ (ii)(dash-dotted green). In the subluminal and luminal theory types (i) and (ii), the displacement memory remains finite for any values of $\phi_\text{p}$. For $\beta_\text{S}=1$ (ii) the memory at $\phi_\text{p}=0$ (corresponding to $\vec{n}=\vec{n}_\text{p}$) the divergence in $\mathcal{V}^\text{p}_\text{S}$ is canceled by the angular structure of the memory formula. On the other hand, in the superluminal type (iii), the direction of unbounded memory sourcing $\vec{n}\cdot\vec{n}_\text{p}=1/\beta_\text{S}$ is not protected.}\label{fig:MemPulse}
\end{figure}

Alternatively, one can also fix the point of evaluation of the memory $\{\theta,\phi\}$ and vary the direction of emission of the pulse $\{\theta_\text{p},\phi_\text{p}\}$. Doing so for the choices $\Omega_x=\{\pi/2,0\}$ and $\theta_\text{p}=\pi/2$ results in a memory of the form
\begin{subequations}\label{eq:MemPulseSpecialIo}
\begin{align}
    _{(\text{I})}h_{L}^{+}\left(u,r,\Omega_x,\left\{\frac{\pi}{2},\phi_\text{p}\right\}\right)&=\frac{-\kappa_{0}E_\text{p}}{4\pi r}\frac{H_\text{G}(u,u'_\text{p})\sin^2\phi_\text{p}}{1-\beta_\text{S} \cos\phi_\text{p}}\,,\\
    _{(\text{I})}h_{L}^{\times}\left(u,r,\Omega_x,\left\{\frac{\pi}{2},\phi_\text{p}\right\}\right)&=0\,.
\end{align}
\end{subequations}
In the last two curves of Fig.~\ref{fig:MemPulse} we depict this dependence on $\phi_\text{p}$ of the total memory offset, as follows
\begin{align}\label{eq:TotalMemOffset}
    \Delta h_L^+\equiv&\;h_L^+(\{t\rightarrow\infty\})-h_L^+(\{t\rightarrow-\infty\})\,.
\end{align}
These curves correspond to the theory types (i) and (ii), that is, for a subluminal $\beta_\text{S}<1$ and luminal $\beta_\text{S}=1$ speed of scalar radiation, for which the condition (I) $\mathcal{V}^\text{p}_\text{S}>0$ is always satisfied.\footnote{Recall that the constraints from Cherenkov radiation discussed in Sec.~\ref{ssSec:Existing Constraints} effectively excludes the subluminal theories. For illustrative purposes, we still discuss this case here.}

On the other hand, if $\mathcal{V}^\text{p}_\text{S}<0$ (Case II) the conclusion from retardation in Eq.~\eqref{eq:rpC} is exactly the opposite and the memory at a given retarded time $u$ is sourced by all values of $u'_\text{S}>u$. This translates into the memory formula
\begin{align}\label{eq:PulseMem2}
     _{\text{(II)}}h_{Lij}^{TT}(u,r,\Omega,\Omega_\text{p})&=\frac{\kappa_{0}E_\text{p}}{2\pi r}[H_\text{G}(u,u'_\text{p})-1]\left[\mathcal{V}^\text{p}_\text{S}\,n^p_in^p_j\right]^{TT}.
\end{align}
At first sight, the statement that the memory at a given instant $u$ depends on all future source retarded times $u'>u$ might be surprising. However, it does not represent a violation of causality of the full spacetime. To further understand how such an inversion of the relationship between source and asymptotic observer is to be interpreted correctly, we provide a short context discussion on causality in Lorentz-breaking theories in Appendix~\ref{App:Causality}. As we show there, the statement that in some cases $u'_\text{S}>u$ can readily be understood in the context of the emission of a single energy pulse.

Similar to Eqs.~\eqref{eq:MemPulseSpecialI} and \eqref{eq:MemPulseSpecialIo}, we can evaluate the expression of the memory in Case II [Eq.~\eqref{eq:PulseMem2}] in the special choice $\theta_\text{p}=\pi/2$, as follows
\begin{subequations}\label{eq:MemPulseSpecialII}
\begin{align}
    _{(\text{II})}h_{L}^{+}(u,r,\Omega,\{\frac{\pi}{2},\phi_\text{p}\})&=\frac{\kappa_{0}E_\text{p}}{4\pi r}\frac{[1-H_\text{G}(u,u'_\text{p})]\sin^2\phi_\text{p}}{1-\beta_\text{S} \cos\phi_\text{p}}\,,\\
    _{(\text{II})}h_{L}^{\times}(u,r,\Omega,\{\frac{\pi}{2},\phi_\text{p}\})&=0\,.
\end{align}
\end{subequations}
Recall that this Case II only arises in superluminal theories (iii) with scalar velocity $\beta_\text{S}>1$.
With the two solutions in Eqs.~\eqref{eq:PulseMem1} and \eqref{eq:PulseMem2} at hand, we can now also assess the characteristic behavior of the final memory for this theory space.  As can be seen in the plot of Fig.~\ref{fig:MemPulse}, where the emission angle $\phi_\text{p}$ is varied at a fixed evaluation point $\Omega_x=\{\frac{\pi}{2},0\}$ and $\theta_\text{p}=\frac{\pi}{2}$ of the memory, the transition between the two regimes at $\beta_\text{S}=1/\vec{n}_\text{p} \cdot \vec{n}$ involves the expected divergence of the memory signal. In the following, we want to understand the origin of this apparent divergence and draw associated conclusions. 

\subsubsection{Understanding the directions of large memory buildup}\label{ssSec:UnderstandingLarge}

We are now in a place to physically understand the divergence in the memory computation that relates the two regimes $\mathcal{V}^\text{p}_\text{S}>0$ (I) and $\mathcal{V}^\text{p}_\text{S}<0$ (II) identified above, and deduce a conjecture on a severe constraint on superluminal E\AE{} theory. Through Eq.~\eqref{eq:GreensFunctionFinalParticle}, the magnitude of this factor determines the magnitude of the Green's function, which describes the propagated influence of the source on its effect.

For this purpose, it is useful to further analyze the behavior of the velocity factor $\mathcal{V}^\text{p}_\text{S}$ within the subluminal and luminal theory types (i) and (ii), in the particular direction $\vec{n}_\text{p}=\vec{n}$ in which the source pulse is emitted in the same direction as we want to evaluate the memory. In this direction, the velocity factor [Eq.~\eqref{eq:VspecialP}] becomes $\mathcal{V}^\text{p}_\text{S}|_{\vec{n}_\text{p}=\vec{n}}=1/(1-\beta_\text{S})$, and will thus grow significantly as the scalar velocity is approaching the luminal speed. 

\begin{figure}[H]
  \centering
    \includegraphics[scale=0.68]{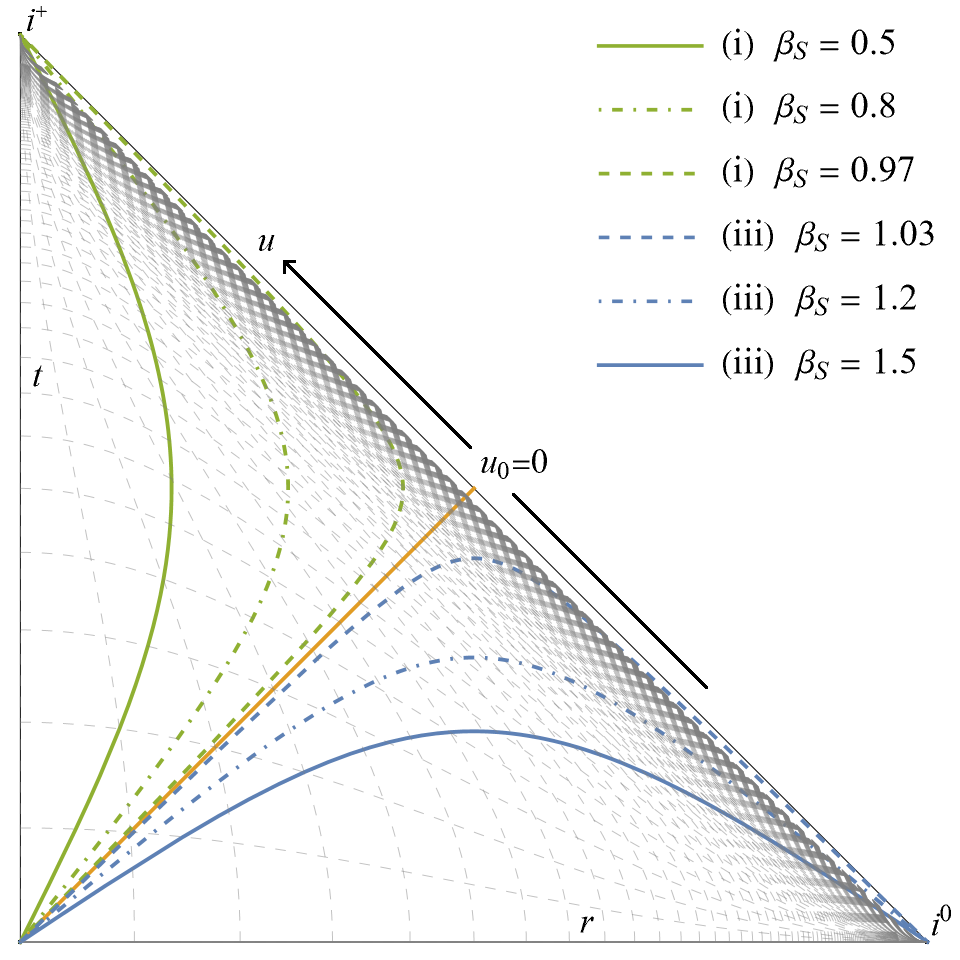}
  \caption{\small Penrose diagrams of conformally compactified asymptotically flat spacetimes in the source-centered coordinates, in which the null direction is chosen to be the one set by the gravitational tensor modes that are assumed to propagate luminally $\beta_\text{T}=1$ with asymptotic retarded time $u\equiv t-r$. Only the temporal and radial coordinates $\{t,r\}$ are represented and one may read the diagram as depicting one particular angular direction in which the spatial direction of the energy pulse emission $\vec{n}_\text{p}$ and the direction of evaluation of the memory $\vec{n}$ coincide. Shown are trajectories of radially outward pulses of scalar wave energy-momentum labeled by a constant asymptotic source retarded time $u'_\text{p}=t_0=0$ [Eq.~\eqref{eq:defup}] parametrizing their time of emission at the spatial coordinate origin. Pulses of theory type (i) with subluminal group velocity $\beta_\text{S}<1$ (green) reach timelike infinity $i^+$, while in theory type (iii) with superluminal group velocity $\beta_\text{S}>1$ (blue) the asymptotic trajectories reach spatial infinity $i^0$ of this particular compactification. In both cases, the pulse emission directions asymptote towards the spacetime direction set by the past light cone (orange) of an event of memory production at $u_0=0$ at future null infinity.}\label{fig:PenroseF}
\end{figure}

The geometrical reason for this behavior can be understood from Fig.~\ref{fig:PenroseF}. In this figure, we depict a portion of a conformally compactified Penrose diagram of asymptotically flat spacetime and choose an event of memory production at future null infinity characterized by $u_0=0$. The diagram can be thought of as precisely showing the particular angular direction given by $\vec{n}_\text{p}=\vec{n}$. Shown are energy pulses of scalar waves with different magnitudes of group velocity $\beta_\text{S}$.\footnote{Recall that the trajectories of the center of energy of such pulses follow trajectories that are characterized by a constant asymptotic ``source regarded time'' $u'_\text{S}=u'_\text{p}=t_0$ that parametrizes its emission time in terms of coordinate time $t$.} At subluminal speeds $\beta_\text{S}<1$, the pulses trace out a curve in spacetime (green) that asymptotes towards the temporal infinity $i^+$. As the speed is increased, the emission lines of course tend to the luminal line (orange line in Fig.~\ref{fig:penroseD}), which in the direction $\vec{n}_\text{p}=\vec{n}$ characterizes the past null cone of the event of memory evaluation at $u=0$. Hence, the source-cause relation of memory production characterized by $\mathcal{V}^\text{p}_\text{S}$ is significantly enhanced as the spacetime direction of the pulse emission aligns with the past null cone of the memory event. 

Indeed, if the scalar source itself is propagating at the speed of the tensor waves $\beta_\text{S}=1$ (ii) one formally has that $\mathcal{V}^\text{p}_\text{S}\rightarrow \infty$ as we approach the critical spacetime direction along the past null cone of the event $u_0=0$ at future null infinity. 
 In the luminal case, this divergence in the Green's function factor is however counterbalanced by the transverse-traceless nature of the signal. As already discussed above, the structure of the memory source is such that it vanishes in the particular direction $\vec n_\text{p}=\vec n$. In fact, the TT projection in Eq.~\eqref{eq:PulseMem1} precisely cancels the divergence in $\mathcal{V}^\text{p}_\text{S}$, as can clearly be seen in Fig.~\ref{fig:MemPulse} where the luminal memory (ii) approaches a finite value as $\phi_\text{p}\rightarrow \phi=0$. 

 In conclusion, the Green's function of memory production is significantly enhanced if the spacetime direction of emission of the energy pulse is approaching a past null cone of the corresponding spacetime point of memory evaluation. This provides an understanding of the origin of the divergence in $\mathcal{V}^\text{p}_\text{S}$. For propagation speeds of $\beta_\text{S}\leq 1$, this can only happen as the speed of the source waves is approaching the luminal speed of the tensor waves $\beta_\text{T}=1$.

If the scalar speed of propagation is faster than the ones of the tensor waves $\beta_\text{S}>1$ (iii), this situation changes drastically. Note first of all that in the direction $n^p_i=n_i$ plotted in Fig.~\ref{fig:PenroseF}, hence in case $\mathcal{V}^\text{p}_\text{S}<0$ (II), the conclusions parallel with the discussion of the subluminal theory type (i) above. The factor $\mathcal{V}^\text{p}_\text{S}$ remains finite, but is significantly enhanced as the scalar velocity tends towards luminality, while the total memory is, however, set to zero through the $TT$ projection. Yet, there is also a crucial difference. In the superluminal case, the corresponding memory Green's function associated to the emission of energy pulses is not only enhanced as $\beta_\text{S}$ approaches luminality, but equally as the direction of the pulse is being moved away from the direction $n^p_i=n_i$ at fixed speed $\beta_\text{S}$. Indeed, there always exists a set of pulse directions 
\begin{equation}\label{eq:ConditionMemEnhancement}
    \vec{n}_\text{p} \cdot \vec{n}=1/\beta_\text{S}\,,
\end{equation}
in which a divergence $\mathcal{V}^\text{p}_\text{S}\rightarrow \infty$ occurs. In contrast to the luminal case, this divergence in the memory is, however, not protected anymore.

\begin{figure}[H]
  \centering
    \includegraphics[scale=0.59]{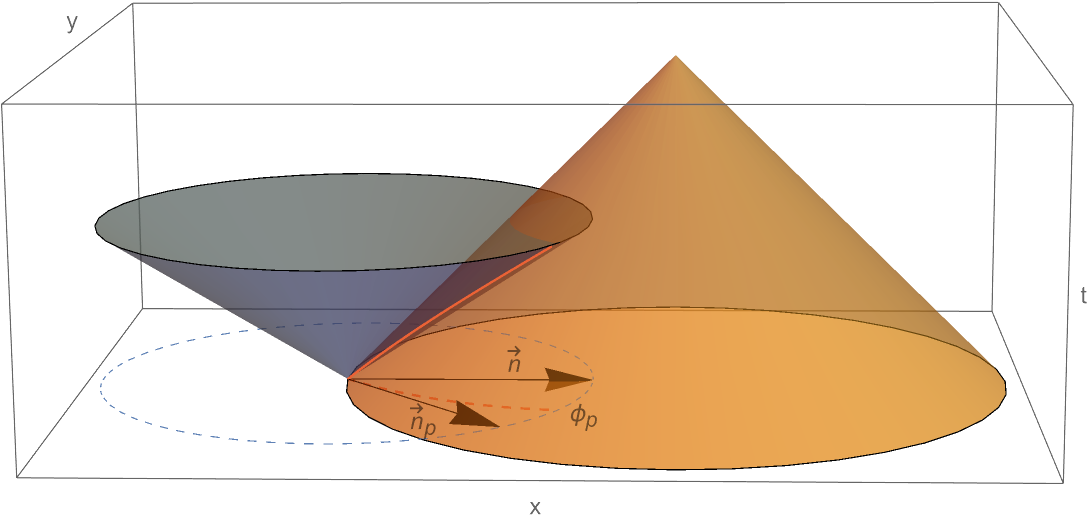}
  \caption{\small Causal cone (blue) of emission of superluminal scalar waves at a speed $\beta_\text{S}=1.5$. Shown are only two spatial directions $x$  and $y$ and the time direction $t$. The cone intersects a past null cone of a memory event (orange) admitting critical directions of emission that give rise to a formally unbound memory buildup. Shown are also the spatial projections of the cone (dashed blue) and the intersection line (dashed red), as well as representatives of the directions $\vec{n}$ of memory evaluation and $\vec{n}_\text{p}$ of pulse emission. The angle $\phi_\text{p}$ between the two directions precisely corresponds to the critical direction identified in Eq.~\eqref{eq:ConditionMemEnhancement} $\vec{n}\cdot\vec{n}_\text{p}=\cos\phi_\text{p}=1/\beta_\text{S}$.}\label{fig:LCones}
\end{figure}

Intuitively, the single direction of infinite memory sourcing in the luminal case is opened up for superluminal propagation to an entire line of directions characterized by Eq.~\eqref{eq:ConditionMemEnhancement}. As already understood above, these directions precisely correspond to the ones in which the scalar energy pulse is emitted along the past light cone of a given spacetime point of memory evaluation. This statement can readily be checked by computing the direction of the intersection between the appropriate light cones at a given point of pulse emission. Reducing the parameter space by fixing $\theta=\theta_\text{p}=\pi/2$ and $\phi=0$, this intersection of causal cones can be visualized as in Fig.~\ref{fig:LCones}. The superluminal emission of energy pulses provides nontrivial directions of emission along the past null memory cone which are precisely characterized by an angle\footnote{This can readily be checked by computing the hyperbolas of intersection of the two cones (red line in Fig.~\ref{fig:LCones}), which also provides the spacetime direction vector of an emission along the past null cone of the memory production.}
\begin{equation}
    \phi_\text{p}=\arccos 1/\beta_\text{S}.
\end{equation}
As expected, this equation is equivalent to the condition of unbound memory enhancement given by Eq.~\eqref{eq:ConditionMemEnhancement}. 

In summary, the geometrical picture provided above clearly shows that the unprotected enhancement in memory production through a drastic increase in the corresponding Green's function factor $\mathcal{V}^\text{p}_\text{S}$ will only exist if $\beta_\text{T}<\beta_\text{S}$.
In other words, the causality structure of theory type (iii) is such that for every given $\vec{n}$ there exists an angular direction $\vec{n}_\text{p}$ in the transition between regions (II) and (I), where the energy pulse will be emitted along the past null cone of a memory event, a priori resulting in an unbounded amplitude of displacement memory.

\subsubsection{Conjecture of new constraint on E\AE{} Gravity}\label{ssSec:ConstrainigSuperluminalEA}

Given the physical origin of this divergence in the memory formula, it represents first of all an interesting feature of Lorentz-breaking theories that might provide a characterizing signature in GW data. Our computation is strictly speaking not valid at the diverging point, since basic assumptions are not satisfied anymore, and we therefore do not provide a quantitative calculation of how big the memory offset will truly be. Yet, the geometrical understanding of the phenomenon as well as the clear trend of the increasing memory value within the neighboring regimes in which our assumptions still hold (see Fig.~\ref{fig:MemPulse}), are strong indications that the phenomenon is physical and displacement memory sourced by superluminal scalar waves indeed admit a causal structure with directions of very large enhancement. As such, we actually want to argue that this phenomenon can already be used to effectively rule out the possibility of superluminal propagation speeds within E\AE{} gravity. 

Returning to a physical situation of a continuous and smoothly spread out emission of scalar waves from a source in E\AE{} theory, for any given spacetime point of memory evaluation with direction $\vec{n}$, there will always be an emission of scalar wave energy-momentum in the critical direction $\vec{n}_\text{p} \cdot \vec{n}\rightarrow 1/\beta_\text{S}<1$. Thus, assuming that E\AE{} gravity is the theory of gravity governing our reality, any emission of spherically asymmetric scalar radiation with a propagation velocity greater than the tensor speed will source an a priori unbounded tensor displacement memory that presumably would already have shown up in current GW observations.

More precisely, in the context of compact binary coalescences, the mere presence of a Lorentz-breaking \ae{}ther field in which compact objects are moving, will give rise to nontrivial sensitivities of the strongly gravitating objects that in turn source the emission of the scalar as well as the vector degrees of freedom \cite{Foster:2007gr,Gupta:2021vdj,Schumacher:2023cxh,Taherasghari:2023rwn}. As explicitly shown in Ref.~\cite{Foster:2007gr} such radiation will inevitably carry away an asymptotic energy-momentum flux from an isolated binary system, which in turn will always source a tensor displacement memory as presented above. While the sensitivities of black holes in E\AE{} theory have not been computed yet due to the computational complexity, the sensitivities of neutron stars are known and have already been used to put constraints on the theory based on current GW data \cite{Gupta:2021vdj,Schumacher:2023cxh}. Moreover, at the merger of observed neutron star binaries, the cutoff frequency of the memory is generally expected to be high enough to reach the current detector frequency bands, such that in combination with an exceedingly large amplitude, one should expect the presence of a detectable signal. Thus, at the very least, the absence of memory within the observed binary neutron star coalescences \cite{Tiwari:2021gfl} poses a theoretically well-founded constraint on unexpectedly large values of displacement memory. The constraints are, of course, even more stringent as soon as also an emission of E\AE{} scalar and vector radiation from binary black hole systems is expected \cite{Cheung:2024zow}.\footnote{The precise nature of E\AE{} black holes and especially their capacity to support additional \ae{}ther charges is however not yet completely settled \cite{Eling:2006ec,Konoplya:2006ar,Barausse:2011pu,Barausse:2013nwa,Lin:2014ija,Adam:2021vsk,Cardoso:2024qie}.}

We therefore take the results of this work as reason enough to conjecture an exclusion of the superluminal theory space of Einstein-\AE{}ther gravity. Indeed, even though the present subsection focused entirely on the scalar waves, the exact same arguments also hold in the vector sector of the theory. In combination with the Cherenkov constraints discussed in Sec.~\ref{ssSec:Existing Constraints}, the parameter space of Einstein-\AE{}ther gravity is therefore reduced to the luminal theory with $\beta_\text{T}=\beta_\text{V}=\beta_\text{S}=1$.

Such a restriction to the propagation at the speed of light of all the dynamical dofs of the theory has important consequences, in particular as concerns the gravitational polarization spectrum of the theory. As discussed in Sec.~\ref{ssSec:GravPolAE}, luminality forbids the presence of any vector polarizations within a direct GW detector response, while the scalar longitudinal and breathing polarizations are trivially related. Interestingly, the Lorentz-violating nature of E\AE{} gravity still allows for the presence of a luminally propagating longitudinal polarization. This is opposite to what is observed in local Lorentz-preserving metric theories, even if one allows for a spontaneous breaking of the Lorentz symmetry. In Sec.~\ref{sec:MemoryinLuminalEA} we will present the full results of tensor null displacement memory in luminal Einstein-\AE{}ther gravity.

However, it is important to stress that the strong memory-based constraint on E\AE{} gravity considered above should be regarded as a conjecture, since several caveats are to be discussed. We already pointed out the breakdown of the assumptions that do not allow us to draw any quantitative conclusions from the computations along the diverging directions. Concretely, the localized source assumption $r'\ll r$ seems not valid anymore as pointed out in Ref.~\cite{Garfinkle:2022dnm}.\footnote{Note, however, that strictly speaking, the memory of individual pulses at a given instant $u$ can be thought of as being ``created'' at $u'_\text{S}=u$, such that the unboundedness in the source radius $r'$ in Eq.~\eqref{eq:rpC} is not conclusive.} Moreover, not knowing the true amplitude of memory in these cases, formally also the perturbation theory assumption Eq.~\eqref{eq:PertAssumption} is to be questioned. And finally, any finite-size effects have largely been neglected throughout this work, which might represent a necessary ingredient in assessing a quantitative result. A final concrete answer would therefore require a significant generalization of the computational methods presented here, which we leave as an interesting but challenging task of future work. 

In addition, the modeling of a memory signal from a realistic compact binary coalescences (or other concrete GW sources) would actually require the knowledge of the full waveforms of the emitted tensor, vector and scalar radiation, in particular as concerns the merger dynamics. A full-fledged numerical simulation of such an event within Einstein-\AE{}ther gravity is not yet available, although first steps in this direction are already being considered \cite{Sarbach:2019yso}. Such a careful case-based investigation would, in particular, also settle the important question on the precise characteristics of the memory signal in terms of frequency content that is crucial for an assessment of detectability through current detectors. Nevertheless, despite this lack of knowledge in several aspects of the problem, already the statement that any superluminal emission of energy-momentum content inevitably possesses unprotected directions of large causal memory buildup is, from our point of view, a notable new mechanism with expectedly important consequences.

\subsection{\label{sec:MemoryinLuminalEA}Displacement memory in luminal Einstein-\AE{}ther gravity}

In the luminal case, $\beta_\psi=1$ for $\psi =$ S, V, T, imposed by the parameter values in Eqs.~\eqref{eq:LuminalityT}, \eqref{eq:LuminalityV} and \eqref{eq:LuminalityS}, the velocity ratio $\mathcal{V}_\psi$ is always strictly positive and the general solution of the displacement memory [Eq.~\eqref{eqn:SolForm}] in the limit to null infinity remains always finite.\footnote{Recall that the divergence in $\mathcal{V}_\psi$ at $\vec n'=\vec n$ is compensated by the fact that in this case also $\perp_{ij,kl}n'_k n'_l = 0$.} 

By plugging Eq.~\eqref{eq:GreensFunctionFinal} into Eq.~\eqref{eqn:SolForm} and performing the appropriate limit $t,r\rightarrow\infty$ at constant $u$ we arrive at the full memory formula of Einstein-\AE{}ther gravity, as follows
\begin{align}
    h_{L ij}^{TT} (u,r,\Omega) &= \frac{ \kappa_0 }{2\pi r} 
    \int_{-\infty}^{u} du'  \int d\Omega' 
     \, \left[\frac{n'_i n'_j}{1-\vec n \cdot \vec n'} \right]^{TT} \nonumber\\
    & \times \left\{F_T(u',\Omega')+F_V(u',\Omega')+F_S(u',\Omega')\right\}\label{eq:LuminalMemory}
    \,,
\end{align}
where the energy flux scalars read
\begin{subequations}
    \begin{align}
    F_\text{T}(u',\Omega')&=\frac{1}{4\kappa_0}r'^2\langle \dot h^{TT}_{Hab}\dot h_{TT}^{Hab}\rangle\,,\\
     F_\text{V}(u',\Omega')&=\frac{ C_\text{V}}{4\kappa_0}r'^2\langle \dot \Sigma^{H}_a\dot \Sigma_{H}^{a}\rangle
     \,,\\
     F_\text{S}(u',\Omega')&=\frac{ C_\text{S}}{4\kappa_0}r'^2\langle \dot \Theta^{H}\dot \Theta^{H}\rangle \,,
\end{align}
\end{subequations}
that now depend on the luminal asymptotic retarded time $u' = t'-r'$. Moreover, the coefficients $ C_\text{V}$ and $ C_\text{S}$ of the luminal theory become
\begin{subequations}
\begin{align}
    C_\text{V}&=4c_1\,,\\
    C_\text{S}&=\frac{4}{\bar{A}^2c_{1}}-2 \,.
\end{align}
\end{subequations}
Recall that the asymptotic high-frequency waves fall off as $1/r$ near null infinity, such that the fluxes $F_\psi$ do not depend on $r'$ as they should.
The nontrivial theory dependence in the memory formula if luminal Einstein-\AE{}ther gravity in Eq.~\eqref{eq:LuminalMemory} therefore lies entirely in the coefficients $C_\text{V}$ and $C_\text{S}$ as well as the proper identification of the gauge-invariant dynamical degrees of freedom of the theory in Sec.~\ref{sec:BasicsEAtheory}.


\section{Discussion and Conclusion}\label{sec:Discussion}

Building on the perturbative Isaacson approach to study gravitational memory effects, we have explicitly analyzed tensor memory in Einstein-\AE{}ther theory. Our first key result is the derivation of the second-order perturbed action in a manifestly gauge-invariant form. Using this action, we obtained the linearized equations for the propagating modes and further reduced the second-order action to depend solely on the dynamical degrees of freedom.
We verified that the polarization modes of E\AE{} theory align with previous literature and derived the self-stress-energy tensor directly from the dynamical second-order action. This enabled us to formulate and solve the general tensor memory equation sourced by the self-stress-energy of the propagating modes, without relying on asymptotic symmetries.

Our analysis shows that, in general, the tensor memory diverges in certain directions. The physical origin of such divergence is the causal structure induced by a propagation of the memory source that is greater than the velocity of gravitational tensor waves. To study this behavior in detail, we assume an empirically imposed luminal propagation of gravitational tensor waves and consider a simplified scenario with individual energy pulse emissions as the memory source.
For luminal and subluminal propagation of the source, the tensor memory behaves consistently with known results from massless SVT and massive Horndeski theories as in Refs.~\cite{Heisenberg:2023prj, Heisenberg:2024cjk}.
However, when the source propagates superluminally, the displacement memory amplitude formally diverges at a critical angle, in which the energy pulse’s light cone intersects the observer’s past null cone.

This divergence is an indication of the breakdown of our methods in providing a quantitative assessment of displacement memory in the neighborhood of the critical spacetime directions. Nonetheless, the underlying geometrical understanding of the causal structure that parallels with the well understood behavior in the luminal case, as well as the increasing trend of the memory signal in regimes around the critical spacetime directions - in which our methods are expected to be valid - strongly suggest that the phenomenon is physical in nature. 
The emission of energy-momentum carrying radiation at a propagation speed greater than the speed of tensor gravitational waves leads to a large enhancement in the amplitude of the induced tensor displacement memory. This qualitative insight forms the basis of our central conjecture: Einstein-\AE{}ther theory with superluminal propagation speeds is incompatible with current gravitational wave data due to the absence of such large memory effects.

A natural extension of our results will therefore be a development of a formalism capable of performing a quantitative analysis in the region where the memory goes unbound under the assumptions used in this work. We in particular want to note the striking resemblance of the criticality condition in Eq.~\eqref{eq:ConditionMemEnhancement} with the condition on the emission angle of Cherenkov radiation (see, for instance, the first equation in Ref.~\cite{Ratcliff:2021ofp}). While the concrete physical manifestation is, of course, rather distinct, some of the conceptual aspects of the phenomenon bear interesting similarities too. Based on a connection of the two mechanisms, it can be envisaged that the a priori unbound memory buildup might be rendered finite through finite-size effects, in a similar way as in the case of Cherenkov radiation. 

We also want to emphasize that the phenomenon of large memory buildup has a priori nothing to do with superluminality per se. The bound is placed at superluminal speeds solely due to the multimessenger constraint that bounds the propagation velocity of the tensor modes to be luminal, as discussed in Sec.~\ref{ssSec:Existing Constraints}. In general, the phenomenon of tensor memory is present for any radiation emission at speeds larger than the speed of tensor waves. The same is in principle true for Cherenkov radiation \cite{Babichev:2024uro}. It is to our knowledge not a settled question whether the emission of superluminal scalar and vector \ae{}ther waves would in turn also lead to the emission of Cherenkov-type radiation.

Although this work focused on the tensor memory effect in Einstein-\AE{}ther theory, the framework developed here opens several important avenues for future research. In particular, a straightforward continuation is to extend the analysis to vector and scalar memory in E\AE{} gravity. These contributions could provide further theoretical insights to help distinguish between different gravitational frameworks.
Additionally, since Einstein-Æther theory is a specific example within the broader class of scalar-vector-tensor metric theories, it would be valuable to apply similar methods to other Lorentz-violating theories. A particularly interesting candidate theory for this purpose is Generalized Proca gravity \cite{Heisenberg:2014rta,BeltranJimenez:2016rff}, in which the local Lorentz symmetry can be broken spontaneously in the asymptotic limit. Computing the tensor memory in this context would not only test the robustness of our approach, but would also yield an interesting opportunity for a comparison of the two scalar-vector-tensor theories \cite{Zosso:2024xgy}.

\begin{acknowledgments}
We would like to thank Kristen Schumacher, Kent Yagi and Nico Yunes for useful discussions in the comparison to their previous works. Moreover we are grateful for the constructive comments of the anonymous referee, in particular on the subtleties on the causal structure in Appendix~\ref{App:Causality}. J.Z. is supported by funding from the Swiss National Science Foundation
Grant No. 222346. The Center of Gravity is a Center of Excellence funded by the Danish National Research Foundation under Grant No. 184.

\end{acknowledgments}



\appendix

\section{Definition and iInterpretation of the Einstein-\AE{}ther action}\label{Appendix:Action}

Naively, the most general action of a metric theory including an aether vector field $A$ coupled to the physical metric $g$ up to two powers of derivative operators and up to integrations by parts, is given by
\begin{align}
    S =&\, \frac{1}{2\kappa}\int d^4x \sqrt{-g} \left ( \sum_{i=2}^{4} \mathcal{L}_i[g,A]+\lambda (X+\mathbb{A}^2)\right ) \nonumber \\
    &+S_\text{m}[g,\sfPsi_\text{m}]\,,
    \label{eqn:EAgeneralAction}
\end{align}
with $\kappa\equiv 8\pi G$,  $\mathbb{A}$ a constant and where the Lagrangian's $\mathcal{L}_i$ read
\begin{align}
    \mathcal{L}_2  =& \text{  } G_2(X)  \,, \nonumber \\
    \mathcal{L}_3  =& \text{  } G_3(X) \nabla_\mu A^\mu \,, \nonumber \\
    \mathcal{L}_4  =& \text{  } G_4(X)R- K\ud{\alpha\beta}{\mu\nu}\nabla_\alpha A^\mu\nabla_\beta A^\nu\,,
\end{align}
where
\begin{align}
    K\ud{\alpha\beta}{\mu\nu} =&\, c_1(X)\, g_{\alpha\beta}g_{\mu\nu}+c_2(X) \,\delta^\alpha_\mu\delta^\beta_\nu+c_3(X) \,\delta^\alpha_\nu\delta^\beta_\mu\nonumber\\
    &-c_4(X)\,A^\alpha A^\beta g_{\mu\nu}\,.
\end{align}
Here, the $G_i$ and $c_i$ are general functionals of the scalar value $X\equiv g_{\mu\nu}A^{\mu}A^{\nu}$.
Note that a term of the form $\hat G_4(X)A^\mu A^\nu R_{\mu\nu}$ can indeed always be absorbed into the definitions of the $c_2$ and $c_3$ through integrations by parts and commutations of the covariant derivatives. As discussed in the introduction, the rationale behind this metric theory is that even though Lorentz symmetry violations are practically ruled out in the matter sector, the same is not true in the gravity sector. Thus, by respecting the assumptions at the basis of metric theories of gravity \cite{poisson2014gravity,papantonopoulos2014modifications,Will:2018bme,YunesColemanMiller:2021lky,Zosso:2024xgy}, this theory captures a purely gravitational breaking of the local Lorentz symmetry via a nontrivial value of a nonminimal vector field that does not explicitly interact with matter.

However, due to the holonomic constraint $X=-\mathbb{A}^2$ in the Lagrange multiplier, it can be show that the equations of motion of Eq.~\eqref{eqn:EAgeneralAction} are equivalent to the action defined in Eq.~\eqref{eq:Action EA}, hence with $G_2=G_3=0$, $G_4=1$ and $c_i(X)=c_i$, where $c_i$ are constants (see also \cite{Will:2018bme}). Hence, up to two powers of derivatives, the E\AE{} action actually does not require an explicit nonminimal coupling between the \ae{}ther and the physical metric. This also implies that the energy-momentum tensor of the aether field $S_{\mu\nu}$ defined in Eq.~\eqref{eq:EAenergymomentum} is properly conserved $\nabla^\mu S_{\mu\nu}=0$ and from this perspective, the aether field behaves just as an additional matter field of the theory. In that sense, one could think that the gravitational nature of the aether field is merely reflected in the fact that it does not explicitly couple to ordinary matter so as to only introduce a Lorentz-breaking in the gravity sector. 

Yet, this perspective is not entirely accurate. This is because the clear gravitational aspect of the \ae{}ther is hidden within its special structure. Despite the absence of explicit nonminimal couplings, the presence of a nontrivial vector background, that is enforced through the Lorentz-breaking constraint, ensures that E\AE{} gravity still admits nontrivial gravitational polarizations as well as modified propagation speeds, which indicate a nontrivial mixing between the metric and aether degrees of freedom. 

Finally, we want to remark that from a fundamental theory building perspective, E\AE{} gravity is not the most general vector-tensor metric theory of its type. Indeed, it is possible to introduce higher-order derivative operators into the theory that would not alter the number of propagating degrees of freedom as it is the case for a Lorentz-preserving massive Proca theory \cite{Heisenberg:2014rta,Allys:2015sht,BeltranJimenez:2016rff}. In other words, E\AE{} gravity could be generalized through higher-order operators without introducing destabilizing ghost degrees of freedom, such that the resulting theory could still be considered as \emph{exact} \cite{Zosso:2024xgy}. And these higher-order derivative operators would reintroduce nonminimal couplings of the aether field, underpinning its gravitatonal nature. In fact, because of the constraint that ensures the nondynamicality of the temporal vector component, the allowed higher-order derivative operators are expected to go beyond the Generalized Proca interactions.

\section{Asymptotic plane waves and transverse projections}\label{App:TTProjection}
In Sec.~\ref{sSec:TensorMemoryGen}, we defined a \emph{TT} projection operator which is local in physical space. On the other hand, the TT part of the metric perturbation arising from the $3+1$ SVT decomposition in Sec.~\ref{sec:SVTdec} is local in momentum space and a priori the two definitions do not coincide \cite{Racz:2009nq,Frenkel:2014cra,Ashtekar:2017wgq}. In this appendix, we will however explicitly show that in the plane-wave limit the two notions are completely equivalent, regardless of the exact form of the propagation equation.

Let us first choose a source-centered inertial reference system $(t,\vec{x})$ in the asymptotic Minkowski background, as we have done throughout this work. For simplicity, we consider the leading-order $1/r$ asymptotic value of a spatial vector $\mathcal A_i(t,\vec{x})$ - we will comment on the case of a symmetric $h_{ij}$ tensor below - and we Helmholtz-decompose it into its longitudinal and transverse part
\begin{equation}
    \mathcal{A}_i=\mathcal{A}_i^\text{T}+\partial_i\mathcal{A}\,,
\end{equation}
where $\mathcal{A}$ is a scalar function of the coordinates and the transverse part satisfies the divergence-free condition $\partial^i\mathcal{A}^\text{T}_i=0$. Formally, the divergence-free part of the vector can be defined using a projector involving the inverse Laplace operator \cite{Jackson:1998nia, Racz:2009nq},
\begin{equation}
    P\du{i}{j}\equiv \delta\du{i}{j}-\partial_i\frac{1}{\nabla^2}\partial^j\,,
\end{equation}
such that
\begin{equation}\label{eq:TTnonlocal}
    \mathcal{A}_i^\text{T}\equiv P\du{i}{j}\mathcal{A}_j=\delta\du{i}{j}\mathcal{A}_j+\frac{1}{4\pi}\partial_i^x\int \frac{\partial_{x'}^j\mathcal{A}_j(\vec{x}')}{|\vec{x}-\vec{x}'|}d^3x\,.
\end{equation}
It appears clearly that this definition is not local in the physical space, while in Fourier space
\begin{equation}\label{eq:FourierSpace1}
    \mathcal{A}_i(t,\Vec{x})=\frac{1}{r}\int d^3k\,[\mathcal{a}_i(t,\Vec{k})e^{i\Vec{k}\cdot\Vec{x}}+c.c]\,,
\end{equation}
to leading-order in $1/r$ the divergence-free condition reduces to
\begin{equation}
    k^i\mathcal{a}^\text{T}_i=0\,,
\end{equation}
and the projection operator simplifies to
\begin{equation}\label{eq:projector1App}
    \tilde{P}\du{i}{j}\equiv \delta\du{i}{j}-\frac{k_ik^j}{k^2}\,.
\end{equation}

An alternative transverse projector, which is local in the physical space but not in the momentum space, is built by projecting the vector field onto the 2-sphere orthogonal to the radial direction $n_i=\partial_i r$ by defining an orthonormal spatial basis $\{n_i, u_i, v_i\}$ as defined in Sec.~\ref{sSec:polmetricth}. The corresponding projection operator then simply takes the form of the definition in Eq.~\eqref{eq:ProjectionT} back in Sec.~\ref{sSec:TensorMemoryGen}, as follows
\begin{equation}\label{eq:projector2App}
    \perp\du{i}{j}\equiv\,\delta\du{i}{j}-n_in^j\,.
\end{equation}
The spatially transverse part of the vector is then given by
\begin{equation}
    \mathcal{A}_j^T\,\equiv\,\perp\du{i}{j}\mathcal{A}_j\,.
\end{equation}

It is now straightforward to show that in the plane-wave limit of the asymptotic radiation the two notions of transversality coincide. This is because a radially outward wave sufficiently far away from a localized astrophysical source at each instant in time is solely characterized by a unique and well-defined direction of propagation $\vec{n}=\vec\nabla r$. This precisely defines the plane-wave limit. Importantly, for this statement to hold, we do not need to assume anything about the precise form of the asymptotic propagation equation and the correspondence is therefore also universally valid in any metric theory beyond GR. 

Let's therefore assume the plane-wave limit, in which, at a given instant in time, the asymptotic radiation is constant across the plane defined by the direction of travel $\vec{n}$. 
The Fourier space component in Eq.~\eqref{eq:FourierSpace1}, which generally depends on a magnitude $k\equiv|\vec{k}|$ and a direction $\hat{k}\equiv\vec{k}/k$, then naturally only depends on the single direction $\vec{n}$ imposed by the physical geometry \cite{maggiore2008gravitational}
\begin{equation}
    \mathcal{a}_i(t,\vec{k})=\delta^2(\hat{k}-\vec{n})\mathcal{a}_i(t,k)\,.
\end{equation}
Effectively, the two notions of directions both in physical and momentum space therefore coincide $\hat k=\vec n$ and the Fourier transform can be reduced to a one-dimensional integral over $k$ that can be related to a frequency component by considering the asymptotic propagation equation.
Comparing Eqs.~\eqref{eq:projector1App} and \eqref{eq:projector2App}, the equivalence between the two notions of transversality is then immediate. Concretely
\begin{align}
    \mathcal{A}^T_i&=\,\perp\du{i}{j}\mathcal{A}_j=\perp\du{i}{j}\mathcal{A}^\text{T}_j+\perp\du{i}{j}\partial_j\mathcal{A}\nonumber\\
    &=\mathcal{A}^\text{T}_i+\partial_i\mathcal{A}-n_in^j\partial_j\mathcal{A}=\mathcal{A}^\text{T}_i+kn_i\mathcal{A}-n_ik\mathcal{A}\nonumber\\
    &=\mathcal{A}^\text{T}_i\,.
\end{align}
We want to stress again that this equivalence holds regardless of the precise form of the propagation equations of motion.

To proceed, we can now assume a specific form of these asymptotic equations of motion, such as the nondispersive equations of the transverse components considered in this work, which in a specific gauge reads
\begin{equation}
    -\ddot{\mathcal{A}}^T_i+V_\mathcal{A}^2\partial^2\mathcal{A}^T_i=0\,,
\end{equation}
for some constant phase velocity $V_\mathcal{A}$. The general solution 
can then be written as
\begin{align}
    \mathcal{A}^T_i(t,\vec{x})=&\frac{1}{r} \int_0^\infty dk\,k^2\,[\mathcal{a}^T_i(k)e^{-i\omega t+ik\vec{n}\cdot\vec{x}}+c.c] \\
    =&\frac{1}{r} \int_0^\infty d\omega\,\frac{\omega^2}{V_\mathcal{A}^3}\,[\mathcal{a}^T_i(k)e^{-i\omega (t-r/V_\mathcal{A})}+c.c]
    \,,
\end{align}
where 
\begin{equation}
    \omega=V_\mathcal{A}\,k\,.
\end{equation}
For a superposition of plane waves, for each given angular direction $\Omega$ it suffices to consider each frequency mode separately, such that we can finally write
\begin{equation}
    \mathcal{A}^T_i(t,\vec{x})=\frac{1}{r}f_i^T(u_\mathcal{A},\Omega)\,,
\end{equation}
for some function $f$ with
\begin{equation}
    u_\mathcal{A}=t-\frac{r}{V_\mathcal{A}}\,.
\end{equation}

The same line of reasoning can be trivially extended to the case of a spatial, symmetric tensor $h_{ij}$, where a plane-wave limit assures the equivalence of a $TT$ projection along a given spatial direction $\vec{n}$ 
\begin{equation}
    h_{ij}^{TT}=\,\perp\du{ij}{k l} h_{kl}
\end{equation}
where [Eq.~\eqref{eq:Projector}]
\begin{align}\label{eq:ProjectorApp}
    \perp_{i j, k l} \,\equiv\, \perp_{i k} \perp_{j l}-\frac{1}{2} \perp_{i j} \perp_{k l}\,,
\end{align}
and the TT component within a $3+1$ Helmholz decomposition [Eq.~\eqref{eq:Decombmetric}]
\begin{equation}
h_{ij}=h_{ij}^\text{TT}+\partial_{(i}E^\text{T}_{j)}+\left(\partial_i\partial_j-\frac{1}{3}\delta_{ij}\partial^2\right)E+\frac{1}{3}\delta_{ij}D\,.
\end{equation}

\section{Causality in Lorentz-breaking theories}\label{App:Causality}

In its crudest form, causality is the principle that if an event at some time instant $t_1$ causes an effect on another event at $t_2$, then $t_1\leq t_2$. Hence, present events cannot influence the past and time ordering naturally distinguishes cause and effect. Combined with Lorentz invariance and the principle of relativity on the equivalence of inertial frames, the concept of causality results in the proposition that no cause may ``propagate'' faster than the speed of light. Indeed, due to the relativity of simultaneity in special relativity that allows for time-order reversal of spacetime separated events, an inertial observer-independent notion of causality can only be ensured if any effect at spacetime event $x_2$ lies within the future light cone of its cause at $x_1$. In that sense, superluminal propagation would violate causality. In metric theories of gravity that preserve local Lorentz symmetry, in principle the same notion of causality holds.

However, if the local Lorentz symmetry is broken, as it is the case in Einstein-\AE{}ther gravity, one may loosen the above consideration and actually allow for superluminal propagation. This is because by introducing a preferred frame of reference one can argue that since the principle of relativity is lost, causality ought only to hold in that specific preferred frame \cite{Jacobson:2005bg,Lin:2014ija}. In other words, we are back to the initial statement of causality attached to a single universal time variable $t$. It is however important to stress that the causal structure of E\AE{} theory is not precisely that of Newtonian global time. First of all, such a universal time coordinate $t$ is only guaranteed whenever the \ae{}ther field is hypersurface orthogonal. While this is the case in spherically symmetric backgrounds considered in this work, in E\AE{} theory it is not true in general \cite{Bhattacharyya:2015gwa}. Moreover, as shown in this work, for a unit timelike \ae{}ther field that introduces a preferred frame, the propagating degrees of freedom are still confined within fixed causal cones set by their respective propagation speeds, such that the fastest of the velocities can be used to define causality in a similar way as in GR. Only for theories with true \emph{preferred foliation} including mode excitations that are not contained within any causal cone, as for instance in Ho\v{r}ava-Lifshitz gravity \cite{Horava:2009uw}, the causal structure is truly Newtonian-like. Yet, the preferred frame of Einstein-\AE{}ther gravity still results in a departure from the causal structure associated to boosts with respect to the propagation speed of electromagnetic radiation. 

In physics, causality is in particular also incorporated in the idea of sourced equations such as the sourced wave equation in Eq.~\eqref{eq:MemEqGenFinal} with general solution Eq.~\eqref{eqn:SolForm}. The field value at a spacetime instant $\{t,\vec{x}\}$ is determined by the value of the source at all $\{t',\vec x'\}$, weighted by the retarded Green's function. First of all, choosing the retarded Green's function precisely imposes the time ordering $t>t'$. Moreover, in the case of a wave equation, the Green's function additionally accounts for the ``propagation'' of the source influence from $x'$ to $x$ by incorporating the appropriate retardation characterized by the finite propagation velocity of the field satisfying the wave equation. In the context of E\AE{} gravity, these considerations imply that any solution to the wave equation weighted by the appropriate retarded Green's function will not violate causality as defined within the preferred coordinate frame.

Yet, in the asymptotic limit, the null fields naturally depend on a different notion of time characterized by the asymptotic retarded time $u$. Coming back to the discussion below Eq.~\eqref{eq:PulseMem2} this leads to a different notion time ordering between time evolution of the source $u'$ and the asymptotic retarded time, which is precisely what is reversed if $\mathcal{V}<0$. In the context of an emission of individual energy-momentum pulses, the statement that in some directions we have $u'>u$ can readily be understood.

\begin{figure}[H]
  \centering
  \begin{subcaptiongroup}
    \centering
    \parbox[b]{.24\textwidth}{%
    \centering
    \includegraphics[scale=0.46]{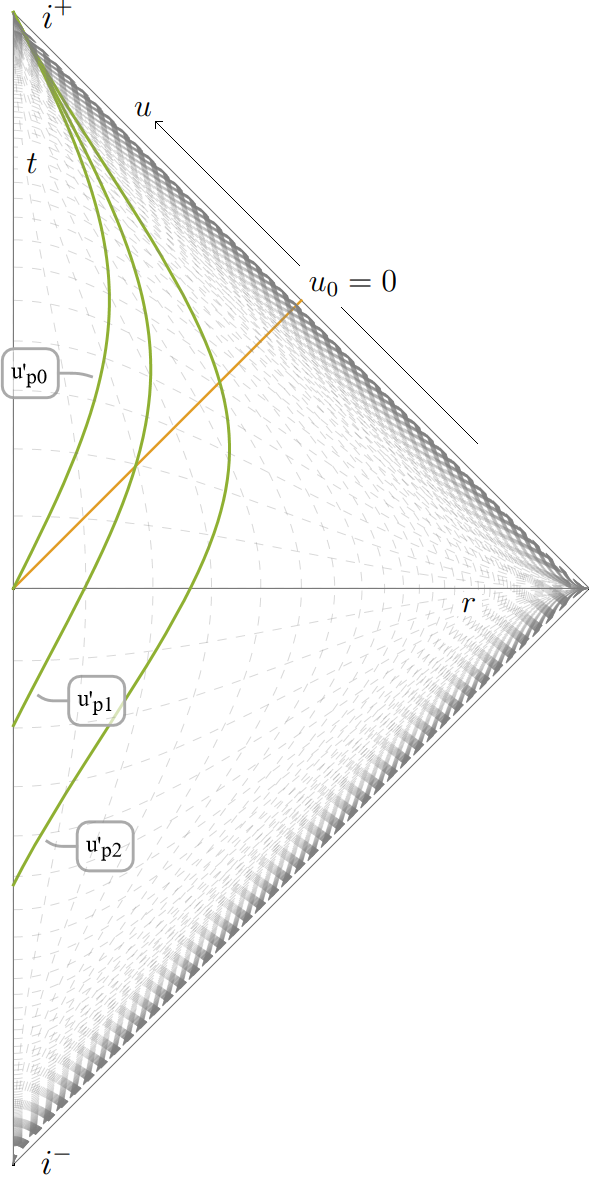}
    \caption{}
  }%
    \parbox[b]{.24\textwidth}{%
    \centering
    \includegraphics[scale=0.46]{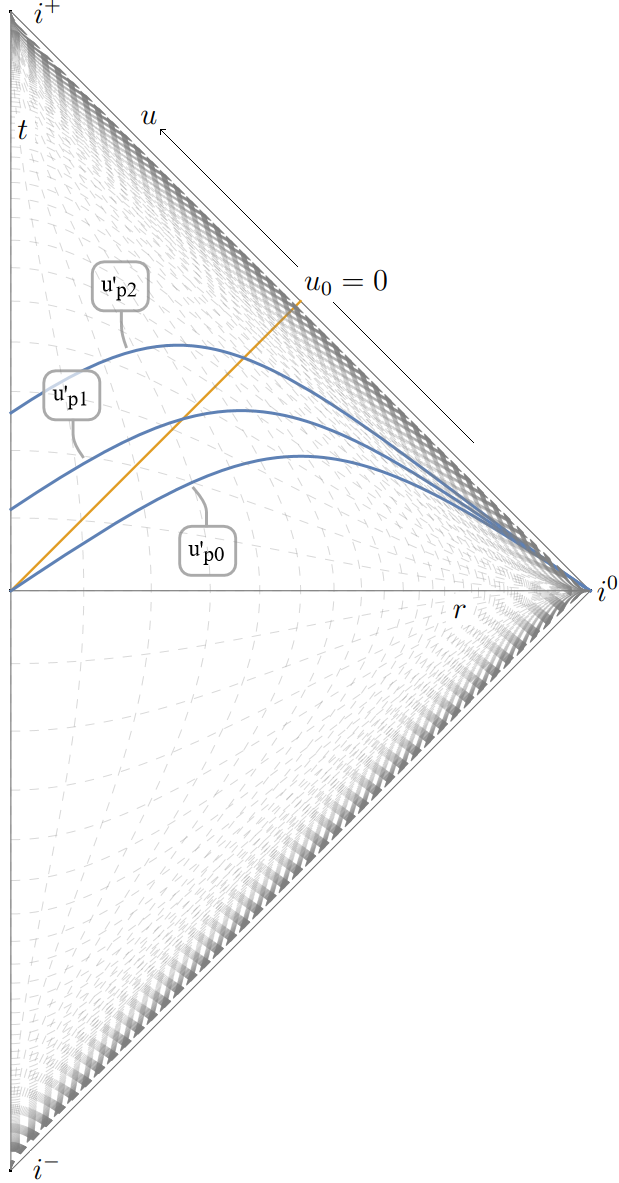}
    \caption{}
    }%
  \end{subcaptiongroup}
  \caption{Penrose diagrams of conformally compactified asymptotically flat spacetimes in the source-centered coordinates, in which the null direction is chosen to be the one set by the gravitational tensor modes that are assumed to propagate luminally $\beta_\text{T}=1$ with asymptotic retarded time $u\equiv t-r$. Shown are trajectories of radially outward pulses of scalar wave energy-momentum labeled by a constant asymptotic source retarded time $u'_\text{p}=t_0$ [Eq.~\eqref{eq:defup}] parameterizing their time of emission at the spatial coordinate origin. (a) For theory type (i) with subluminal group velocity $\beta_\text{S}<1$ only energy pulses (green) emitted at $u'_\text{p}\leq u_0$ will influence the memory at $u_0$. (b) On the other hand, for theory type (iii) with superluminal group velocity $\beta_\text{S}>1$ only energy pulses (blue) emitted at $u'_\text{p}\geq u_0$ will influence the memory at $u_0$. A luminal direction at $u_0=0$ is depicted in orange. Only the temporal and radial coordinates $\{t,r\}$ are represented and one may read these diagrams as depicting one particular angular direction in which the direction of the energy pulse $\vec{n}_\text{p}$ and the direction of evaluation of the memory $\vec{n}$ coincide.}\label{fig:penroseD}
\end{figure}

 For this purpose, let us first of all understand the statement of $u'_\text{p}<u$ in the subliminal case in which $\mathcal{V}^\text{p}_\text{S}>0$ (Case I) at all angles. For concreteness, we will fix the event at future null infinity given by $u_0=0$ and try to understand what type of asymptotic energy pulses influence the asymptotic region at this particular instance. 
This situation is depicted in a conformally compactified Penrose diagram of asymptotically flat spacetime in Fig.~\ref{fig:penroseD} (a). An energy pulse of scalar waves with group velocity $\beta_\text{S}<1$ sends out at a retarded source time of $u'_{\text{p}1}< 0$\footnote{Recall that the trajectories of the center of energy of such pulses follow trajectories that are characterized by a constant asymptotic ``source retarded time'' $u'_\text{S}=u'_\text{p}=t_0$ that parametrizes its emission time in terms of coordinate time $t$.} will trace out a curve in spacetime (green) of a subluminal particle asymptoting towards the temporal infinity $i^+$. Due to the retardation of the source to its effect, its influence of the memory at the point $u_0=0$ at null infinity is given at the instant the curve crosses the luminal line (orange line in Fig.~\ref{fig:penroseD}) that characterizes the past null cone of the event at $u=0$. Any pulse emitted at $u'_{\text{p}}\leq 0$ will cross this line and therefore influence the memory at $u_0=0$. The limiting case is of course given by $u'_{\text{p}0}=0$.

Conversely, in a superluminal theory $\beta_\text{S}>1$ in directions in which $\mathcal{V}^\text{p}_\text{S}<0$ (Case II), the conclusion is the direct opposite as depicted in the Penrose diagram of Fig.~\ref{fig:penroseD} (b). The trajectories of the superluminal energy pulses (blue) now reach spatial infinity $i^0$ within the given compactification, such that only pulses emitted at $u'_{\text{p}}\geq 0=u$ at the coordinate origin will cross the line of influence of the memory at $u=0$. 
This also implies that as the emitted pulse of energy is radially propagating outwards at increasing coordinate time $t'$, it will gradually influence the memory of decreasing asymptotic retarded time $u$.

Note that, as discussed above, this is not a violation of the principle of causality. Indeed, in terms of the global time $t$ in the preferred frame set by the \AE{}ther field, or more precisely, with respect to the causal structure set by the widest propagation cone of the theory, causality is never violated. A similar effect described above actually resembles very much a supersonic phenomenon known from mediums with subluminal propagation speeds. However, in the context of memory, such a reversal of time advancement between the source and its effect is unquestionably unique to Lorentz-violating theories.

\section{Energy-momentum tensor of localized pulses}\label{App:EMTensorGen}

In this appendix we derive a general expression for the energy-momentum tensor associated to a momentum four-covector $p^\text{p}_\mu(t)$ of an energy pulse or ``particle''. This is a generalization of the derivation in Ref.~\cite{Weinberg1972} to Lorentz-violating cases and recovers the general results for instance in Ref.~\cite{PhysRev.121.18}. For simplicity, we will restrict ourselves here to a flat Minkowski background metric $\eta_{\mu\nu}$.

The starting point is an energy pulse of energy $E_\text{p}$ and trajectory $\vec{x}_\text{p}(t)$ of its ``center of energy''. In order to assign a general momentum four-vector to this energy pulse, one needs to define a fundamental relation between the energy and momentum, which is provided by the relativistic equations of motion. Formally, given a high-frequency wave with Eikonal phase $S(x)$, whose constant surfaces define the characteristic wavefronts, one can define an associated momentum covector
\begin{equation}
    p_\mu\equiv \partial_\mu S(x)\,.
\end{equation}
In the canonical case of a plane wave $S(x)=-\omega t+\vec{k}\cdot\vec{x}$ with fixed spatial direction $k_i/|\vec{k}|=n_i$ the transition to momentum space implies
\begin{equation}
    p^\mu=(\omega,k^i)=\left(\omega,\frac{\omega}{V}n^i\right)\,,
\end{equation}
where in the last equality we have used the defining relation of the phase velocity
\begin{equation}
    V\equiv \frac{\omega}{|\vec{k}|}\,.
\end{equation}

Translated to the pulse of energy $E_\text{p}$ this implies that the general four-momentum reads
\begin{equation}
    p_\text{p}^\mu=(E_\text{p},\frac{E_\text{p}}{V_\text{p}}n_\text{p}^i)\,.
\end{equation}
where $V_\text{p}$ is the formal phase velocity. 
We therefore naturally define an associated energy-momentum density \cite{Weinberg1972}
\begin{equation}
    t_\text{p}^{\mu0}(x')\equiv  p_\text{p}^\mu\, \delta^{(3)}(\vec{x}'-\vec{x}_\text{p}(t'))\,.
\end{equation}
On the other hand, very generally, energy-momentum is carried at the spatial speed defined as the group velocity \cite{PhysRev.105.1129} which in this work we denote as $\beta_\text{p}^i=\beta_\text{p} \,n_\text{p}^i$. Hence, the energy-momentum current associated to $ p^\text{p}_\mu$ is naturally given by 
\begin{equation}
    t_\text{p}^{\mu i}(x')\equiv  p_\text{p}^\mu\,\beta_\text{p}^i\, \delta^{(3)}(\vec{x}'-\vec{x}_\text{p}(t'))\,.
\end{equation}
Defining the group four-velocity
\begin{equation}
    \beta_\text{p}^\mu \equiv (1,\beta_\text{p}^i)\,,
\end{equation}
one can combine these two definitions into the familiar concept of energy-momentum tensor
\begin{equation}\label{eq:defEmGen}
    t_\text{p}^{\mu\nu}(x')= p_\text{p}^\mu\,\beta_\text{p}^\nu\, \delta^{(3)}(\vec{x}'-\vec{x}_\text{p}(t'))\,,
\end{equation}
which is indeed a proper tensor \cite{Weinberg1972}. However, note that in this general form, the energy-momentum tensor is a priori not symmetric. This is in fact generally expected for Lorentz symmetry-breaking backgrounds \cite{PhysRev.121.18}.

In particular, in the case of a Lorentz-breaking via the presence of a preferred time direction while preserving spatial isotropy, leading to the possibility of nondispersive and nonluminal group velocities that are equivalent with the phase velocity $V_\text{p}=\beta_\text{p}$, as it is the case in E\AE{} gravity discussed in this work, the energy-momentum tensor in components of the energy density, energy current, momentum density and momentum current has the following form
\begin{align}
    t_\text{p}^{00}&=E_\text{p}\delta^{(3)}(\vec{x}'-\vec{x}_\text{p}(t'))\,,\\
    t_\text{p}^{0i}&=E_\text{p}\,\beta_\text{p}\,n^i\delta^{(3)}(\vec{x}'-\vec{x}_\text{p}(t'))\,,\\
     t_\text{p}^{i0}&=\frac{E_\text{p}}{\beta_\text{p}}\,n^i\delta^{(3)}(\vec{x}'-\vec{x}_\text{p}(t'))\,,\\
      t_\text{p}^{ij}&=E_\text{p}\,n^in^j\delta^{(3)}(\vec{x}'-\vec{x}_\text{p}(t'))\,.
\end{align}
In particular, note that we have the relation
\begin{equation}\label{eq:RelEMTApp}
    t_\text{p}^{ij}=t_\text{p}^{00}\,n^in^j\,.
\end{equation}
Moreover, for a radial outward trajectory of the pulse [Eq.~\eqref{eq:outwardTrajectory}]
\begin{equation}\label{eq:outwardTrajectoryApp}
    x^\text{p}_i(t')=\beta_\text{p}\, n_i\,(t'-t_0)\,.
\end{equation}
we have that
\begin{align}
   t_{ij}^\text{p}(u'_\text{p},r',\Omega')=\frac{1}{r'^2\beta_\text{S}}E_\text{p}\,\delta(t_0-u'_\text{p})\delta^2(\Omega'-\Omega_p)\,n^p_i n^p_j\,.\label{eq:spatialEMTpfApp}
\end{align}
where
\begin{equation}\label{eq:AsVApp}
    u'_\text{p}=t'-\frac{r'}{\beta_\text{p}}\,.
\end{equation}

It is interesting to compare these results to a massive but Lorentz-preserving wave with dispersion relation
\begin{equation}
    \omega(k)=\sqrt{k^2+m_\text{m}^2}=V_\text{m}(k)k\,.
\end{equation}
and corresponding group velocity
\begin{equation}\label{eq:SpeedMassiveDOF}
    \beta_\text{m}(k)\equiv\frac{d\omega}{dk}=\frac{k}{\sqrt{k^2+m_\text{p}^2}}=\frac{1}{V_\text{m}(k)}\,.
\end{equation}
one recovers the familiar totally symmetric form of the energy-momentum tensor
\begin{align}
    t_\text{m}^{00}&=E_\text{m}\delta^{(3)}(\vec{x}'-\vec{x}_\text{m}(t))\,,\\
    t_\text{m}^{0i}&= t_\text{m}^{i0}=E_\text{m}\,\beta_\text{m}\,n^i\delta^{(3)}(\vec{x}'-\vec{x}_\text{m}(t))\,,\\
      t_\text{m}^{ij}&=E_\text{m}\,\beta_\text{m}^2\,n^in^j\delta^{(3)}(\vec{x}'-\vec{x}_\text{m}(t))\,.
\end{align}
with the relation
\begin{equation}
    t_\text{m}^{ij}=\beta_\text{m}^2\,t_\text{m}^{00}\,n^in^j\,.
\end{equation}
Note that in this case, the phase velocity is not equal to the group velocity, but crucially, a pulse of energy still travels at the physical group velocity \cite{PhysRev.105.1129} as in Eq.~\eqref{eq:outwardTrajectoryApp}. This directly implies that also the energy-momentum fluxes
\begin{align}
   t_{ij}^\text{m}(u'_\text{p},r',\Omega')=\frac{1}{r'^2}\beta_\text{p}E_\text{m}\,\delta(t_0-u'_\text{m})\delta^2(\Omega'-\Omega_p)\,n^p_i n^p_j\,,\label{eq:spatialEMTpfAppM}
\end{align}
depend on the retarded time with respect to this group velocity [Eq.~\eqref{eq:AsVApp}]
\begin{equation}\label{eq:AsVApp2}
    u'_\text{m}=t'-\frac{r'}{\beta_\text{m}}\,.
\end{equation}

\section{Derivation of the gauge-invariant leading-order propagation equations}\label{Appendix:lineareq}
In this appendix, we comment on the leading-order high-frequency propagation equations. These can be derived from the 3+1 SVT decomposition of the metric and the field linear equations of motion. 

\subsubsection{\AE{}ther gauge-invariant linear equation of motion decomposition}
Let us first analyse the \ae{}ther vector field equation of motion. The temporal component reads
\begin{align}
   {}_{(1)}\mathcal{I}_0=c_{1234}(-\bar{A}\ddot{\Phi}+\ddot{\Omega})
   -c_1\nabla^2(-\bar{A}\Phi+\Omega) =0\,,
   \label{eq1}
\end{align}
while the longitudinal and transverse parts of the spatial component are
\begin{align}
   {}_{(1)}\mathcal{I}_i^L=& -2c_{14}(-\bar{A}\dot{\Phi}+\ddot{\Upsilon}) + \bar{A}(c_{123}+2c_2) \dot{\Theta} \nonumber \\
   &+2c_{123}\nabla^2\Upsilon=0 \,,
    \label{eq2}
\end{align}
\begin{align}
     {}_{(1)}\mathcal{I}_i^T=&-2c_{14}(\bar{A}^2\ddot{\Xi^a}+\bar{A}\ddot{\Sigma}^a)+\bar{A}^2 (c_1-c_3)\nabla^2 \Xi^a\nonumber \\
    &+2\bar{A}c_1 \nabla^2\Sigma^a=0 \,;
    \label{eq3}
\end{align}

\subsubsection{Metric gauge-invariant linear equation of motion decomposition}
As regards the metric equation of motion, the temporal-temporal component reads
\begin{align}
    {}_{(1)}\mathcal{G}_{00}&=\bar{A}c_{1234}(\ddot{\Omega}-\bar{A}\ddot{\Phi})-\bar{A} c_1 \nabla^2\Omega  +\bar{A}c_{14}\nabla^2\dot{\Upsilon}\nonumber\\
&-\bar{A}^4c_4\nabla^2\Phi+\nabla^2\Theta  =0\,,
    \label{eq4}
\end{align}
whereas the longitudinal and the transverse parts of the temporal-spatial component are respectively
\begin{align}
    {}_{(1)}\mathcal{G}_{0i}^L=&\bar{A}c_{14}(\partial_i \ddot{\Upsilon}-\bar{A}\partial_i\dot{\Phi})+\partial_i\dot{\Theta}=0\,,
    \label{eq5}
\end{align}
\begin{align}
    {}_{(1)}\mathcal{G}_{0i}^T=&\bar{A}c_{14}(\bar{A}\ddot{\Xi}_i+\ddot{\Sigma}_i) -\frac{1}{2}\bar{A}(c_1-c_3)\nabla^2\Sigma_i  \nonumber\\
    &-\frac{1}{2}(\bar{A}^2(c_1-c_3)-1)\nabla^2\Xi_i=0 \,,
    \label{eq6}
\end{align}
Finally, let us consider the transverse-traceless part of the spatial-spatial component of the metric equation of motion,
\begin{align}
   {}_{(1)}\mathcal{G}^{TT}_{ij}=-\frac{1}{2}(1-\bar{A}^2c_{13})\ddot{h}^{TT}_{ij}+\frac{1}{2}\nabla^2 h^{TT}_{ij}=0 \,,
   \label{eq7}
\end{align}
its longitudinal and transverse parts, 
\begin{align}{}_{(1)}\mathcal{G}_{ij}^L=&\left(1+\frac{1}{2}\bar{A}^2(c_{123}+2c_2)\right)\partial_i\partial_j\ddot{\Theta} \nonumber \\
&+\bar{A}c_{123}\nabla^2\partial_i\partial_j\dot{\Upsilon}=0  \,,
   \label{eq8}
\end{align}
\begin{align}
{}_{(1)}\mathcal{G}_{ij}^T&=\partial_{(i}\dot{\Xi}_{j)}+\bar{A}c_{13}\partial_{(i}\dot{\Sigma}_{j)}=0 \,.
   \label{eq10}
\end{align}
and the trace component
\begin{align}
     {}_{(1)}\mathcal{G}_{ij}\eta^{ij}=& 3\left(1+\frac{1}{2}\bar{A}^2c_{123}\right)\ddot{\Theta}+\bar{A}(c_{123}+2c_2)\nabla^2\dot{\Upsilon}\nonumber\\
    &+ 2\nabla^2\Phi-\nabla^2\Theta =0\,,
    \label{eq9}
\end{align}

It is possible to show that Eq. \eqref{eq6} can be derived from Eqs. \eqref{eq3} and \eqref{eq10}, and Eqs. \eqref{eq5} and \eqref{eq8} from Eqs. \eqref{eq1}, \eqref{eq2}, \eqref{eq4} and \eqref{eq9}. Therefore, the independent equations are Eqs. \eqref{eq1}, \eqref{eq2},\eqref{eq4},\eqref{eq6}, \eqref{eq9} and \eqref{eq10}.

\subsubsection{Tensor modes}
The tensor equation of motion reads
\begin{align}
   \frac{1}{2}(1-\bar{A}^2c_{13})\ddot{h}^{TT}_{ij}-\frac{1}{2}\nabla^2 h^{TT}_{ij}=0 \,.
   \label{TTeq}
\end{align}
 When $(1-\bar{A}^2c_{13})= 0$ the above equation is not dynamical, so we only consider the physically interesting case when $(1-\bar{A}^2c_{13}) \neq 0$.

\subsubsection{Vector modes}
The independent equations for the vector modes are the transverse part of ${}_{(1)}\mathcal{G}_{0i}$ and of ${}_{(1)}\mathcal{G}_{ij}$.
From Eq. \eqref{eq10}, it follows the relation 
\begin{align}
    \Xi_i = -\bar{A}c_{13} \Sigma_i\,.
\end{align}
We use this constraint equation to simplify Eq. \eqref{eq6}, which turns out to be
\begin{align}
    2(1-\bar{A}^2c_{13})c_{14}\ddot{\Sigma}_i -(2 c_1 -\bar{A}^2(c_1^2-c_3^2))\nabla^2\Sigma_i = 0 \,,
\end{align}
which is a propagation equation for the vector mode $\Sigma$ if $((1-\bar{A}^2c_{13})c_{14}) \neq 0$.

\subsubsection{Scalar modes}
The independent equations for the scalar modes come from the temporal part  ${}_{(1)}\mathcal{J}_0$ and the spatial longitudinal part ${}_{(1)}\mathcal{J}_i^L$ of the vector equation of motion, the temporal-temporal component of metric equation of motion ${}_{(1)}\mathcal{G}_{00}$ and the trace part of the metric equation of motion ${}_{(1)}\mathcal{G}_{ij}\eta^{ij}$.

From the difference between Eqs. \eqref{eq1} and \eqref{eq4}, we assume that $c_{14}\neq0$ and express $\Phi$ in terms of $\Theta$ and $\dot{\Upsilon}$,
\begin{equation}
    \Phi=-\frac{\Theta}{\bar{A}^2c_{14}}-\frac{\dot{\Upsilon}}{\bar{A}}.
\end{equation}
We replace this solution in Eqs. \eqref{eq1},\eqref{eq2} and \eqref{eq9}. For $c_{123}\neq0$ we obtain
\begin{equation}
    \Omega=\dot{\Upsilon}+\frac{\Theta}{\bar{A}c_{14}},
\end{equation}
\begin{equation}
    \nabla^2\Upsilon=-\frac{2+\bar{A}^2(c_{123}+2c_2)}{2\bar{A}c_{123}}\dot{\Theta}.
\end{equation}
Finally, we plug these solutions into Eqs. \eqref{eq9} to find the dynamical equation for $\Theta$,
\begin{align}
    &-\frac{(1-\bar{A}^2c_{13})(2+\bar{A}^2(c_{123}+2c_2))}{\bar{A}^2c_{123}}\ddot{\Theta}\nonumber\\
    &+\left(\frac{2}{\bar{A}^2c_{14}}-1\right)\nabla^2\Theta=0\,,
\end{align}
and the constraint equations in terms of $\Theta$ follow
\begin{subequations}
    \begin{align}
        \Phi&=-\frac{c_{13}+c_3-\bar{A}^2c_4}{2c_{14}(1-\bar{A}^2c_{13})} \Theta\, ,\\
        \Omega&=-\frac{\bar{A}(c_{13}+c_3-\bar{A}^2c_4)}{2c_{14}(1-\bar{A}^2c_{13})} \Theta\, ,\\
        \dot{\Upsilon}&=-\frac{2-\bar{A}^2c_{14}}{2\bar{A}c_{14}(1-\bar{A}^2c_{13})} \Theta.
    \end{align}
\end{subequations}
If furthermore we require that $(1-\bar{A}^2c_{13})\neq0$ and $(2+\bar{A}^2(c_{123}+2c_2))\neq0$, the propagating equation for $\Theta$ can be cast into a wave equation, as in Eq. \eqref{theq}.

\section{Comparison to existing literature}\label{Appendix:Literature}

Most of the literature in Einstein-\AE{}ther gravity carries out computations by explicitly choosing a specific gauge. In this appendix, we will demonstrate that we can recover the previous results by restricting our general gauge-invariant quantities, to those gauge conditions. We will specifically compare to the  results in Refs.~\cite{Saffer:2017ywl} and \cite{Foster:2006az}. 

In Ref.~\cite{Saffer:2017ywl}, the Helmholtz decomposition of the temporal-spatial part of the metric and the vector field perturbation (Eq.(95a) and (95b) in Ref.~\cite{Saffer:2017ywl}) have different notations, namely
\begin{subequations}\label{eq:matching1}
\begin{align}
    h_{00}&\equiv2S\,,\\
    B &\equiv \gamma \,, \\
    B_i^T &\equiv \gamma_i \,, \\
    a^0 &\equiv-\omega_0 \,, \\
    a &\equiv \nu \,, \\
    a_i^T &\equiv \nu_i \,.
\end{align}
\end{subequations}
The special-special part of the metric  perturbations is decomposed as 
\begin{align}
    h_{ij} = \phi_{ij}^{TT} +2 \partial_{(i}\phi_{j)}^T+\partial_i\partial_j\phi + \frac{1}{2}(\delta_{ij}\partial^2 f- \partial_i\partial_j f) \,,
\end{align}
assuming that the background metric is flat (Eq.(95c) in Ref.~\cite{Saffer:2017ywl}).
The correspondence between their notation and ours is as follows: 
\begin{subequations}\label{eq:matching2}
\begin{align}
    h_{ij}^{TT}&\equiv \phi_{ij}^{TT}\,,\\
    E_{j}^{T}&\equiv 2\phi_{j}^{T}\,,\\
    E&\equiv\phi-\frac{1}{2}f \,, \label{eq: E-phif}\\
    D&\equiv \partial^2 (\phi+f)\,.
\end{align}
\end{subequations}
Moreover, it is standard to define $F \equiv \partial^2 f$.

In this setting, the following gauge is chosen
\begin{equation}
    a=0\,,\quad B=0\,,\quad E_i^T=0\,.
\end{equation}
In this gauge, we see that our gauge-invariant variables reduces to 
\begin{subequations}\label{eq:gauge-fixing}
\begin{align}
    h_{ij}^{TT} &\Rightarrow  h_{ij}^{TT} \,,\\
    \Xi_i&\Rightarrow   B_i^T\,,\\
    \Sigma_i& \Rightarrow    a^T_i \,,\\
    \Phi&\Rightarrow   S +\frac{1}{2}\ddot{E}\,,\\
    \Theta&\Rightarrow   \frac{1}{3}(D-\partial^2E) \,,\\
    \Omega &\Rightarrow    a^{0} + \frac{1}{2}\bar{A}\ddot{E} \,, \label{eq:omgaugefix} \\
    \Upsilon &\Rightarrow    \frac{1}{2}\bar{A} \dot{E} \,. \label{eq:psigaugefix}
\end{align}
\end{subequations}
Together with the change of notation, we therefore arrive at the following dictionary to switch from the general gauge-invariant formulation to the popular gauge used in the literature in particular in Refs.~\cite{Saffer:2017ywl,Foster:2006az}
\begin{subequations}\label{eq:gauge-fixed}
\begin{align}
    h_{ij}^{TT} &\Rightarrow  \phi_{ij}^{TT} \,,\\
    \Xi_i&\Rightarrow  \gamma_i^T\,,\\
    \Sigma_i& \Rightarrow   \nu^T_i \,,\\
    \Phi&\Rightarrow \frac{1}{2}h_{00} +\frac{1}{2}\ddot{\phi}-\frac{1}{4}\ddot f\,,\\
    \Theta&\Rightarrow  \frac{1}{2}\partial^2 f = \frac{1}{2}F \,,\\
    \Omega &\Rightarrow   -\omega_{0} + \frac{1}{2}\bar{A}\ddot \phi -\frac{1}{4}\bar{A}\ddot f \,, \label{eq:omgaugefix2} \\
    \Upsilon &\Rightarrow   \frac{1}{2}\bar{A}\dot \phi -\frac{1}{4}\bar{A}\dot f \,.\label{eq:psigaugefix2}
\end{align}
\end{subequations}

\subsection{Equations of motion}

\emph{Constraint equation.}
If we directly vary the action with respect to the coefficient $\lambda$, we will obtain a constraint equation in terms of the gauge-invariant quantities
\begin{align}
    -2\bar{A} \Omega +2\bar{A}^2 \Phi=0\,,
\end{align}
which in the gauge in Eq.~\eqref{eq:gauge-fixed} reads
\begin{align}\label{eq:constraintgaugefix}
    2\bar{A}\omega_0+\bar{A}^2 h_{00}=0\,.
\end{align}
This equation agrees with the corresponding Eq.(96) in Ref.~\cite{Saffer:2017ywl} when choosing $\bar{A}=1$.

\emph{Tensor modes.}
Under this gauge, the linear tensor mode equation remains unchanged,
\begin{align}
    -(1-\bar{A}^2 c_{13})\ddot h_{ij}^{TT} +\,\partial^2h_{ij}^{TT}  = 0\,,
\end{align}
which agrees with Eq.(117) in Ref.~\cite{Saffer:2017ywl} and Eq.(41) in Ref.~\cite{Foster:2006az} while taking $\bar{A}^2=1$.

\emph{Vector modes.}
The linear equations for vector modes now become
\begin{subequations}
    \begin{align}
        B_i^T = -\bar{A}\,c_{13}\,a_i^T \label{eq:vector_relation}\,,\\
        -\frac{-2(1-\bar{A}^2 c_{13})c_{14}}{2c_1-\bar{A}^2(c_1^2-c_3^2)}\ddot{a}_i^T+\nabla^2 a_i^T=0\,.
    \end{align}
\end{subequations}
The first equation matches the vector mode relation Eq.(104) in Ref.~\cite{Saffer:2017ywl} while taking $\bar{A} = 1$.
The dynamical vector equation agrees with Eq.(119) in Ref.~\cite{Saffer:2017ywl} and Eq.(45) in Ref.~\cite{Foster:2006az} when $\bar{A}^2=1$.

\emph{Scalar modes.}
To compare the scalar equations, we first realize that combining the constraint equation Eq.\eqref{eq:constraintgaugefix} and Eq.\eqref{eq:omgaugefix2}, we have
\begin{align}\label{eq:constrainteq}
    \Omega = \bar{A} S + \frac{1}{2}\bar{A}\ddot{E}\,.
\end{align}
Therefore, we naturally have $\Omega = \bar{A} \Phi$, which corresponds to Eq.(96) in Ref.~\cite{Saffer:2017ywl} after gauge fixing, and we only need to compare the equations of motion for the other three scalar modes.

First, we see that
\begin{align}\label{eq:thF}
    \Theta = \frac{1}{3}(D-\partial^2 E) 
    = \frac{1}{2}\partial^2 f = \frac{1}{2}F \,,
\end{align}
so the dynamical equation [Eq.\eqref{theq}] is equivalent to 
\begin{align}
    -\frac{1}{V_S^2} \ddot F + \partial^2 F =0\,,
\end{align}
with 
\begin{align}
    \frac{1}{V_S^2} = \frac{(1-\bar{A}^2 c_{13})(2+\bar{A}^2(c_{123}+2c_2))c_{14}}{ c_{123}(2-\bar{A}^2 c_{14})}\,.
\end{align}
This agrees with Eq.(122) in Ref.~\cite{Saffer:2017ywl} and Eq.(51) in Ref.~\cite{Foster:2006az}. Moreover, since $F=\partial^2 f$, $f$ satisfies the same dynamical equation $-\frac{1}{V_S^2} \ddot f + \partial^2 f =0$. Equivalently,
\begin{align}\label{eq:F-ddtf}
    F = \partial^2 f = \frac{1}{V_S^2} \ddot f\,.
\end{align}

Next, we look at $\Phi$ and $\Upsilon$ modes. Observe that
\begin{align}
    S = \Phi - \frac{1}{\bar{A}}\dot\Upsilon \,.
\end{align}
Plugging in Eq.\eqref{eq:phith} and Eq.\eqref{eq:psith}, we have
\begin{align}
    S &= \left(\frac{2 c_{13} - c_{14}}{2(1-\bar{A}^2 c_{13})c_{14}}-\frac{2 -\bar{A}^2 c_{14}}{2\bar{A}^2(1-\bar{A}^2 c_{13})c_{14}}\right)\Theta \,\nonumber\\
    &=\frac{-1}{\bar{A}^2 c_{14}}\Theta \,.
\end{align}
Since $S = \frac{1}{2}h_{00}$ and $\Theta = \frac{1}{2}F$, we find 
\begin{align}
    h_{00}=\frac{-1}{\bar{A}^2 c_{14}}F\,,
\end{align}
which aligns with Eq.(121a) in \cite{Saffer:2017ywl} when setting $\bar{A}^2=1$.

Finally, the relationship between $\Upsilon$ and $\Theta$ reveals the last scalar relationship. Substituting Eq.\eqref{eq:psigaugefix}, Eq.\eqref{eq: E-phif} and Eq.\eqref{eq:thF} in Eq.\eqref{eq:psith}, we obtain
\begin{align}
    \frac{1}{2}\bar{A} (\ddot{\phi}-\frac{1}{2}\ddot{f}) = \frac{2-\bar{A}^2 c_{14}}{2\bar{A} c_{14}(-1+\bar{A}^2 c_{13})} \frac{1}{2}F \,.
\end{align}
Rearranging the equation and rewriting $F$ in terms of $f$ using Eq.\eqref{eq:F-ddtf}, we have
\begin{align}
    \ddot{\phi}&=\frac{1}{2}\ddot{f} + \frac{2-\bar{A}^2 c_{14}}{2\bar{A}^2 c_{14}(-1+\bar{A}^2 c_{13})} \frac{1}{V_S^2} \ddot{f} \nonumber\\
    &=- \frac{1+\bar{A}^2 c_2}{\bar{A}^2 c_{123}} \ddot f \,.
\end{align}
This is exactly Eq.(121b) in Ref.~\cite{Saffer:2017ywl} taking $\bar{A}^2=1$.

In summary, in choosing $\bar A=1$ our results align with the linear dynamical equations both in Refs.~\cite{Foster:2006az} and \cite{Saffer:2017ywl}. However, strictly speaking Ref.~\cite{Saffer:2017ywl} assumes a value of $\bar A=-1$, as defined in their Eq.(92), that would flip the signs of the constraint equations Eqs.~\eqref{eq:constraintgaugefix} and \eqref{eq:constrainteq} and the vector relation Eq.~\eqref{eq:vector_relation}. We could not find a convention shift that might explain this sign discrepancy, which, given our agreement with Ref.~\cite{Foster:2006az}, suggests a typographical error.

\subsection{Gravitational polarizations}

Expanding the polarization modes [Eq.~\eqref{eq:polarizations}] in terms of the SO(3)-invariant components [Eq.~\eqref{eq:ggaugepot}] of the perturbation fields, then switching the notation using Eq.~\eqref{eq:matching1} and ~\eqref{eq:matching2} we have
\begin{subequations}
\begin{align}
P_+  &=  \frac{e^{ij}_+}{2}E^{TT}_{ij} = \frac{e^{ij}_+}{2}\phi^{TT}_{ij} \,,\\
P_\times  &= \frac{e^{ij}_\times}{2}E^{TT}_{ij} = \frac{e^{ij}_+}{2}\phi^{TT}_{ij} \,,\\
 P_u  &=-   \frac{\bar{A}c_{13}}{\beta_V} (a_i^T +\frac{1}{2}\bar{A} E_i^T)u^i \nonumber\\
    &= -   \frac{\bar{A}c_{13}}{\beta_V} (\nu_i +\bar{A} \phi_i^T)u^i \,, \\
 P_v  &= -\frac{\bar{A}c_{13}}{\beta_V}  (a_i^T +\frac{1}{2}\bar{A} E_i^T)v^i \nonumber\\
    &= -\frac{\bar{A}c_{13}}{\beta_V}(\nu_i +\bar{A} \phi_i^T)v^i \,, \\
 \quad P_l  
    &= \left(1- \frac{2c_{13}-c_{14}}{\beta_S^2(1-\bar{A}^2c_{13})c_{14}}\right)P_b \\
 P_b& = \frac{1}{3}(D-\partial^2 E) =\frac{1}{2}F \,.
\end{align}
\end{subequations}
And we see that if we fix the gauge as in Eq.~\eqref{eq:gauge-fixing}, the polarization modes are 
\begin{subequations}
\begin{align}
P_+  &= \frac{e^{ij}_+}{2}\phi^{TT}_{ij} \,,\\
P_\times  & = \frac{e^{ij}_+}{2}\phi^{TT}_{ij} \,,\\
 P_u  &= -   \frac{\bar{A}c_{13}}{\beta_V} \nu_i u^i \,, \\
 P_v  &= -\frac{\bar{A}c_{13}}{\beta_V}\nu_i v^i \,, \\
 \quad P_l 
    &= \left(1- \frac{2c_{13}-c_{14}}{\beta_S^2(1-\bar{A}^2c_{13})c_{14}}\right)P_b \\
 P_b& = \frac{1}{3}(D-\partial^2 E) =\frac{1}{2}F \,.
\end{align}
\end{subequations}
The gauge-invariant and gauge-fixed polarization modes we calculated agree with Eq.(3.9) in Ref.~\cite{Schumacher:2023jxq} after taking $\bar{A}=1$.

\subsection{Asymtotic energy-momentum tensor}

To compare the self-stress-energy tensor, we perform gauge fixing on Eq.~\eqref{eq:self_stress_energy} using Eq.~\eqref{eq:gauge-fixed}, as follows
\begin{subequations}\label{eq:Momentum Flux app}
\begin{align}
    {}_{\mys{(2)}}t^\text{T}_{ij}&=\frac{1}{4\kappa_0} \,\Big\langle\partial_i \phi^{TT}_{ab}\partial_j \phi_{TT}^{ab}\Big\rangle=\frac{n_in_j}{4\kappa_0\beta_\text{T}^2}\Big\langle\dot \phi^{TT}_{ab}\dot\phi_{TT}^{ab}\Big\rangle\,,\\
    {}_{\mys{(2)}}t^\text{V}_{ij}&=\frac{\bar C_\text{V}}{4\kappa_0}  \Big\langle\partial_i \nu_{a}\partial_j \nu^{a}\Big\rangle=\frac{\bar C_\text{V}n_in_j}{4\kappa_0\beta_\text{V}^2}\Big\langle\dot \nu_{a}\dot \nu^{a}\Big\rangle\,,\\
     {}_{\mys{(2)}}t^\text{S}_{ij}&=\frac{\bar C_\text{S}}{4\kappa_0}  \frac{1}{4} \Big\langle\partial_i F \partial_j F \Big\rangle=\frac{\bar C_\text{S}n_in_j}{4\kappa_0\beta_\text{S}^2}\Big\langle\dot F \dot F \Big\rangle \,,
\end{align}
\end{subequations}
with coefficients
\begin{subequations}
\begin{align}
\bar C_\text{V}&=2(1-\bar{A}^2c_{13})(2c_1-\bar{A}^2 c_{13}(c_1-c_3))\,,\\
\bar C_\text{S}&=\frac{4-2\bar{A}^2c_{14}}{\bar{A}^2c_{14}}\,.
\end{align}
\end{subequations}
These expressions match with the spatial components of the results for instance in Eq. (113) of Ref.~\cite{Saffer:2017ywl} if $\bar{A}^2=1$. 

Using the relation in Eq.~\eqref{eq:RelEMTApp} we can furthermore relate these results to the total energy loss rate of the system by by multiplying with a velocity factor of $-\beta_\psi$ and integrating the expressions over all solid angles, as follows
\begin{subequations}
\begin{align}
    \dot{E}_\text{T}&=-\frac{1}{4\kappa_0\beta_\text{T}}\int d^2\Omega r^2\Big\langle\dot \phi^{TT}_{ab}\dot\phi_{TT}^{ab}\Big\rangle\,,\\
    \dot{E}_\text{V}&=-\frac{\bar C_\text{V}}{4\kappa_0\beta_\text{V}}\int d^2\Omega r^2\Big\langle\dot \nu_{a}\dot \nu^{a}\Big\rangle\,,\\
    \dot{E}_\text{S}&=-\frac{\bar C_\text{S}}{4\kappa_0\beta_\text{S}}\int d^2\Omega r^2\Big\langle\dot F \dot F \Big\rangle\,.
\end{align}
\end{subequations}
This agrees with Eq. (140) in Ref.~\cite{Saffer:2017ywl}  and Eq. (101) in Ref.~\cite{Foster:2006az}.
However, we do not agree with the total momentum loss of the system reported in Eq. (141) in Ref.~\cite{Saffer:2017ywl} which is simply given by integrating the momentum fluxes in Eqs.~\eqref{eq:Momentum Flux app} over all directions $dS^j=n^jd^2\Omega$, as follows
\begin{subequations}
\begin{align}
    \dot{P}_i^\text{T}&=-\frac{1}{4\kappa_0\beta_\text{T}^2}\int d^2\Omega \, n_i\, r^2\Big\langle\dot \phi^{TT}_{ab}\dot\phi_{TT}^{ab}\Big\rangle\,,\\
    \dot{P}_i^\text{V}&=-\frac{\bar C_\text{V}}{4\kappa_0\beta_\text{V}^2}\int d^2\Omega\, n_i\,r^2\Big\langle\dot \nu_{a}\dot \nu^{a}\Big\rangle\,,\\
    \dot{P}_i^\text{S}&=-\frac{\bar C_\text{S}}{4\kappa_0\beta_\text{S}^2}\int d^2\Omega\, n_i\, r^2\Big\langle\dot F \dot F \Big\rangle\,.
\end{align}
\end{subequations}
Indeed, in each expression above we differ by a factor of the velocities $\beta_\psi$. This discrepancy arises due to a typographical error in Eq. (49) in \cite{Saffer:2017ywl},\footnote{Confirmed by the corresponding authors.} which states an erroneous transformation of the temporal derivatives.

\section{Constraints on E\AE{} parameters}\label{App:Existing Constraints}
In Sec.~\ref{sSec:linearEOM}, we set some constraints on the E\AE{} parameters to obtain wavelike linear equations of motion for the tensor, vector and scalar modes, namely
\begin{subequations}
    \begin{align}
        \bar{A}^2c_{13}&\neq 1 \,,\\
        c_{14}&\neq 1 \,, \\
        c_{123}&\neq 0\,.
    \end{align}
\end{subequations}
Moreover, we required real radial velocities in each sector, which translates into additional assumptions on the couplings,
\begin{subequations}\label{eq:Vconstraints}
\begin{align}
    \bar{A}^2c_{13}&> 1 \,,\\
    \frac{c_1-\frac{1}{2}\bar{A}^2(c_1^2-c_3^2)}{c_{14}}&>0 \,,\\
    \frac{c_{123}(2-\bar{A}^2)c_{14}}{2+\bar{A}^2(c_{123}+2c_2)c_{14}}&>0\,.
\end{align}
\end{subequations}
\newline
On top of these assumptions, in this last appendix we wish to gather the most important current observational and theoretical constraints on the parameter space spanned by the coefficients of the Einstein-\AE{}ther theory. First of all, the requirement of having stable gravitational radiation is met by imposing the so-called \emph{stability conditions} \cite{Schumacher:2023cxh}
\begin{subequations}
\begin{align}
      \beta_\psi^2&>0\,, \label{eq:beta_constraints}\\
      c_1-c_3& \geq -\frac{c_{13}}{1-c_{13}}\,,\\
      0\leq c_{14}&\leq 2\, ,
\end{align}
\end{subequations}
where $\beta_\psi^2$ indicates the square speed of the gravitational tensor, vector and scalar polarizations, respectively $\psi=\text{T,V,S}$. We notice that Eq. \eqref{eq:beta_constraints} coincides with the requirements in Eqs. \eqref{eq:Vconstraints}. \newline
Moreover, as anticipated in Sec.~\ref{ssSec:Existing Constraints}, additional stringent bounds were placed by the observation of GWs from a neutron star binary merger, GW170817, and the simultaneous short gamma-ray burst, GRB170817A \cite{LIGOScientific:2017zic}. As explained in more details in the main text, such an event put an experimental constraint on the speed of the tensor polarization,
\begin{equation}
    -3\times 10^{-15}<\beta_\text{T} -1 < 7\times 10^{-16}\,.
\end{equation}
Following from the relation Eq.~\eqref{eq:velT}, the E\AE{} parameter $c_{13}$ is restricted as well,
\begin{equation}
    c_{13}\approx \mathcal{O}(10^{-15})\,.
\end{equation}
It is then reasonable to assume $c_{13}=0$, or equivalently that the tensor modes propagate luminally with $\beta_\text{T}=1$. 
\newline
Next, the observation of ultra-high-energy cosmic rays implies the absence of \emph{Gravi-Cherenkov radiation} within E\AE{} gravity \cite{Elliott_2005} that effectively constrains any subluminal propagation of the gravitational degrees of freedom. More precisely, the emission rate of gravitational Cherenkov radiation has been calculated in Ref.~\cite{Elliott_2005}: knowing that the highest energy cosmic rays should travel astronomical distances to reach us without substantial energy loss, which means at least $10$ Kpc $\sim 10^{36}\text{ GeV}^{-1} $, the following \emph{Cherenkov constraints} on the speed of the gravitational polarizations can be derived
\begin{subequations}
    \begin{align}
    \beta_\psi^2 \gtrsim 1-\mathcal{O}(10^{-15})\,,
    \end{align}
\end{subequations}
for $\psi=\text{T,V,S}$, so that a subluminal propagation speed of the degrees of freedom can effectively be ruled out.

Finally, along with the bounds just discussed, the observation of primordial ${}^4$He from big bang nucleosynthesis puts an important cosmological constraint on Einstein-\AE{}ther parameter space \cite{Carroll_2004}, as follows
\begin{subequations}
    \begin{align}
        c_{123}+2c_2+\frac{8c_{14}}{7}&\lesssim \frac{2}{7}\,,\\
        c_{123}+2c_2+\frac{8c_{14}}{9}&\gtrsim-\frac{2}{9}\,.
    \end{align}
\end{subequations}



\newpage
\bibliographystyle{utcaps}
\bibliography{references}

\clearpage

\end{document}